\documentclass[
 amsmath,
 amssymb,twocolumn,
 aps,prx]{revtex4-2}

\usepackage{graphicx}
\usepackage{dcolumn}
\usepackage{bm}
\usepackage{hyperref}
\usepackage{float}
\usepackage[normalem]{ulem}

\begin{document}

\preprint{APS/123-QED}

\title{Scalable photonic platform for real-time quantum reservoir computing}

\author{Jorge Garc{\'\i}a-Beni}
\author{Gian Luca Giorgi}
\author{Miguel C. Soriano}
\author{Roberta Zambrini}%
\email{roberta@ifisc.uib-csic.es}

\affiliation{%
 Instituto de F{\'\i}sica Interdisciplinar y Sistemas Complejos (IFISC), UIB–CSIC \\
 UIB Campus, Palma de Mallorca, E-07122, Spain
}%

\date{\today}
\begin{abstract}

Quantum Reservoir Computing (QRC) exploits the information processing capabilities of quantum systems to solve non-trivial temporal tasks,  improving over their classical counterparts. Recent progress has shown the potential of QRC exploiting the enlarged Hilbert space, but real-time processing and the achievement of a quantum advantage with efficient use of resources are prominent challenges towards viable experimental realizations. In this work, we propose a photonic platform suitable for real-time QRC based on a physical ensemble of reservoirs in the form of identical optical pulses recirculating through a closed loop. While ideal operation achieves maximum capacities, statistical noise is shown to undermine a quantum advantage. We propose a strategy to overcome this limitation and sustain the QRC performance when the size of the system is scaled up. The platform is conceived for experimental implementations to be viable with current technology.

\end{abstract}

\maketitle

\section{Introduction} \label{Intro}
Quantum photonic technologies are currently being exploited for quantum communications, quantum computing and quantum information processing \cite{doi:10.1063/1.5115814,Takeda2019}, for their speed-of-light propagation, ultrafast operations/gates and weak interactions with the environment even at room temperature.
In the context of computation, photonic quantum computing for boson sampling has shown
a time advantage of 14 orders of magnitude  over today's classical supercomputers \cite{Wang2019BosonSpace,Zhong2020QuantumPhotons}. In measurement-based one-way quantum computing in \textit{continuous variable} (CV) regimes cluster states up to one million modes have been reported using time multiplexing \cite{Yoshikawa2016} and reconfigurable cluster states achieved with frequency multiplexing \cite{Cai2017b}. 
Variational eigensolvers  implemented in photonic quantum processors \cite{Peruzzo2014} have achieved  much more efficient use of quantum resources than alternative algorithms, such as quantum phase estimation. 
Quantum states of light have also been used in Ising Machines, proving more efficient than current algorithms in certain scenarios \cite{McMahon2016, Honjo2021} and complementing the results obtained for alternative classical approaches \cite{pierangeli2019large,bohm2019poor}.
In this work, we propose a quantum photonic approach to time series processing designing and addressing the scalability, memory and performance for time-series prediction of an optical setup in the quantum reservoir computing (QRC) framework.

Reservoir computing (RC) is a paradigm of Machine Learning in which the information processing capabilities of dynamical systems are exploited for solving temporal tasks, with real-world examples such as the prediction of monthly electricity production \cite{WYFFELS20101958}, and financial \cite{LIN20097313,Ilies07steppingforward} or water level forecasting  \cite{RCbook,COULIBALY201076}. In addition, RC can also be applied to solve static tasks, such as the classification of phonemes \cite{NIPS2010_2ca65f58} or the detection of human finger movements from EEG data \cite{WANG2016237}. In practice, RC can solve these information processing tasks without the need for an external memory thanks to the fading memory present in the internal state of the reservoir itself \cite{VERSTRAETEN2007391}. Since RC exploits generic dynamical systems for computing, the concept of RC has been successfully transferred to physical substrates \cite{tanaka2019}, with the prominent example of high-speed photonic and optoelectronic implementations \cite{brunner2013parallel,vandoorne2014experimental,larger2017high,VanDerSande2017}. Ultimately, RC has been generalized to the quantum regime in order to benefit from the large number of degrees of freedom available in quantum systems \cite{Mujal2021,ghosh-RC-review}. In order to experimentally achieve time series processing with superior performance in quantum reservoir computing with respect to classical approaches, several challenges need to be addressed, identifying the most promising applications, efficient platform designs and dealing with quantum measurement retaining quantum advantage \cite{neurom_review_grollier, Mujal2021,time-series-QRC-measurements}.

With respect to classical reservoir computing, where single measurements on the reservoir produce the relevant information at the output layer, when moving into the quantum realm, one usually extracts the expectation values of observables at the output, from large ensembles of experiment copies. In the pioneering experimental implementation of quantum reservoir computing in \textit{noisy intermediate-scale quantum} (NISQ) platforms  \cite{PhysRevApplied.14.024065}, this was achieved by repeating the input processing sequence several times, being this an obstacle towards viable real-time technological realizations.
Our goal is to show a strategy to move to temporal signal processing, operating in a continuous way and without buffering inputs in external memories. The approach can be adapted to different photonic platforms and takes advantage of light-speed propagation and fast operation to monitor the reservoir processor. Optical sources producing high repetition pulses and optical fibers allow one to design an ensemble of identical reservoirs inside a closed loop, which removes the necessity for an external classical memory. The reservoir signal is continuously driven by the external inputs and monitored through a beam splitter and homodyne detection, thus obtaining expected values of observables without external buffers. We are restricting the analysis here to vacuum Gaussian states, as we know they can provide universal RC \cite{Nokkala2021}. These reservoir states can be engineered as complex networks in the frequency domain \cite{Araujo2014,Roslund2014,Cai2017b,NokkalaNJP}.

After introducing the photonic platform and main tools for quantum reservoir computing in section \ref{section-II},  we address its memory capabilities (Sect. \ref{section-III}) both in the ideal case of an infinite ensemble (Sect. \ref{results-ideal-case}) where statistical errors vanish  and in a realistic scenario of a finite ensemble assessing the limitations of statistical noise  (Sect. \ref{finite-pulses-case}). The analysis of noise detrimental effects in the resolution of past inputs allows  identification of strategies to improve the performance when the reservoir size is scaled up (Sect. \ref{strategy-section}). We also address the performance of our proposal for chaotic time-series prediction (Sect. \ref{prediction-section}).

\section{Photonic platform} \label{section-II}
\subsection{Platform description}
Before introducing the photonic setup for QRC, let us start by reminding the main features of reservoir computing for time series processing. RC schemes \cite{RCbook} consist of three main layers: the input, the reservoir, and the output. First, an input signal, typically belonging to a time series, is injected into the reservoir,  a dynamical system (often a recurrent neural network) that performs a complex nonlinear transformation to the injected data. Then, in the readout layer, a certain number of reservoir observables are measured and their combination  (usually linear) is optimized to match the desired target, depending on the temporal task, like e.g. linear memory or chaotic series prediction (see App. \ref{appendix-A} and \ref{appendix-B} for further details).  
\begin{figure*}[tb!]
    \centering
    \includegraphics[width=\linewidth]{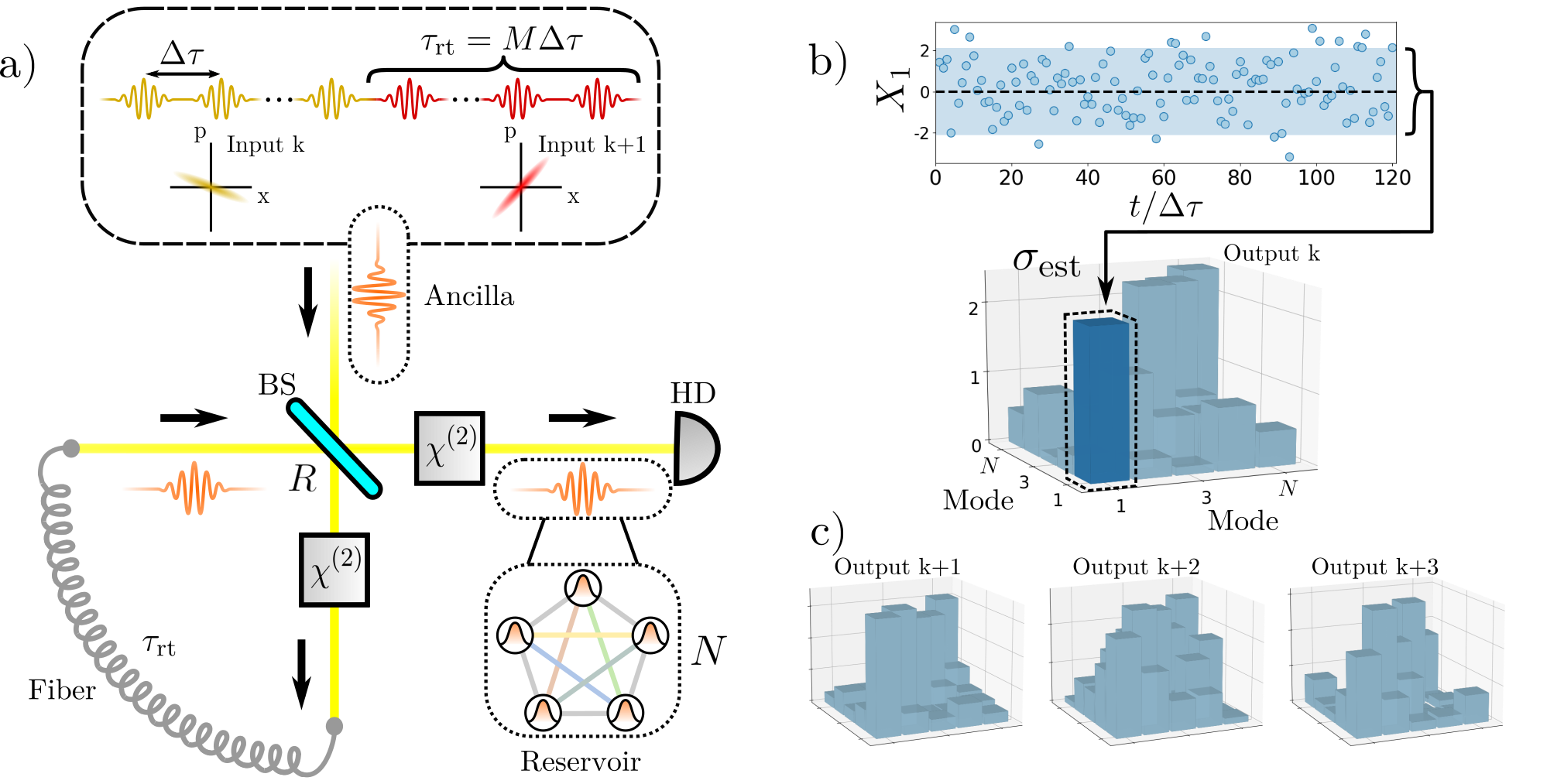}
    \caption{\textbf{Scheme of the process:} (a) Schematical drawing of the whole platform; (b) example of a series of the measured $x$-quadrature of the first mode for each pulse (horizontal axis) and a 3D representation of the estimated covariance matrix for the $k$-th round-trip (covariance matrix representations for the following round-trips are depicted below it). In this case we have taken $N = 5$, $M = 5000$ and $R = 0.75$. The horizontal axes represent the position of each term in $\sigma_{\text{est}}$.  (c) Sequence of estimated $\sigma_{\text{est}}$ for three consecutive inputs.}
    \label{figure-1}
\end{figure*}

The RC paradigm has been recently extended to quantum reservoirs \cite{Nakajima2017,Mujal2021, rodrigo_PRL, Yelin_PRX-Quantum} showing a quantum advantage due to the enlarged Hilbert space and therefore to the increased processing capability. Nevertheless, this approach also presents several challenges to be viable experimentally. In order to act as an online time series processor, the reservoir needs to continuously receive the input data and produce output extracted for the desired task. The relevant information for the readout layer is generally encoded in the expected values of the observables and due to the stochastic nature of quantum measurements, several detections are needed. Furthermore, the act of measuring yields back-action into the system that may also negatively affect the performance of the reservoir. Two strategies to overcome this last issue in qubits platforms have been recently proposed in Ref. \cite{time-series-QRC-measurements} using weak measurements and partial sequence repetition (rewinding). Otherwise one needs to restart the protocol for each input injection, buffering the input sequence as in Refs.  \cite{PhysRevApplied.14.024065}.  In the following, we propose a photonic platform design for QRC offering fast and easily scalable operation accounting for  measurements, in view of experimental implementation with state-of-the-art technologies.  

The proposed design is based on continuous variables and revolves around the use of optical pulses, whose individual dynamics along the feedback loop defines the reservoir computer (see scheme in Fig. \ref{figure-1}a). Indeed the reservoir is a traveling pulse, containing  $N$ modes whose interaction is mediated by a nonlinear ($\chi^{(2)}$) crystal and is modeled as a complex network (bottom inset in Fig. \ref{figure-1}a). 
The input is encoded in each ancilla pulse (see top inset in Fig. \ref{figure-1}a) coupled with the reservoir through a beam splitter (BS) and is prepared in a product state of $N$ Gaussian squeezed vacuum states. We will consider their squeezing angles as the classical inputs as this guarantees a good QRC performance \cite{Nokkala2021}; alternatively, the ancilla state could be directly treated as a quantum input itself to be processed, providing quantum (instead of classical) information \cite{johannes,Mujal2021}.  The reservoir pulse travels in a closed feedback loop, that can be enabled by an optical fiber. 
This feedback allows information from previous inputs to be retained, an essential requirement for temporal series processing.
At the other output arm of the BS, the reflected input  and transmitted reservoir signals are instead detected, after interacting with a nonlinear crystal. We are considering for this purpose a \textit{homodyne detector} (HD in \ref{figure-1}a), common in CV quantum optical experiments whose bandwidth continuous improvements allow for much faster detections \cite{Shaked2018,Takanashi:20}. 

A crucial feature of the setup presented in Fig. \ref{figure-1} is the possibility to delay several reservoir pulses enabling to run the experiment with $M$ copies of the pulses. This provides the needed ensemble to effectively realize quantum measurements without restarting the protocol by reinjecting the input sequence. Indeed, the same input is encoded in $M$ ancilla states before being updated (upper inset in Fig. \ref{figure-1}a) and  every pulse  interacts with a matching reservoir pulse at a given round-trip. This creates a physical ensemble of $M$ copies of the same reservoir in real-time, whose expected values are obtained via homodyne detection. 
High optical repetition rates and fast processing determine the potential of this scheme for sequential input processing while the signal is continuously injected. 
Beyond the reservoir size (the number of modes $N$ in each pulse) and the ensemble size ($M$ input repetitions), other crucial parameters determining the reservoir features are the BS reflectivity $R$ (we assume partial transmission, i.e. $R\neq0,1$). and  the time interval between pulses $\Delta \tau$ (the time interval that a whole round-trip lasts is $\tau_{\text{rt}} = M \Delta \tau$). At the $k$-th round-trip, the injected pulses are squeezed vacuum states that have the input $s_{k}$ encoded in their squeezing angle. We will focus on CV Gaussian states and then   model the effect of each element through the \textit{covariance matrix} of the external pulses (see App. \ref{appendix-C} for details). In particular, the BS couples the feedback and the external signal, generally entangling the feedback loop pulses with the ones traveling to the detector. The $\chi^{(2)}$ crystals induce squeezing and correlations between the modes in each pulse and are modeled by two (in principle different even if we omit a distinguishing label) Hamiltonians
$\hat{H}_{\chi^{(2)}} = \hat{H}_{\text{free}} + \hat{H}_{\text{int}}$ with 
\begin{align} 
        \hat{H}_{\text{free}} &= \sum_{i=1}^{N} \omega_{i} \hat{a}^{\dagger}_{i} \hat{a}_{i} \\ \label{hamiltonian}
    \hat{H}_{\text{int}} &= \sum_{j>i} \left(g_{ij}\hat{a}^{\dagger}_{i} \hat{a}_{j} + i h_{ij} \hat{a}^{\dagger}_{i} \hat{a}^{\dagger}_{j} + \text{h.c.} \right) 
\end{align}
where $\hat{a}_{i}$ ($\hat{a}_{i}^{\dagger}$) is the annihilation (creation) operator of mode $i$; $\omega_{i}$ is its frequency and parameters $g_{ij}$ and $h_{ij}$ depend on the second order non-linearities of the crystals and are not tuned but rather assumed to be random real numbers and different for each one of the two crystals. We assume that these crystals have a high enough bandwidth to keep the entering pulses independent of one another, and only couple the modes inside each pulse. The time each pulse spends inside a $\chi^{(2)}$ material is labeled as $\Delta t$ (equal for both crystals). 

For the measurement scheme we propose multimode homodyne detection: we measure, at the same time, the same quadrature operator for every mode. That is, for every incoming pulse the detector measures all $x$-quadratures $
    \hat{x}_{i} = \hat{a}_{i} + \hat{a}_{i}^{\dagger} \quad \left(i = 1, \dots, N \right) \ $, obtaining a set of stochastic outcomes, $X_{i}$, defining a quantum trajectory \cite{Wiseman_Milburn_quadrature,Wiseman_Milburn_homodyne}. %
Thus, the measurement output of the $m$-th pulse at the $k$-th round-trip is given by the vector $\mathbf{X}^{(k,m)} = ( X_{1}^{(k,m)}, \dots, X_{N}^{(k,m)} )^{\top}$ and, from these quadratures measurement, we can access the expected values of the covariance matrix, fully characterizing the squeezed vacuum states at the output for our readout layer. That is, for each round-trip (and thus, each input in the sequence) we extract $N(N+1)/2$ outputs for the readout layer  
$   
O_{ij}^{(k)} = \langle X_{i}^{(k)} X_{j}^{(k)} \rangle_{M} - \langle X_{i}^{(k)} \rangle_{M} \langle X_{j}^{(k)} \rangle_{M}$ with $j \geq i = 1, \dots, N \quad$ and with $\langle \cdot \rangle_{M} $ corresponding to the ensemble average over $M$ pulses.
These outputs give the $estimated$ covariance matrix $\left[ \sigma_{\text{est}}^{(k)} \right]_{ij} = O_{ij}^{(k)}$ at each $k$-th round-trip, in which the input $s_{k}$ is injected.
While here we will restrict to the $N$ position quadrature (and homodyne detection), one could also consider momenta (i.e. $2N$ quadratures) through heterodyne  detection. Higher-order moments could also be considered as features for the readout layer to improve the performance in some temporal tasks (see Sect. \ref{prediction-section}). For Gaussian states, these higher-order moments are functions of the first- and second-order ones. Here we restrict our analysis to the covariance matrix for the linear and nonlinear memory.
In Fig. \ref{figure-1}b, a visual representation of this readout protocol is shown. The outcome measurements of the $M$ sequence of $X_{1}^{(k,m)}$ at an unspecified step $k$ are shown and the black dashed line is the mean, vanishing because we are working with vacuum states. The (blue shaded) standard deviation of $X_{1}$ provides $\left[ \sigma_{\text{est}} \right]_{1, 1}$ shown below, and, considering all $N$ modes, one estimates the covariance matrix.  The covariance matrix at each time step $k$ constitutes the output layer (Fig. \ref{figure-1}c),  optimized by a simple linear regression (see App. \ref{appendix-A}) to achieve the best performance in the desired task. We notice that a measurement in one of the beam splitter outputs generally influences the conditional state of the other beam (being these generally entangled), but we will consider here only ensemble-averaged quantities for Gaussian states when these expected values correspond to the unconditional quantities.  

The design proposed in Fig. \ref{figure-1}(a) is reminiscent of other photonic approaches for RC based on time-delay feedback and time-multiplexing \cite{brunner2013parallel,VanDerSande2017}. In analogy to these works, the feedback loop provides the mechanism for fading memory in the optical implementation. The time-multiplexing  instead has a different role: in our platform it has the distinctive function to provide the ensemble for quantum measurements of the covariance matrix, while in classical approaches it serves as a way to increase the dimensionality of the reservoir. The reservoir size in the current proposal depends instead on the number of (frequency or spatial) modes contained in each temporal pulse.

\subsection{Feasibility of the photonic design} \label{feasibility}

In this section, we will discuss the viability of experimental implementations of our proposal with state-of-the-art technology. The main aspects of interest are the engineering of the input ancilla states, the generation of complex networks using $\chi^{(2)}$ nonlinearities, the multimode detection of the output pulses (measuring all the modes at once) and the losses due to the fiber propagation.

\textit{Input states generation:} in the frequency domain, squeezed states can be deterministically generated at high rates for a wide spectrum band. Specifically, over $10^{8}$ pulses per second containing up to 21 squeezed spectral modes each have been recently attained \cite{Kouadou2022}. The ability to encode the inputs in the squeezing phase of each pulse is also an important \textit{requirement} for the feasibility of our platform. Highly accurate and versatile phase setting devices have been demonstrated using programmable phase shifters in Xanadu's Borealis platform \cite{Madsen2022}. Accurately changing the squeezing levels and phases of pulses at time intervals below 100 ns has also been attained recently by continuously modulating the pump light of a waveguide optical parametric amplifier \cite{Tomoda2022}.

\textit{Optical networks:} mode couplings similar to the ones shown in Eq. \eqref{hamiltonian} are experimentally attainable in the frequency domain \cite{Chen2013,Araujo2014,Cai2017b,Roslund2014,Kouadou2022}. Reconfigurability in the spectral network shapes can also be reached by changing the measurement basis of the homodyne detection \cite{Cai2017b,Kouadou2022}.

\textit{Homodyne detection:} being able to measure several frequency modes at once is another key requirement for the real-time processing capabilities of our platform. Multimode homodyne detection of several frequency bands has been demonstrated in experimental setups \cite{Plick2018,Cai2021}.

\textit{Fiber losses:} retaining the ensemble pulses inside a long optical fiber without great losses is another relevant aspect to tackle for the experimental implementation of our real-time processing proposal. 
Low-loss fiber delay lines have been successfully implemented in several CV optical setups \cite{Yoshikawa2016,McMahon2016,Honjo2021,Larsen2019,Madsen2022}, ranging from single pulse delay lines \cite{Yoshikawa2016,Larsen2019} to over $10^{5}$ recirculating pulses in a 5km fiber \cite{Honjo2021}. So fiber losses do not constrain the viability of our platform with state-of-the-art technology.

In summary, the key aspects of our suggested platform have already been demonstrated experimentally. The main challenge ahead is to combine them in a common realization.

\section{Results} \label{section-III}
In this section, we will present a detailed analysis of the information processing capabilities of the quantum photonic platform for time series processing. RC systems require the ability to retain the memory of previous signal inputs, the capacity to reproduce nonlinear functions of the inputs and high dimensionality, which requires that the readout observables are linearly independent functions of the inputs. For the evaluation of the reservoir memory, we numerically compute the \textit{linear capacity} of the photonic QRC, which measures how accurately can the system reproduce inputs in the past. It is a normalized output-target correlation that ranges from 0 to 1. The higher the value of the linear capacity, the better the target input is reproduced by the reservoir.
To test both the nonlinearity as well as the high dimensionality, we will make use of the total \textit{Information Processing Capacity} (IPC) \cite{dambre2012information}, which generalizes the linear capacity memory to nonlinear contributions (see App. \ref{appendix-B} for a detailed explanation). It quantifies the expressivity of a wide variety of dynamical systems and has recently been used in quantum settings \cite{Martinez2020,Nokkala2021}.
The IPC ranges from 0 to the number of output observables (upper bound). When this upper bound is reached, it means that all the readout observables are linearly independent. In our case, as we are using the covariance matrix of the measured $x$-quadratures, the number of observables ($\langle\hat x_i\hat x_j\rangle$) is equal to $N(N+1)/2$ ($N$ denoting the reservoir size, or the number of modes in each pulse).

In the following, we will consider the  memory capacity of the proposed photonic QRC both in ideal conditions and when explicitly accounting for statistical noise. In particular, we will consider the $N$ mode frequencies to be equal and scale time so that $\omega_{i} = 1 \quad \forall i$ in Eq. \eqref{hamiltonian}. %
 This is a reasonable approximation when considering modes in the frequency domain, as usually in experiments with frequency combs the detuning is much smaller than the coupling strengths between modes \cite{Chen2013,Cai2017b}. %
The coupling parameters $g_{ij}$ and $h_{ij}$ have been chosen from a uniform distribution in the intervals $\left[ \langle g \rangle - \Delta g, \langle g \rangle + \Delta g \right]$ and $\left[ \langle h \rangle - \Delta h, \langle h \rangle + \Delta h \right]$, respectively, where $\langle g \rangle = 0.2$, $\langle h \rangle = 0.3$ and $\Delta g = \Delta h = 0.1$, and the interaction time inside each crystal is set to one. Inputs $s_{k}$ are encoded in the ancilla as squeezing angles, providing a nonlinear input encoding \cite{Nokkala2021}; each one of the $M$ pulses in the ancilla train is a squeezed vacuum state with covariance matrix equal to $\sigma_{\text{anc}}^{(k)} = \bigoplus_{i=1}^{N} \sigma_{\text{sq}}^{(k)}$, where $\sigma_{\text{sq}}^{(k)}$ is the covariance matrix of a single mode squeezed state. In particular, squeezing strength is set to $r_{k} = 1$ while the squeezing angle encoding is $\phi_{k} = 3\pi s_{k}/4$ (further details in App. \ref{appendix-C}). It was recently reported how the encoding choice affects the degree distribution of the IPC \cite{Nokkala2021} and our choice provides a balanced contribution of the linear terms and the nonlinear ones. Different angle encodings would provide different linear to nonlinear contributions (see App. \ref{app-encoding}). Inputs and parameters, even when random, are set to lead to squeezing levels currently viable with state-of-the-art technology, which is around 15 dB \cite{PhysRevLett.117.110801}, equivalent to  $r_{k} \simeq 1.7$.

\subsection{Ideal case} \label{results-ideal-case}
In the ideal case of an infinite number of pulses, the covariance can be obtained as the limit
\begin{equation} \label{covariance matrix-ideal-limit}
    \sigma_{\text{ideal}}^{(k)} = \lim_{M \to \infty} \sigma_{\text{est}}^{(k)}.
\end{equation}
While not experimentally attainable,  this ideal covariance matrix sets the performance of our reservoir in the absence of statistical noise, providing an important insight to quantify the effect of this noise in realistic (finite ensemble) scenarios.  %
Furthermore, the ideal performances we will present here can be compared with most results in QRC in the literature \cite{Mujal2021}, generally not accounting for experimental limitations.

As a consequence of the linearity of our photonic  platform, by inspecting the inputs dependence of $\sigma_{\text{ideal}}$ (see App. \ref{appendix-D}), we observe that inputs at different times contribute additively to the covariance matrix, so we can decompose the covariance as
\begin{equation} \label{ideal-decomposition}
    \sigma_{\text{ideal}}^{(k)} = \sum_{d=0}^{\infty} \gamma_{d}^{(k)},
\end{equation}
where the matrices $\gamma_{d}^{(k)}$ are functions of a single input with delay $d$. Hence, if we are in the $k$-th round trip, the covariance is a sum of nonlinear functions $\gamma_{d}^{(k)}$ of the previous inputs $s_{k-d}$ and  $\gamma_{d}^{(k)}: \Re \rightarrow \Re^N\times\Re^N$ turns each input into an $N \times N$ positive symmetric matrix ($N$ denoting the size of the reservoir). 
The reflectivity $R$ of the BS is found to be the physical parameter determining the relative size of these terms (see App. \ref{appendix-D})
\begin{equation} \label{eq-gamma-decay}
    \gamma_{d}^{(k)} \propto (1-R)^{2} R^{d-1}, \quad \text{for } d \geq 1.
\end{equation}
with an exponential decay $R^{d-1}$ in the delay for the average magnitude of the observables. %
Indeed, each time a feedback pulse gets reflected by the BS in the fiber loop, it is  scaled by $R$. If the signal recirculating in the loop is weak (small reflectivity $R$),
a much faster decay in the magnitude of delayed matrices $\gamma_{d}^{(k)}$ is observed (when the delay $d$ increases). The BS reflectivity is then expected to have a strong influence on the fading memory in this setup.
As for the term $(1-R)^{2}$ in Eq. (\ref{eq-gamma-decay}), it is due to the repeated transmission through the BS of each delayed term: first as an  ancilla signal when it enters the loop and then when it leaves the loop towards the detector.
Then, on the one hand, increasing the value of the reflectivity $R$ reduces the amount of feedback light reaching the detector; on the other hand, it also increases the memory retention of the reservoir, as the magnitude decay of the delayed matrices in Eq. \eqref{ideal-decomposition} is slower.

\begin{figure}[h!]
    \centering
    \includegraphics[width=\linewidth]{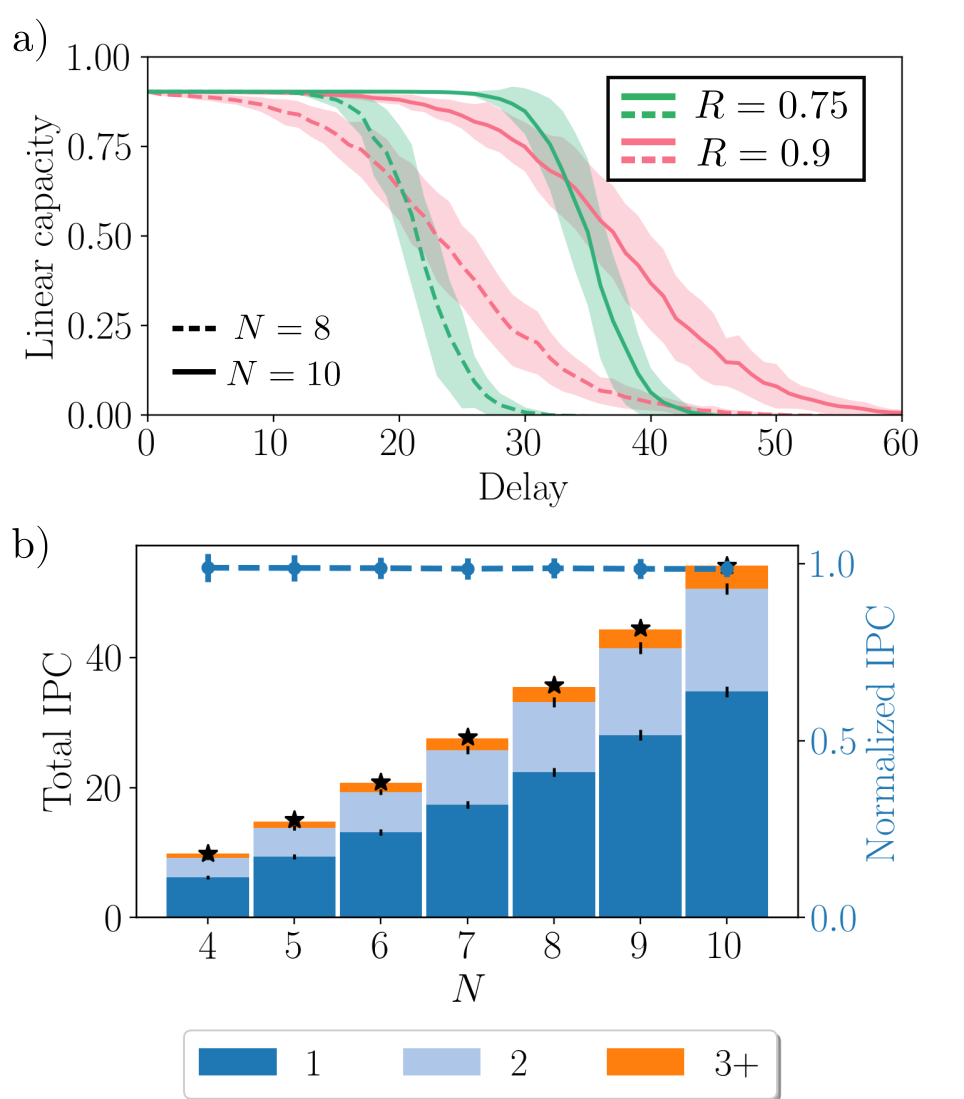}
    \caption{\textbf{Capacity in the ideal case:} (a) Linear capacity as a function of the delay of the input for different values of the BS reflectivity $R$ (color) and different values of the reservoir size $N$ (line); (b) IPC as a function of $N$ for $R = 0.9$. The bars follow the vertical axis at the left, which shows the total absolute capacity. Each bar is also split into different degree contributions. The discontinuous blue line follows the vertical axis at the right, which depicts the normalized IPC. Black stars follow the left axis and correspond to the value of the total IPC for $R = 0.75$, as a function of $N$. Every plot is the result of averaging over 100 realizations with different random networks in the coupling crystals and different input strings (this average has been performed in every figure of the manuscript).}
    \label{figure-2}
\end{figure}

We now present the numerical results for the performance of the QRC in the ideal limit, starting with the  \textit{linear capacity}, which %
brings information about the ability of our system to reproduce the encoded inputs at different times in the past (linear memory). 
In Fig. \ref{figure-2}a, the linear capacity is shown as a function of the delay, $d$, of the target input. By increasing the number of modes (from $N = 8$ to $N = 10$) and consequently the size of the output layer, the system can reproduce more delayed inputs, extending the memory.
In contrast, the capacity for low delays does not change. Interestingly, tuning the BS reflectivity $R$ alters the shape of the linear capacity curve. Consistently with Eq. \eqref{eq-gamma-decay},
for higher values of the BS reflectivity $R$, further terms into the past can be resolved, although the curve starts to descend earlier from its maximum value, achieved with smaller reflectivity.

The total IPC for different values of the reservoir size $N$ is shown in Fig. \ref{figure-2}b as a bar plot, where each bar is split into different  linear and  nonlinear memory contributions (identified by their polynomial degree). 
Concretely, the linear contributions (blue bars in Fig. \ref{figure-2}b) correspond to areas below the lines in Fig. \ref{figure-2}a. %
The dashed blue line -representing the normalized capacity equal to one- shows that the capacity saturates its maximum theoretical value $N(N+1)/2$  for every $N$, corresponding to the number of output degrees of freedom (terms of the covariance matrix).
Actually, our QRC platform is found to display both fading memory and echo state property, which are required for good RC performance (see proof in App. \ref{RC-conditions-proof}). This explains why the normalized IPC is saturated, as the readout observables are linearly independent and the above properties are fulfilled \cite{dambre2012information}.
Furthermore, this occurs for every value of the reflectivity $R$ that has been tested (e.g. we show the total IPC for both $R=0.9$ and $R = 0.75$ in Fig. \ref{figure-2}b). The degree distribution shown in Fig. \ref{figure-2}b, displaying significant contributions up to cubic degree, also does not change significantly with $R$,
being mainly determined by the encoding choice (see App. \ref{app-encoding} for details).

\subsection{Finite measurement ensemble} \label{finite-pulses-case}

In this subsection, a more realistic scenario is considered, when the physical ensemble is limited by the finite number of pulses, $M$, inside the loop. Then, any estimation of the observables is affected by statistical errors, whose average magnitude depends on  $M$. Additive noise in the readout layer has been considered in Gaussian CV models for RC with coherent states \cite{9525045}, and for non-temporal tasks \cite{PhysRevResearch.3.013077}. Here we explore statistical noise due to a physical ensemble of reservoirs for temporal tasks and for QRC with squeezed vacuum states. Hence the performance based on estimated covariance elements will be quantified and used to determine how the resources scale in this platform when the reservoir size increases. 
\begin{figure}[h!]
    \centering
    \includegraphics[width=\linewidth]{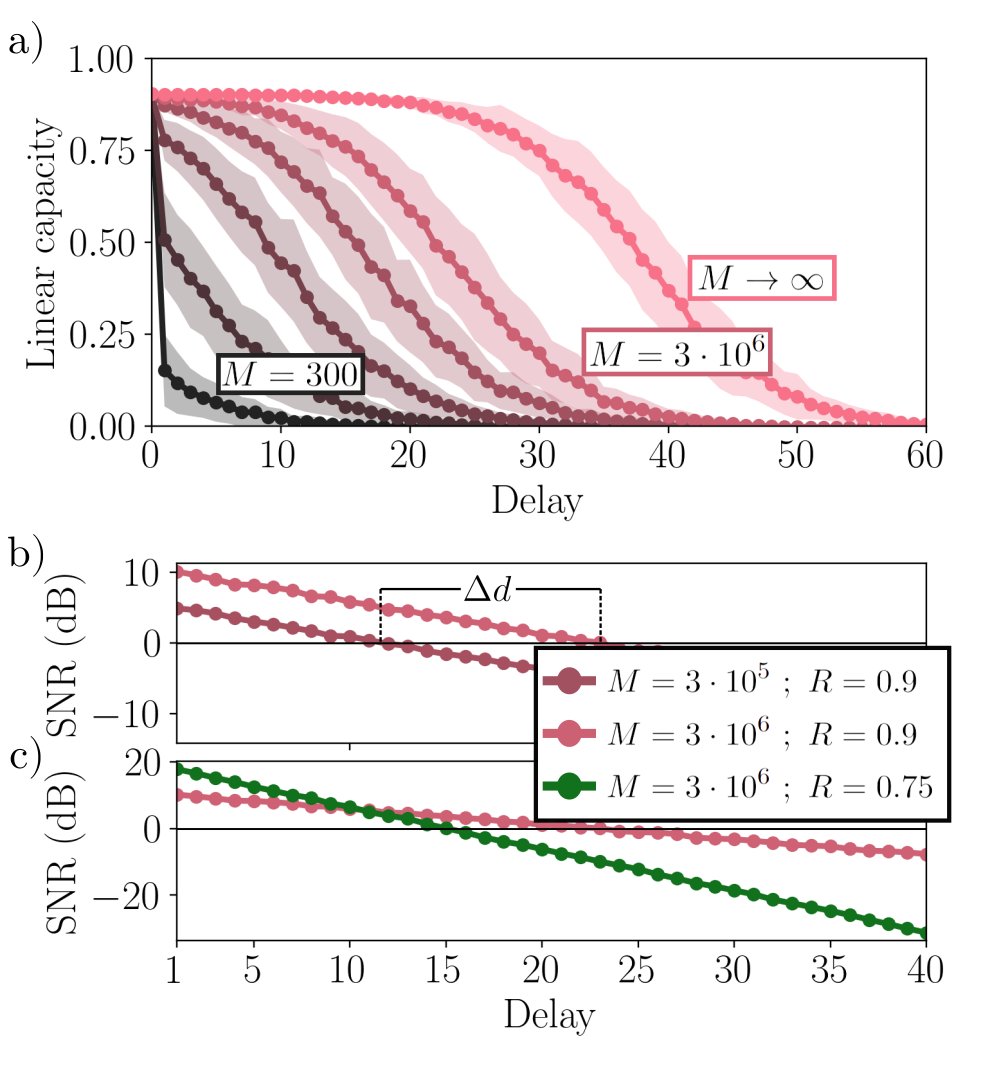}
    \caption{(a) Linear capacity as a function of the delay for different values of the ensemble size $M$ ranging in powers of 10 from $300$ to $3 \cdot 10^{6}$, as well as the curve from the ideal case ($M \to \infty$). In all curves the reservoir size is equal to $N = 10$ and the reflectivity is $R = 0.9$; (b-c) SNR (in decibels) for the delayed terms $\gamma_{d}$ as a function of the delay ($d$). In (b) this is shown for fixed $R = 0.9$ and different values of $M$ and in (c) it is shown for fixed $M$ and two values of $R$. In both figures $N = 10$.}
    \label{figure-3}
\end{figure}
In Fig. \ref{figure-3}a, the linear capacity as a function of the delay is shown in the ideal case and for different values of the ensemble size $M$, while the reservoir size $N$ and the reflectivity $R$ are kept fixed. The number of delayed inputs a $10$-node  reservoir can ideally reproduce  with good capacity  is high ($\sim30$ injection steps). %
The collapse of the linear capacity for finite samples is clearly seen in Fig. \ref{figure-3}a for finite ensemble sizes $M$. Even though we show a progressive improvement up to $3 \cdot 10^{6}$ fiber pulses (with a memory of up to 10 further delays times when increasing the ensemble size by factors of 10), we also see that the ability to reproduce inputs further into the past of the ideal system is hard to reach. The origin of this performance scaling can be traced back to the exponential decay of the delayed terms, $\gamma_{d}^{(k)}$, in Eq. (\ref{eq-gamma-decay}) and is quantified in the following. 

In order to be able to reproduce a given input term with delay $d$, the ratio between its corresponding term $\gamma_{d}$ in the   ideal covariance and the statistical error should be large enough on average.
As the magnitude of $\gamma_{d}$ decays exponentially with $d$, it becomes harder to keep the statistical noise below it,
actually requiring an exponential increase of pulses with the delay to resolve further inputs in the past.
Scaling with the reservoir size $N$ will also be severely affected  by  statistical errors as the root of performance improvements by increasing $N$ is in the ability of the system to reproduce further delayed terms (see Sect. \ref{strategy-section}).

The performance  presented in Fig. \ref{figure-3}a can be quantified distinguishing in the measured observables, $\sigma_{\text{est}}$, the ideal case contribution, $\sigma_{\text{ideal}}$, and an added stochastic noise term, $\xi_{M}$, which depends on the number of pulses circulating the fiber, $M$:  %
\begin{equation}
    \sigma_{\text{est}}^{(k)} = \sigma_{\text{ideal}}^{(k)} + \xi_{M}^{(k)} \ .
\end{equation}
The latter for large ensembles has a variance  $\text{Var}\left( \xi_{M}^{(k)} \right) \propto M^{-1/2}$.
In contrast, the ideal covariance terms  $\gamma_{d}^{(k)}$ decay exponentially with the delay $d$ (Eq. \eqref{eq-gamma-decay}). The delay resolution will then be determined by the number of terms $\gamma_{d}^{(k)}$ whose magnitude is greater than the noise magnitude. How much do we need to increment the ensemble size $M$ in order to maintain equivalent signal resolution at larger delays? This can be determined by constraining the signal-to-noise ratios, as shown in App. \ref{App-F1}, and leads to an exponential factor in the increment in the number of the pulses
\begin{equation} \label{SNR-m-scaling}
    \left\langle \left| \frac{\gamma_{d}^{(k)}}{\xi_{M}^{(k)}}\right| \right\rangle_{\mathbf{s},\hat{H}} = \left\langle \left| \frac{\gamma_{d + \Delta d}^{(k)}}{\xi_{M^{\prime}}^{(k)}} \right| \right\rangle_{\mathbf{s},\hat{H}} \ \longrightarrow \ \frac{M^{\prime}}{M} = R^{-2\Delta d},
\end{equation}
where $\langle \cdot \rangle_{\mathbf{s},\hat{H}}$ stands for the average among realizations of the reservoir (Eq. \eqref{hamiltonian}) and random input strings. $\gamma_{d}^{(k)}\xi_{M}^{(k) -1}$ is a ($N \times N$) matrix whose elements are $\left[\gamma_{d}^{(k)}\xi_{M}^{(k) -1}\right]_{ij} \equiv \left[\gamma_{d}^{(k)}\right]_{ij}\left[\xi_{M}^{(k)}\right]_{ij}^{-1}$. This ratio matrix represents the \textit{signal-to-noise ratio} (SNR) of $\gamma_{d}^{(k)}$. The SNR constraint (Eq. \eqref{SNR-m-scaling}) establishes that good visibility is obtained by incrementing the number of pulses by a factor exponential in the delay, $\Delta d$. Equivalently, the delay resolution enhancement $\Delta d$ is a logarithmic function of the ensemble size ratio $M'/M$ (see Eq. \eqref{Delta_d_eq}). Consistently with the ideal case Eq. (\ref{eq-gamma-decay}), the dependence on the reflectivity through the scaling factor, $R^{-2\Delta d}$, is reduced by increasing the reflectivity $R$, implying a more convenient implementation (requiring fewer measurements) for large reflectivity BS. 

Some implications of Eqs. (\ref{eq-gamma-decay}) and (\ref{SNR-m-scaling}) are illustrated in Figs. \ref{figure-3}b and \ref{figure-3}c where the SNRs are plotted as functions of the delay, $d$, in decibels. We observe that they are straight lines with height equal to $\langle|\gamma_{1}^{(k)}\xi_{M}^{(k) -1}|\rangle_{\mathbf{s},\hat{H}}$ (at delay $1$)  and slope dependent only on the BS reflectivity, which is due to the $R^{d-1}$ dependency in Eq. (\ref{eq-gamma-decay}).
In Fig. \ref{figure-3}b, we see that the effect of incrementing the ensemble size is a uniform improvement of the SNR at each delay, corresponding to a shift equal to $ 5 \log_{10}\left(  {M^{\prime}}/{M} \right) $ (see App. \ref{App-F2}). In Fig. \ref{figure-3}c instead, the reflectivity  $R$ is changed while keeping $M$ fixed. This $R$ change alters both the SNR of $\gamma_{1}^{(k)}$ (height) and the decay of the SNR for further $d$ (slope). A change from $R$ to $R^{\prime}$ yields a difference of the SNR of $\gamma_{1}^{(k)}$ of
\begin{equation}
\left\langle\left|\frac{\gamma_{1}^{(k)}(R')}{\xi_{M}^{(k)}}\right|\right\rangle_{\mathbf{s},\hat{H}} - \left\langle\left|\frac{\gamma_{1}^{(k)}(R)}{\xi_{M}^{(k)}}\right|\right\rangle_{\mathbf{s},\hat{H}} = 20 \log_{10} \left[\frac{1-R^{\prime}}{1-R} \right] \ .
\end{equation}
Hence for small delays, a large reflectivity corresponds to a reduced feedback light measured in the detector and can be detrimental, while for larger delays the situation is reversed.
Increasing the value of the reflectivity $R$ translates into an enhancement of memory retention improving the reservoir performance. Indeed more delayed terms remain above the noise threshold (zero line for the SNR in Fig. \ref{figure-3}c). %

\subsection{Strategy to improve size scaling performance} \label{strategy-section}
\begin{figure}[h!]
    \centering
    \includegraphics[width=\linewidth]{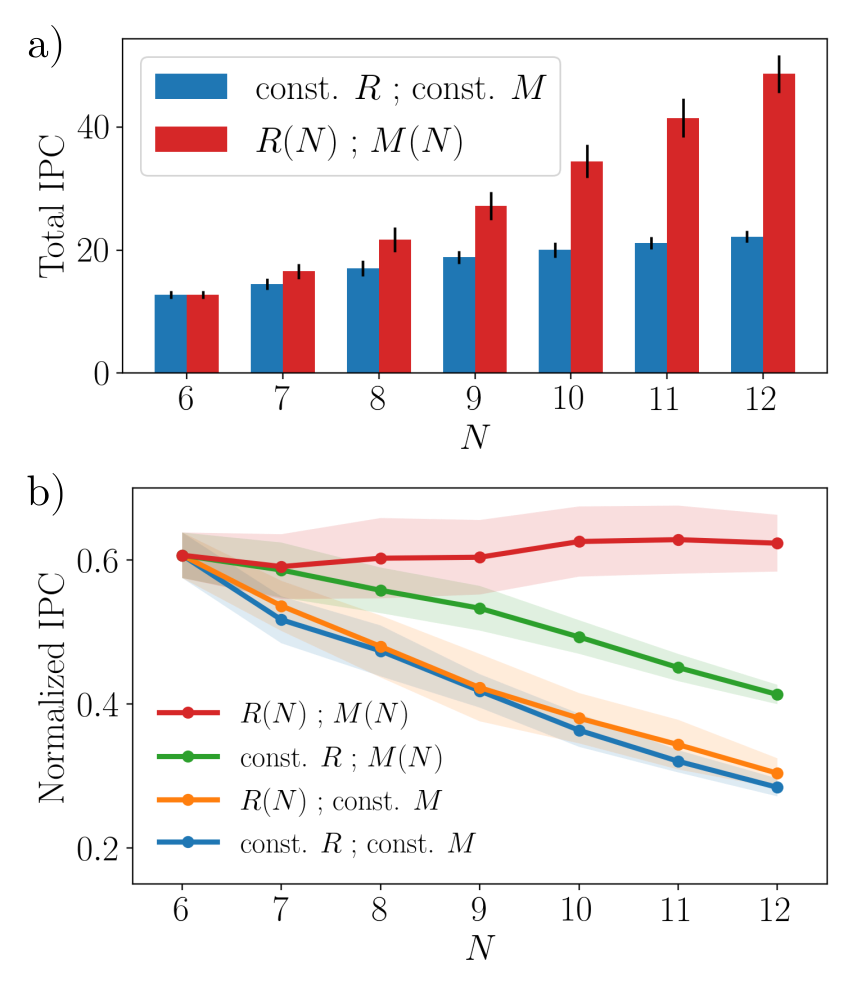}
    \caption{(a) Total IPC as a function of the reservoir size $N$ for the case of constant reflectivity $R = 0.72$ and ensemble size $M = 1.4 \cdot 10^{5}$ (blue) and the case where $R(N) = 1 - 10 N^{-2}$ and $M(N) = 3 N^{6}$. (b) Normalized IPC as a function of $N$ for four different cases: when both $R$ and $M$ are constant (blue), when $M$ is constant and $R$ varies with $N$ (orange), when $R$ is constant and $M$ varies with $N$ (green) and when both $R$ and $M$ vary with $N$ (red). The constant values of the reflectivity and the ensemble size are $R = 0.72$ and $M = 1.4 \cdot 10^{5}$, while the $N$-variable ones follow the functions $R(N) = 1 - 10 N^{-2}$ and $M(N) = 3 N^{6}$.}
    \label{figure-4}
\end{figure}

A main advantage of quantum with respect to classical RC is the possibility to access a large Hilbert space \cite{Nakajima2017, Mujal2021,kalfus2022}. Our aim is to test this potential when including quantum measurement,   addressing how statistical noise affects the performance of the system when increasing the size of the reservoir, $N$. We have seen that in the ideal case, the total IPC grows quadratically with $N$ (Fig. \ref{figure-2}b) due to the delay depth enhancement (Fig. \ref{figure-2}a). In the finite ensemble case, however, the IPC displays a sub-optimal growth with the reservoir size (blue bars in Fig. \ref{figure-4}a). Indeed the resolution of high delay inputs becomes increasingly demanding as the ideal delayed terms decrease exponentially with the delay $d$, Eq. \eqref{eq-gamma-decay}.
It follows that in order to maintain a constant normalized IPC when increasing the reservoir size $N$ one needs to improve the measurement precision, by increasing the ensemble size $M$. Our aim is to quantify how to scale the resources (here the number of feedback loop pulses $M$), to maintain a good IPC for larger reservoirs, as in the ideal case. In principle, if the delay depth is a function $d(N) \sim \alpha N^{2}$ (in the ideal case) we would require (in the finite ensemble case) the ensemble size $M$ to be an exponential function of the network size $N$ of the form $\sim R^{-2\alpha N^{2}}$ for the delay resolution to also improve quadratically with $N$.
In the following, we show the performance of a less demanding use of resources, with the ensemble size $M$ scaling with a polynomial, instead of an exponential, factor in the reservoir size $N$ but still allowing to resolve longer delays.

From our previous discussion (Fig. \ref{figure-3}c), it can be inferred that reservoirs with a smaller size, $N$, will exhibit shorter memory and therefore achieve better performance for smaller values of the reflectivity $R$. In contrast, as $N$ increases, making $R$ larger would be beneficial, improving the resolution in accessing higher delayed input information. 
This suggests a strategy to improve the performance and the scaling of our reservoirs by tuning the reflectivity of the BS, which is typically an accessible parameter in experimental setups. 
We propose to take both the reflectivity $R$ and the ensemble size $M$ as functions of the reservoir size $N$, $R \equiv R(N)$ and $M \equiv M(N)$,
in order to sustain a high normalized IPC when increasing $N$. 
We find  that a good  resolution of delayed terms can be maintained for larger reservoirs when considering
\begin{equation} \label{R-m-functions-scaling}  
R(N) = 1 - \frac{\mathcal{C}}{N^{2}} \quad \text{and} \quad M(N) \propto N^{8} \ ,
\end{equation}
in which $\mathcal{C}$ is an arbitrary constant  (see derivation in App. \ref{App-F3}).
We notice that the convenience of the quadratic dependence in the reflectivity follows from the scaling of the delay depth with the system size. Therefore for non-Gaussian states, where the output layer could grow faster than quadratically with $N$, a different scaling could be needed.

The condition in Eq. \eqref{R-m-functions-scaling} ensures a quadratic scaling of the delay resolution with the network size, $N$. It thus may also guarantee a similar scaling with $N$ of the total IPC.
Actually, in the parameter range explored here, even  numerical results limited to a less demanding number of pulses, choosing the scaling of the ensemble size $M(N) \propto N^{6}$ but for the reflectivity $R(N) = 1 - 10N^{-2}$, succeed in displaying a quadratic growth of the IPC (red bars in Fig. \ref{figure-4}a). In other words, we achieve the ideal performance scaling and a sharp improvement with respect to the suboptimal scaling for the reflectivity and the ensemble size kept constant (blue bars in Fig. \ref{figure-4}a).
In Fig. \ref{figure-4}a, we compare the total IPC scaling with the reservoir size $N$ for the case of constant ensemble size $M$ and reflectivity $R$ (blue bars) and the case in which both $M$ and $R$ scale with $N$ (red bars). We set a $0.6$ target {normalized} IPC obtained with $M(N=6) = 1.4 \cdot 10^{5}$ measurements for the smallest considered reservoir ($N=6$). Of course a higher target could be set with a larger ensemble $M(N=6)$.
A quadratic capacity scaling is achieved with the mentioned growth of reflectivity and ensemble size with $N$ (red bars in Fig. \ref{figure-4}a). In order to assess the respective influences of increasing $R$ and $M$, in Fig. \ref{figure-4}b, we compare the normalized capacities of the two previous cases (blue and red curves respectively), with two more scenarios, when one of these parameters is kept constant. Both the increase of the reflectivity and the ensemble size are found to play a key role in achieving the best performance.
These results illustrate that this photonic quantum platform exhibits the three main ingredients for QRC, namely memory, nonlinearity and high-dimensionality  (quantum advantage), in a realistic scenario.

\subsection{Performance for a time-series prediction task} \label{prediction-section}
In this section, we will analyze the performance of our reservoir in forecasting  a chaotic time series. The concrete task is to predict the next step in the series, so the target function is $\bar{y}(s_{k}) = s_{k+1}$. We use, as common in the literature, the so-called Santa Fe dataset of  experimental measurements of a chaotic laser \cite{santafe1,santafe2}.
The angle encoding chosen to perform the task is $\phi_{k} = 3\pi s_{k}/4$ as it provides high linear memory, as well as nonlinear memory (as shown in Fig. \ref{figure-2}b and App. \ref{appendix-D}). For the output layer, we consider the covariance matrix and also fourth-order moments:
$ \left\{\left\langle X_{i}^{2} X_{j}^{2} \right\rangle_{M}, \left\langle X_{i}^{3} X_{j} \right\rangle_{M}\right\}_{j \geq i} $
for $i, j = 1, \dots, N$ (the $k$ superscripts, denoting the round trip, have been omitted for clarity). These observables have been added to avoid underfitting the data and, as a technical note, provide functions of crossed input terms of the form $s_{k-d} s_{k-d'}$, which are relevant for performing this prediction task. Different input encodings on Gaussian states than the ones used in this article have been shown to provide crossed input functions in their second-order moments as well \cite{NokkalaNJP}.

The dataset analyzed in this work contains a total of 4000 input points, which we divide into three consecutive sequences: the wash-out steps (of length $L_{m}$), the training steps (of length $L$) and the testing steps (of length $L'$). The length of the training step sequence is always fixed to be $L = 3000$, while the wash-out and testing length would depend on the choice of the BS reflectivity $R$. The wash-out sequence length is set to guarantee the echo state property, and from Eq. \eqref{eq-gamma-decay} we require $R^{L_{m}} < 10^{-8}$ (so that $L_{m} \simeq 27$ for $R = 0.5$ and $L_{m} \simeq 175$ for $R = 0.9$). Given that the higher the reflectivity, the higher the memory retention of the reservoir, the number of wash-out steps to forget the initial conditions increases with $R$. The testing phase is done with the remaining data after the wash-out and the training phases.

\begin{figure}[h!]
    \centering
    \includegraphics[width=\linewidth]{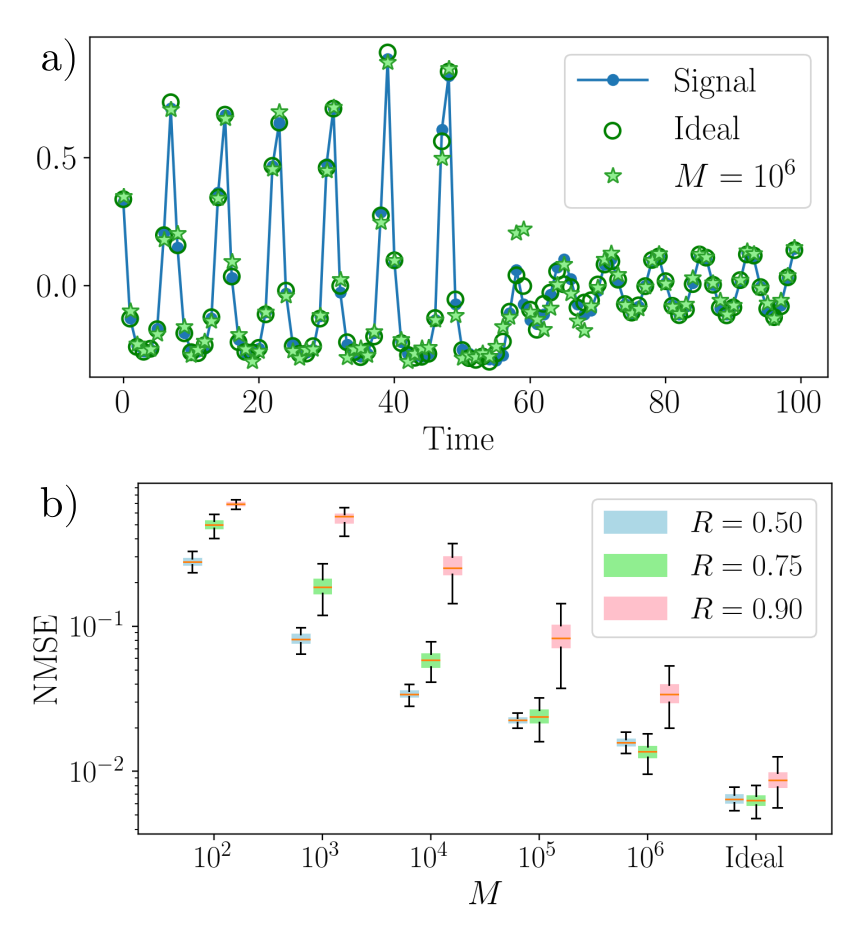}
    \caption{\textbf{Time-series prediction of chaotic signal:} (a) Prediction of chaotic signal (blue lines and dots) using a reservoir of $N = 12$ modes and reflectivity $R = 0.75$ both in the ideal case (green circles) and the finite ensemble case with size $M = 10^{6}$ (green stars). (b) Box plot of the normalized mean squared error (NMSE) as a function of the size of the ensemble $M$ for different values of $R$. For a given value of $M$, the boxes for each reflectivity are split in the x-axis to avoid overlapping. In all cases the number of modes inside each pulse was $N = 12$.}
    \label{santafe-fig}
\end{figure}

In Fig. \ref{santafe-fig}a, we compare a sequence of signal values belonging to the testing phase with the reservoir predictions, both in the ideal case and in the finite ensemble case (with ensemble size $M = 10^{6}$). They provide accurate predictions, although the finite ensemble case shows a higher error when the oscillation amplitudes change abruptly. In Fig. \ref{santafe-fig}b the \textit{normalized mean square error} (NMSE), defined in Eq. \eqref{MSE-eq}, is plotted as a function of the ensemble size $M$ (including the ideal limit $M \to \infty$) for different values of the BS reflectivity $R$. The performance consistently improves when the ensemble size increases, reaching state-of-the-art performance (with NMSE $\sim 10^{-2}$ \cite{inubushi2017reservoir,harkhoe2019delay,kumar2021efficient}) for high ensemble sizes.

In the ideal scenario, errors below the $10^{-2}$ threshold are reached with a relatively small reservoir size ($N = 12$). As shown in Fig. \ref{santafe-fig}b, for relatively small measurement ensembles, the best performance is achieved with $R = 0.5$, while for higher values of $M$ the optimal reflectivity is $R = 0.75$. As we have seen in section \ref{finite-pulses-case}, for smaller values of $R$ the SNR of the nearest past inputs (with a small delay) is greater (see Fig. \ref{figure-3}c). For very small ensembles, and thus a very high statistical noise, these setups with small reflectivity become more robust. When the ensemble size increases, a high SNR of further delayed inputs (higher memory) becomes more relevant, and thus the case of $R = 0.75$ improves its performance in comparison to the $R = 0.5$ case. Still, forecasting and memory performance are often varying differently \cite{marzen}. It seems that, although the $R = 0.9$ case provides the most amount of memory, it is not so relevant for this forecasting task, and a higher SNR for small and intermediate delayed inputs (the case of $R = 0.75$ in Fig. \ref{santafe-fig}b) has the best performance.

\section{Discussion and Outlook}

Optical platforms have been successful in classical RC \cite{brunner2019photonic}  and for  their features, such as fast processing rates and low decoherence, are also promising candidates for the first quantum implementations  \cite{Mujal2021}. In ideal conditions, photonic platforms for  QRC  have been predicted to achieve optimal performance, displaying a quantum advantage, in particular the access to an enlarged Hilbert space and operation with low signals \cite{Nokkala2021,Spagnolo2022}. 
Still, to implement successfully and in real-time temporal tasks with a QRC advantage, open challenges need to be overcome \cite{PhysRevApplied.14.024065,Khan2021a,Spagnolo2022,time-series-QRC-measurements}, namely  the limited experimental precision when estimating the output layer, as well as reducing the needed resources,  also avoiding the use of external memories.  
Here, we have proposed an optical platform suitable for real-time QRC %
based on a physical ensemble of reservoirs, as independent pulses recirculating inside an optical fiber at each input injection. 

The proposed setup displays the needed features for RC, such as fading memory and, in the limit of an infinite ensemble,  it
achieves optimal performance, with quadratic scaling of the IPC with the reservoir size (number of  modes inside each pulse) for \textit{vacuum} Gaussian states. 
This ideal limit is degraded by the effect of statistical noise and, as we increase the size of the reservoir $N$, the number of pulses in the fiber to sustain this quantum performance needs to be boosted.
The strategy proposed here to overcome this limitation in large Hilbert space allows sustaining a quadratic scaling of the IPC, with efficient use of resources, thanks to an increase of the beam splitter reflectivity when the reservoir size is scaled up. Indeed, being the dynamics of our platform linear, we can get analytical insights and find a  bound factor for the number of pulses that is polynomial in the reservoir size. Beyond the performance in terms of memory, we have also checked that our platform can be used for nontrivial temporal tasks such as chaotic series prediction.

Optical devices required to implement our proposal in a laboratory are available with state-of-the-art photonic technology and the proposed architecture for on-line time series processing provides both a tunable fading memory and the possibility to realize large measurement ensembles. With respect to recent experiments in NISQ circuits \cite{PhysRevApplied.14.024065}, our approach overcomes the need for external storage of input and repeating the experiment, just requiring the photonic QRC to be faster than the input rate to achieve real-time processing. For non-temporal classification tasks, a general dynamical framework applicable in circuit QED platforms has been reported in \cite{Khan2021a}, while integrated photonic circuits  \cite{Spagnolo2022} have been proposed using a quantum memristor. Our approach can be extended also to these platforms and going beyond Gaussian states.

Our work paves the way to real-time time series processing with QRC, with state-of-the-art photonics technology and displaying a scalable quantum advantage leveraging limited resources.

\begin{acknowledgments}
We acknowledge the Spanish State Research Agency, through the Severo Ochoa and Mar\'ia de Maeztu Program for Centers and Units of Excellence in R\&D (MDM-2017-0711) and through the  QUARESC project (PID2019-109094GB-C21 and -C22/ AEI / 10.13039/501100011033).
 We also acknowledge funding by CAIB through the QUAREC project (PRD2018/47).
GLG is funded by the Spanish  MEFP/MiU   and  co-funded by the University of the Balearic Islands through the Beatriz Galindo program  (BG20/00085). The CSIC Interdisciplinary Thematic Platform (PTI) on Quantum Technologies in Spain is also acknowledged.
\end{acknowledgments}

\appendix
\section{Reservoir Computing} \label{appendix-A}
Reservoir Computing (RC) is a ML framework that takes advantage of complex dynamical systems to perform learning tasks with a low-cost training protocol. It is especially suited for temporal tasks regarding time series. Every RC scheme can be separated into three steps/layers: the input layer, the reservoir dynamics and the readout layer. For the input layer, a vector $\mathbf{s}_{k}$ from the time series signal, $\{ \mathbf{s}_{1},\mathbf{s}_{2},\dots,\mathbf{s}_{L} \}$, is encoded and fed into the dynamical system, also called \textit{reservoir}, degrees of freedom. While in general each $\mathbf{s}_{k}$ can be a vector, throughout our work we took it as a scalar (the same applies to the output elements). The length of the list, $L$, denotes the number of training samples, or training steps, to perform the learning protocol. After each introduction of an input, the reservoir degrees of freedom, $\mathbf{v}_{k-1}$, evolve in time through a nonlinear mapping for a fixed time $\Delta t$ (reservoir dynamics).
The reservoir observables/degrees of freedom, $\mathbf{v}_{k}$, are measured and taken as outputs. A linear mapping is then applied to these observables to yield an output vector $\mathbf{y}_{k}$ (readout layer), which is built as a function of the weight parameters that are updated in the training stage.
After the last step, the protocol continues for the following input, $\mathbf{s}_{k+1}$.  

We can write the dynamical mapping of the reservoir degrees of freedom as
\begin{equation}
    \mathbf{v}_{k} = \mathcal{H} \left( \mathbf{v}_{k-1}, s_{k} \right) \quad ; \quad k \in \mathbb{Z} \ ,
\end{equation}
in which we need $\mathcal{H}$ to be a nonlinear mapping of the input $s_{k}$. This mapping will remain fixed throughout the whole protocol. Following standard RC practices, we take the output layer function to be a linear function of the reservoir observables
\begin{equation}
\begin{aligned}
    y_{k} &= w_{0} + \mathbf{w}^{\top} \mathbf{v}_{k} \\
    &= \left( 1 , \mathbf{v}_{k}^{\top} \right) \left( \begin{array}{c}
        w_{0}  \\
        \mathbf{w} \ ,
    \end{array} \right)
\end{aligned}
\end{equation}
with the $(D_{\text{R}} + 1)$-dimensional vector $\mathbf{W} = \left( w_{0}, \mathbf{w}^{\top} \right)^{\top}$ containing the training weights (we add a constant bias term $w_{0}$). For a given set of $L$ inputs that we feed to the reservoir, we can define the following matrices
\begin{equation}
    V = \left(
    \begin{array}{cc}
        1 & \mathbf{v}_{1}^{\top} \\
        1 & \mathbf{v}_{2}^{\top} \\
        \vdots & \vdots \\
        1 & \mathbf{v}_{L}^{\top}
    \end{array}
    \right) \quad ; \quad  \mathbf{y} = \left( \begin{array}{c}
        y_{1}  \\
        y_{2} \\
        \vdots \\
        y_{L}
    \end{array} \right) \ ,
\end{equation}
so that the following equation holds
\begin{equation} \label{output-vector-RC}
    \mathbf{y} = V \mathbf{W} \ .%
 \end{equation}
After choosing a given target function that we want our reservoir to reproduce, $\bar{\mathbf{y}}$, we want to find the weights that minimize the normalized mean-square error (NMSE) of the predicted $\mathbf{y}$ and the target
\begin{equation} \label{MSE-eq}
    \text{NMSE}_{L}(\mathbf{y},\bar{\mathbf{y}}) = \frac{\sum_{k=1}^{L} (y_{k} - \bar{y}_{k})^{2}}{\sum_{k=1}^{L} \bar{y}_{k}^{2}} \ .
\end{equation}
The optimal set of weights for this condition to hold are the ones obtained through the following procedure, \cite{LUKOSEVICIUS2009127}:
\begin{equation}
    \mathbf{W}_{\text{opt}} = V^{\text{MP}} \bar{\mathbf{y}} \ ,
\end{equation}
where $V^{MP} = \left(V^{\top} V\right)^{-1} V^{\top}$ is the Moore-Penrose inverse of $V$. The higher the value of $L$ the more precise our estimation of the optimal weights will usually be.

\section{Linear capacity and Information Processing Capacity} \label{appendix-B}
Throughout this article, we tested the performance of our (already trained) reservoirs using the \textit{capacity} to reproduce a function of the inputs. We consider a reservoir that has already finished the learning protocol to reproduce a given target function $\bar{y}$. We thus take a given string of $L'$ inputs $\mathbf{s} = (s_{1}, s_{2}, \dots, s_{L'})$ and a vector $\bar{\mathbf{y}} = (\bar{y}_{1}, \bar{y}_{2}, \dots, \bar{y}_{L'})$ with components $\bar{y}_{k} = \bar{y}(\mathbf{s},k)$. The length of the new input and target function vectors, $L'$, denotes the number of testing steps right after the learning protocol. Then, the capacity of our reservoir to reproduce the given target function $\bar{y}$ is
\begin{equation}
    C_{\bar{\mathbf{y}}} = \text{max} \left[0, 1 - \text{min}_{\mathbf{W}} \  \text{NMSE}_{L'}(\mathbf{y},\bar{\mathbf{y}}) \right] \ ,
\end{equation}
where $\mathbf{y}$ is the output vector of the reservoir, as defined in Eq. (\ref{output-vector-RC}), after the training of the weights has been performed. To check the memory of the system we use the \textit{linear capacity}, in which we set the target functions to be $\bar{y}_{d}(s_{k}) = s_{k-d}$, where the parameter $d$ denotes the delay of the input. As a quantitative measure of how well our reservoir performs in general, we use the \textit{information processing capacity} (IPC) or total capacity \cite{dambre2012information}. The main idea is to compute the capacity of all the different orthogonal functions that our reservoir can approximate. By choosing a complete, orthogonal family of functions, $\{ \bar{y} \}$, the IPC can be estimated as
\begin{equation}
    \text{IPC} = \sum_{\bar{y}} C_{\bar{\mathbf{y}}} \ .
\end{equation}
In our simulations, we have chosen the family of functions to be
\begin{equation}
    \bar{y}_{D}^{(d)}(s_{k}) = \mathcal{P}_{D}(s_{k-d}) \ ,
\end{equation}
where the function $\mathcal{P}_{D}$ is the normalized Legendre polynomial of degree $D$. As we cannot sum over an infinite number of delays and degrees, we have chosen $d \in \left[ 0, d_{\text{max}}  \right]$ and $D \in \left[ 1, D_{\text{max}}  \right]$. For every simulation in this article, it has sufficed to take $d_{\text{max}} = 75$ and $D_{\text{max}} = 5$.

Fading memory is a necessary property for any dynamical system to work as a reservoir computer, \cite{Konkoli2017}. In our platform, short-term memory is ensured by the BS coupling (proof on App. \ref{RC-conditions-proof}). In fact, as it was shown in Eq. \eqref{eq-gamma-decay}, the magnitude of the delayed terms $\gamma_{d}^{(k)}$ dropped as $R^{d}$ ($d$ denoting the delay and $R$ denoting the BS reflectivity). For a large enough delay, the reservoir would not be able to resolve that delayed input information. As we start our reservoir in a completely random state, we first introduce a list of $L_{\text{m}}$ wash-out inputs, before starting the learning protocol, just to ensure the initial conditions have been forgotten by the reservoir. We make $L_{\text{m}}$ large enough to have the prefactor going below the numerical precision. This is equivalent to fulfilling the following condition
\begin{equation}
    R^{L_{\text{m}}} < 10^{-16} \ ,
\end{equation}
so the number of wash-out steps depends on $R$.
For our estimations of the linear capacity and the IPC on section \ref{section-III}, the training steps have been chosen to be $L = 10^{4}$ and the testing steps $L' = 5000$.

\section{Gaussian states} \label{appendix-C}
The evolution of an $N$-mode Gaussian quantum state $\hat{\rho}$ is completely determined by the dynamics of its displacement vector $\mathbf{r}$ of dimension $2N$ and its positive symmetric covariance matrix $\sigma$ of dimension $2N\times 2N$ \cite{adesso2014,serafini2017}. 
Each element of $\mathbf{r}$ ($\sigma$) denotes the mean value (covariance) of each quadrature for each mode.
As we inject vacuum states, the displacements of the external signal are null vectors. The covariance matrix of the ancilla state is the composition of $N$ independent and identical squeezed states. We can thus write its covariance matrix at round-trip $k$ as $\sigma_{\text{anc}}^{(k)} = \bigoplus_{i=1}^{N} \sigma_{\text{sq}}(s_{k})$, in which $\sigma_{\text{sq}}(s_{k})$ is the covariance matrix of each single external mode. It is a nonlinear function of the $k$-th input, $s_{k}$, which can be expanded as
\begin{equation} \label{single_squeezing_covariance matrix}
    \sigma_{\text{sq}}(s_{k}) =  \left(\begin{array}{cc}
        c_{+}(s_{k}) & z(s_{k}) \\
        z(s_{k}) & c_{-}(s_{k})
    \end{array} \right) \ ,
\end{equation}
for the following functions
\begin{align} \label{c_pm_expression}
    c_{\pm}(s_{k}) &= \cosh \left( 2 r_{k} \right) \pm \cos \left(\phi_{k} \right) \sinh \left( 2 r_{k} \right) \\ \label{z_expression}
    z(s_{k}) &= \sin \left(\phi_{k} \right) \sinh \left( 2 r_{k} \right) \ .
\end{align}
The two main parameters to tune are the squeezing strength, $r_{k}$, and the squeezing angle, $\phi_{k}$. As commented at the beginning of section \ref{section-III}, we use the following encoding: $r_{k} = 1$ and $\phi_{k} = 3\pi s_{k}/4$.

\subsection{Gaussian measurements on multipartite systems} \label{gaussian-measurements-appendix}
In this section, we provide an analytical treatment to describe the effect of quantum measurements on the quadrature operators.
We start by defining a composite Gaussian state, which we can separate into subsystems A and B. We can define the displacement vector and covariance matrix of the total Gaussian state as
\begin{equation} \label{subsystem-split}
\begin{aligned}
    \mathbf{r} &= \left(
    \begin{array}{c}
        \mathbf{r}_{A} \\
        \mathbf{r}_{B}
    \end{array}
    \right) \\
    \Gamma &= \left(
    \begin{array}{cc}
        \sigma_{A} & \sigma_{AB} \\
        \sigma_{AB} & \sigma_{B}
    \end{array} \right) \ ,
\end{aligned}
\end{equation}
where $\mathbf{r}_{A}$ ($\mathbf{r}_{B}$) and $\sigma_{A}$ ($\sigma_{B}$) are the displacement vector and covariance matrix of subsystem A (B), while $\sigma_{AB}$ denotes the correlations between subsystem A and B. We now consider measurements that can be performed on subsystem $B$  so that the conditional state of subsystem $A$ remains Gaussian. %
For that we consider the family of general-dyne measurements, \cite{serafini2017,conditional_unconditional_gaussian}, in which homodyne detection is included.
Thus, measuring the subsystem $B$ with a random outcome, $\mathbf{r}_{\text{out}}^{B}$, gives rise to the following conditional state of $A$:
\begin{align} \label{BA-r}
    \mathbf{r}_{A}^{\prime} &= 
        \mathbf{r}_{A} + \sigma_{AB} \left(\sigma_{B} + \sigma_{\text{m}}\right)^{-1} \left(\mathbf{r}_{\text{out}}^{B} - \mathbf{r}_{B} \right) \\
    \label{BA-sigma}
    \sigma_{A}^{\prime} &= \sigma_{A} - \sigma_{AB} \left(\sigma_{B} + \sigma_{\text{m}}\right)^{-1} \sigma_{AB}^{\top} \ ,
\end{align}
where $\sigma_{\text{m}}$ is a positive symmetric matrix that depends on the kind of general-dyne measurement that is being performed. %
The measurement outcome is drawn from a multivariate Gaussian distribution with covariance matrix equal to $\sigma_{B} + \sigma_{\text{m}}$ and mean vector equal to $\mathbf{r}_{B}$. Thus, we can write
\begin{equation} \label{r-out-B}
    \mathbf{r}_{\text{out}}^{B} = \mathbf{r}_{B} + \sqrt{\sigma_{B} + \sigma_{\text{m}}} \mathbf{u}(\mathbf{0},I) \ ,
\end{equation}
where $\mathbf{u}(\mathbf{0},I)$ is a random vector drawn from a Gaussian distribution with 0 mean and covariance matrix equal to the identity, $I$. With this, we can rewrite Eq. \eqref{BA-r} as
\begin{equation} \label{BA-displacement}
    \mathbf{r}_{A}^{\prime} = 
        \mathbf{r}_{A} + \sigma_{AB} \left(\sigma_{B} + \sigma_{\text{m}}\right)^{-1/2} \mathbf{u}(\mathbf{0}, I) \ ,
\end{equation}
which describes how measuring subsystem $B$ has affected the state in $A$.

For the specific case of homodyne detection of the $x$-quadratures of each mode, we have  \cite{serafini2017,adesso2014,PhysRevLett.89.137903} %
\begin{equation} \label{sigma_m-homodyne}
    \sigma_{\text{m}} = \lim_{z \to \infty} \bigoplus_{i=1}^{N} \text{diag}\left( z^{-2}, z^{2} \right) \ ,
\end{equation}
In this case, the term of Eqs. \eqref{BA-r}-\eqref{BA-sigma} $(\sigma_{B} + \sigma_{\text{m}})^{-1}$ in the limit of $z$ tending to infinity, tends towards the following expression:
\begin{equation} \label{MP-inverse-homodyne}
    \lim_{z \to \infty}\left[\sigma_{B} + \bigoplus_{i=1}^{N} \text{diag}\left( z^{-2}, z^{2} \right) \right]^{-1} = \left(\Pi \sigma_{B} \Pi\right)^{\text{MP}} \ ,
\end{equation}
where $\Pi = \bigoplus_{i=1}^{N} \left( \begin{array}{cc}
    1 & 0 \\
    0 & 0
\end{array} \right)$ and MP stands for the Moore-Penrose inverse, \cite{PhysRevLett.89.137903}. The outcome vector from the homodyne measurement can be modeled from Eq. \eqref{r-out-B}, using the $\sigma_{\text{m}}$ matrix from Eq. \eqref{sigma_m-homodyne}. However, special care has to be taken in this case, as in the limit of $z \to \infty$ we are introducing a divergent variance in the $p$-quadrature degrees of freedom. The shape that the inverse matrix from Eq. \eqref{MP-inverse-homodyne} takes in this limit ensures convergence of Eqs. \eqref{BA-r} and \eqref{BA-sigma}, as the $p$-quadrature degrees of freedom of the measurement outcome do not play a role in determining the conditional state of $A$. In practice, to obtain the outcome observables of the homodyne detection, we trace out the diverging degrees of freedom and, thus, only consider the $x$-quadratures. The resulting probability distribution of the outcome becomes:
\begin{equation}
    p(\mathbf{x}^{B}_{\text{out}}) = \frac{\exp\left\{ - \left( \mathbf{x}^{B}_{\text{out}} - \mathbf{x}_{B} \right)^{\top} \sigma_{B,\mathbf{x}}^{-1} \left( \mathbf{x}^{B}_{\text{out}} - \mathbf{x}_{B} \right) \right\}}{\pi^{N} \sqrt{\text{Det}(\sigma_{B,\mathbf{x}})}} \ ,
\end{equation}
where $\sigma_{B,\mathbf{x}} = \text{Tr}_{\mathbf{p}} \left(\sigma_{B}\right)$ and $\mathbf{x}_{B} = \text{Tr}_{\mathbf{p}} \left(\mathbf{r}_{B}\right)$ (the degrees of freedom of the $p$-quadratures have been traced out), \footnote{The notation $\text{Tr}_{\mathbf{p}}\left[ \bullet \right]$ does not actually stand for the usual partial trace of a matrix, it is just a way of writing we are tracing out the components of the covariance matrix and first-moment vector which contain information of the $p$-quadratures of every mode. For a generic $2N$-dimensional covariance matrix, $\sigma$, tracing out these components would yield a $N$-dimensional matrix with components:
\begin{equation}
    \left[ \text{Tr}_{\mathbf{p}}\left( \sigma \right) \right]_{ij} = \left\langle \hat{x}_{i} \hat{x}_{j} \right\rangle - \left\langle \hat{x}_{i} \right\rangle \left\langle \hat{x}_{j} \right\rangle \ ,
\end{equation}
where the mean values stand for the quantum expected values of the observables for a given quantum state.}. This is consistent with the fact that, in experiments, homodyne detection only yields an outcome of one quadrature for each mode.

In our platform, we can easily identify the two subsystems. Before the BS coupling, we have the pulse coming out from the fiber and the external ancilla pulse, which are both independent of one another. After the BS coupling, we have the pulse that belongs to the fiber path as one subsystem and the one that travels to the homodyne detector (HD) as another. In this last case, there is generally entanglement between them (appearing as correlation terms in the $\sigma_{AB}$ matrix). As the $x$-quadratures of each pulse that reaches the detector is measured, the relative pulse (going through the fiber path) is conditioned to the measurement outcomes as detailed in Eqs. \eqref{BA-r} and \eqref{BA-sigma}. The observables averaged over this ensemble of conditional states resemble the unconditional evolution, as it is expected in the case of Gaussian states \cite{conditional_unconditional_gaussian}. Back-action effects in other quantum substrates would generally affect averages performed over conditional ensembles, so the unconditional evolution is not completely obtained through averaging. In the context of QRC this can negatively affect the performance, as it was shown in the case of qubits \cite{time-series-QRC-measurements}.

\subsection{Round-trip dynamics of the reservoir} \label{round-trip_dynamics_appendix}
In this section, we will describe the dynamical evolution of a single pulse at each round-trip. As a physical ensemble of pulses is found in the fiber, the resulting equations will hold true for every pulse. At round-trip $k$, every pulse coming out from the fiber will eventually couple to an external ancilla state through the BS. The quantum state of the whole system prior to the BS coupling can be described with the following displacement vector and covariance matrix
\begin{align}
    \mathbf{r}^{(k)}_{0} &= \left(
    \begin{array}{c}
        \mathbf{r}_{\text{R}}^{(k)}  \\
        \mathbf{0}
    \end{array}
    \right) \\
    \Gamma^{(k)}_{0} &= \left(
    \begin{array}{cc}
        \sigma_{\text{R}}^{(k)} & 0 \\
        0 & \sigma_{\text{anc}}^{(k)}
    \end{array}
    \right) \ ,
\end{align}
where we write it as a composite system in which $\mathbf{r}_{\text{R}}$ and $\sigma_{\text{R}}$ are the displacements and covariance matrix of the pulse coming out from the fiber and $\sigma_{\text{anc}}$ is the covariance matrix of the ancilla pulse; the subindex $0$ is introduced to denote the initial state in which the ancilla pulse and the reservoir pulse have not yet arrived to the BS. We remark that the ancilla pulse has a null displacement vector, as it is a vacuum state. Also, as both pulses are initially independent, the off-diagonal matrices of $\Gamma_{0}$ are null. As there are $N$ modes inside each pulse, with two quadratures each, the displacement vectors of each subsystem are $2N$-dimensional and their covariance matrices are $2N \times 2N$ matrices. In total, $\mathbf{r}_{0}$ is a $4N$-dimensional vector and $\Gamma_{0}$ is a $4N \times 4N$ matrix. The action of the BS can be written in the following matrix form
\begin{equation}
    B_{R} = \left( \begin{array}{cc}
        \sqrt{R} \ I_{2N} & \sqrt{T} \ I_{2N} \\
        -\sqrt{T} \ I_{2N} & \sqrt{R} \ I_{2N}
    \end{array} \right) \ ,
\end{equation}
where $I_{2N}$ is the $2N \times 2N$ identity matrix. We remind that $R$ ($T$) is the reflectivity (transmissivity) of the BS. In turn, the symplectic matrix that describes the evolution inside both non-linear crystals is the following
\begin{equation}
    S(\Delta t) = \left( \begin{array}{cc}
        S_{1}(\Delta t) & 0 \\
        0 & S_{2}(\Delta t) \end{array} \right) \ ,
\end{equation}
in which $S_{1}$ ($S_{2}$) describes the evolution inside the crystal placed at the feedback fiber (detector) path. These matrices describe the evolution of the quadrature operators under the action of quadratic Hamiltonians such as the ones we are considering in Eq. (\ref{hamiltonian}). Concretely, the matrices $S_{1}(\Delta t)$ and $S_{2}(\Delta t)$ are analog to the generic unitary operator $\hat{U}(\Delta t ) = \exp \left[ -i \hat{H}_{\chi^{(2)}} \Delta t \right]$ for each non-linear crystal.
 They act on the quadrature operators in phase-space \cite{serafini2017,adesso2014}.
The symplectic matrix counterpart of any quadratic Hamiltonian can be numerically computed as long as the Hamiltonian is positive \cite{serafini2017}. In our discussion of the parameters at the beginning of section \ref{section-III}, it was mentioned that the materials were not allowed to produce squeezing levels far beyond 15 dB. The squeezing produced by the materials can be computed by the Bloch-Messiah decomposition of matrices $S_{1}$ and $S_{2}$ \cite{bloch-messiah,PhysRevA.93.062115}. We write the symplectic matrix of the whole process (BS + non-linear crystals) as follows
\begin{equation} \label{S-prime-eq}
    S^{\prime}(\Delta t) = S(\Delta t) B_{R} \ .
\end{equation}
After both pulses have come out from the non-linear media, their state parameters have evolved as follows
\begin{align} \label{after-dynamics-r}
    \mathbf{r}^{\prime} &= S^{\prime}(\Delta t) \mathbf{r}_{0}  \\ \label{after-dynamics-sigma}
    \Gamma^{\prime} &= S^{\prime}(\Delta t) \Gamma_{0} S^{\prime}(\Delta t)^{\top} \ ,
\end{align}
where the $k$ labeling has been omitted for clarity. We can now split both $\mathbf{r}^{\prime}$ and $\Gamma^{\prime}$ into two subsystems: one for the pulse that is being reinjected into the fiber and another one for the pulse that travels to the detector. We thus label with the `fiber' (`HD') subscript to the parameters of the pulse that is reinjected in the fiber (traveling to the detector). So the resulting displacement vector and covariance matrix from Eqs. \eqref{after-dynamics-r} and \eqref{after-dynamics-sigma} can be written as
\begin{align}
    \mathbf{r}^{\prime} &= \left(
    \begin{array}{c}
        \mathbf{r}_{\text{fiber}}  \\
        \mathbf{r}_{\text{HD}}
    \end{array}
    \right) \\
    \Gamma^{\prime} &= \left(
    \begin{array}{cc}
        \sigma_{\text{fiber}} & \sigma_{\text{corr}} \\
        \sigma_{\text{corr}}^{\top} & \sigma_{\text{HD}}
    \end{array}
    \right) \ ,
\end{align}
in which $\sigma_{\text{corr}}$ is a $2N \times 2N$ matrix containing the correlations between the fiber pulse and the detected pulse. It is equivalent to $\sigma_{AB}$ in Equ. (\ref{subsystem-split}). As we saw in subsection \ref{gaussian-measurements-appendix}, and equivalent to the result in Eq. (\ref{r-out-B}) the measurement outcome of the measured pulse quadratures is the following
\begin{equation} \label{output-r}
    \mathbf{r}_{\text{out}}^{(k)} = \mathbf{r}_{\text{HD}}^{(k)} + \sqrt{\sigma_{\text{HD}}^{(k)} + \sigma_{\text{m}}} \mathbf{u}^{(k)} \ ,
\end{equation}
where, again, $\mathbf{u}^{(k)}$ is a random $2N$-dimensional vector whose components are drawn from a normal distribution with zero mean a variance equal to one. The feedback pulse coming out from the fiber in the following round-trip will have the following parameters
\begin{align} \label{BA-r-k+1}
    \mathbf{r}_{\text{R}}^{(k+1)} &= \mathbf{r}_{\text{fiber}}^{(k)} + \sigma_{\text{corr}}^{(k)} \left[ \sigma_{\text{HD}}^{(k)} + \sigma_{\text{m}} \right]^{-1/2} \mathbf{u}^{(k)} \\
    \label{BA-sigma-k+1}
    \sigma_{\text{R}}^{(k+1)} &= \sigma_{\text{fiber}}^{(k)} - \sigma_{\text{corr}}^{(k)} \left[ \sigma_{\text{HD}}^{(k)} + \sigma_{\text{m}} \right]^{-1}\left[\sigma_{\text{corr}}^{(k)}\right]^{\top} \ ,
\end{align}
in which the state of the pulse is conditioned to the outcome measurement from the homodyne detector (see Eqs. \eqref{BA-r} and \eqref{BA-sigma}).
\subsection{Recursive equations}
In this subsection, we work from Eqs. \eqref{BA-r-k+1} and \eqref{BA-sigma-k+1} to obtain expressions for the relevant output parameters as functions of the input history. These expressions are going to be useful to obtain the ideal case observables from Eq. \eqref{ideal-decomposition}. We start with the output displacements. From Eq. \eqref{after-dynamics-r} we have the following simple relations:
\begin{align} \label{simple-fiber-PD1}
    \mathbf{r}_{\text{fiber}}^{(k)} &= \sqrt{R} S_{1}(\Delta t) \mathbf{r}_{\text{R}}^{(k)} \\
    \label{simple-fiber-PD2}
    \mathbf{r}_{\text{HD}}^{(k)} &=  -\sqrt{T} S_{2}(\Delta t) \mathbf{r}_{\text{R}}^{(k)} \ .
\end{align}
We can substitute Eq. \eqref{BA-r-k+1} into Eq. \eqref{simple-fiber-PD1} and \eqref{simple-fiber-PD2}. Thus, by recursion, it yields the result
\begin{align} \label{fiber-desarrollada}
    \mathbf{r}_{\text{fiber}}^{(k)} &= \sum_{d=1}^{k} R^{d/2} S_{1}^{d}(\Delta t) \sigma_{\text{corr}}^{(k-d)} \nonumber \\
    &\times \left[ \sigma_{\text{HD}}^{(k-d)} + \sigma_{\text{m}} \right]^{-1/2} \mathbf{u}^{(k-d)} \\
    \label{PD-desarrollada}
    \mathbf{r}_{\text{HD}}^{(k)} &= -T S_{2}(\Delta t) \sum_{d=1}^{k} R^{(d-1)/2} S_{1}^{d}(\Delta t) \sigma_{\text{corr}}^{(k-d)} \nonumber \\ &\times \left[ \sigma_{\text{HD}}^{(k-d)} + \sigma_{\text{m}} \right]^{-1/2} \mathbf{u}^{(k-d)} \ .
\end{align}
Deviations from the origin (vacuum) in the reservoir pulse are originated by the stochastic displacements that the recurrent back-action produces. The same procedure can be performed with the covariance matrices. From Eq. \ref{after-dynamics-sigma}, these relations follow
\begin{align} \label{covariance matrix-fiber}
    \sigma_{\text{fiber}}^{(k)} &= S_{1}(\Delta t) \left[ R \sigma_{\text{R}}^{(k)} + T\sigma_{\text{anc}}^{(k)} \right] S_{1}(\Delta t)^{\top} \\
    \label{covariance matrix-PD}
    \sigma_{\text{HD}}^{(k)} &= S_{2}(\Delta t) \left[ T \sigma_{\text{R}}^{(k)} + R\sigma_{\text{anc}}^{(k)} \right] S_{2}(\Delta t)^{\top} \\
    \label{covariance matrix-corr}
    \sigma_{\text{corr}}^{(k)} &= \sqrt{RT} S_{1}(\Delta t) \left[ \sigma_{\text{R}}^{(k)} - \sigma_{\text{anc}}^{(k)} \right] S_{2}(\Delta t)^{\top} \ .
\end{align}
From the equation for $\sigma_{\text{fiber}}$, we note that similar recursion equations can be obtained by substituting it in Eq. \eqref{after-dynamics-sigma}. By recursion we can, again, get the expression of the covariance matrix in Eq. \eqref{covariance matrix-PD} as functions of the input history. We only write the expression for $\sigma_{\text{HD}}$ below, as it is the only one that is necessary for the mathematical derivations in the next section:

\begin{widetext}
\begin{equation} \label{sigma-PD-long}
\begin{aligned}
    \sigma_{\text{HD}}^{(k)} &= R S_{2} \sigma_{\text{anc}}^{(k)} S_{2}^{\top} \\
    &+ TS_{2}\left[\sum_{d=1}^{k} R^{d-1} S_{1}^{d-1} \left[ T S_{1} \sigma_{\text{anc}}^{(k-d)} S_{1}^{\top} - \sigma_{\text{corr}}^{(k-d)} \left[ \sigma_{\text{HD}}^{(k-d)} + \sigma_{\text{m}} \right]^{-1} \left(\sigma_{\text{corr}}^{(k-d)}\right)^{\top} \right] \left(S_{1}^{d-1}\right)^{\top}\right] S_{2}^{\top} \ .
\end{aligned}
\end{equation}
\end{widetext}
The $\Delta t$ dependency of $S_{1}$ and $S_{2}$ has been omitted for clarity.

\section{Input encoding and nonlinearity} \label{app-encoding}
\begin{figure}[h!]
    \centering
    \includegraphics[width=\linewidth]{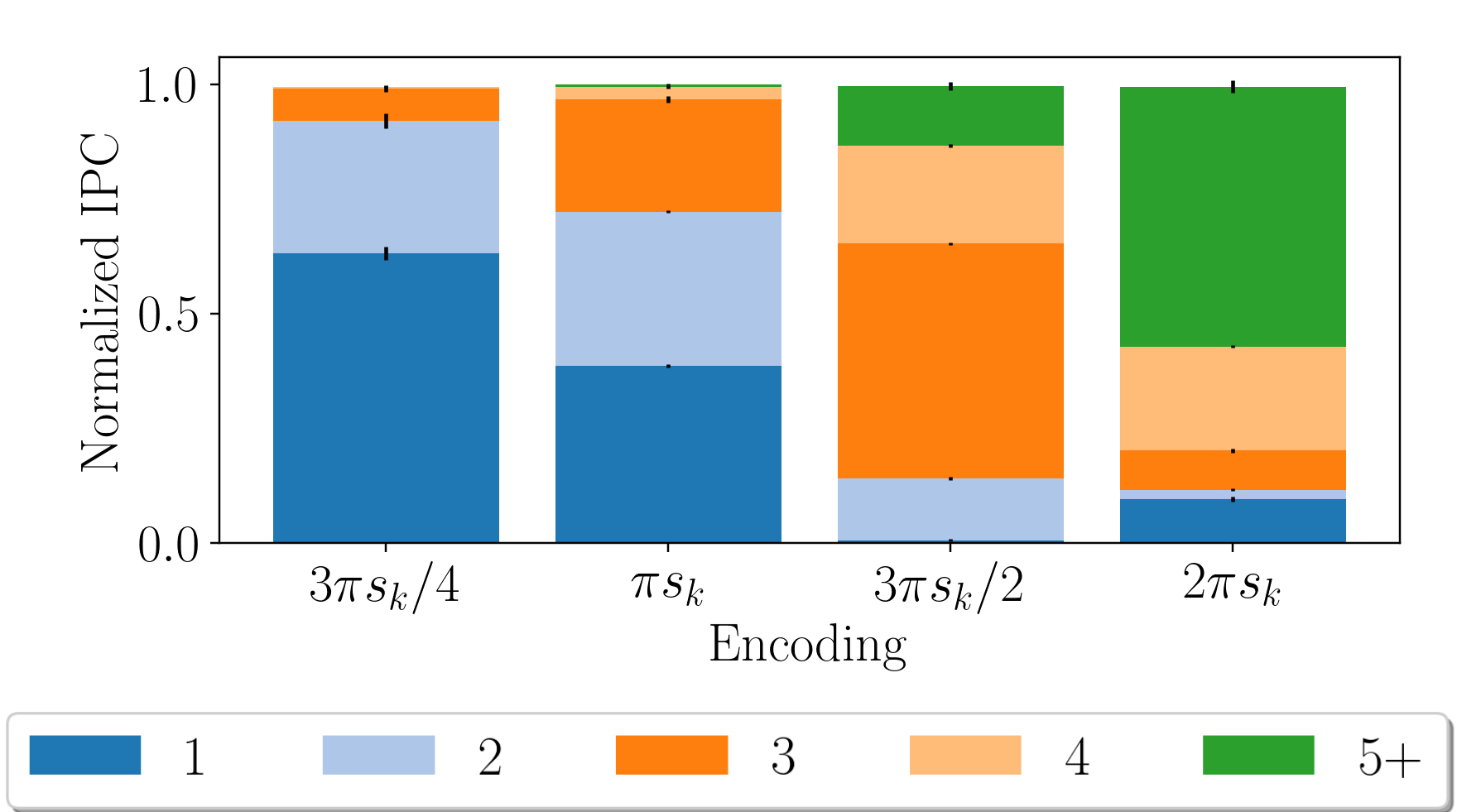}
    \caption{\textbf{Encoding choice effect on the IPC:} bar plot of the normalized IPC as a function of different angle encodings $\phi_{k} = \beta s_{k}$. Each bar is split into different degree contributions. In all cases the reservoir size is $N = 10$ and the reflectivity is $R = 0.5$.}
    \label{figure-encoding}
\end{figure}

In this brief section we are going to elaborate on the degree distribution of the IPC as a function of the encoding choice. We recall that the input ancilla states are single mode squeezed vacuum states with the input signal encoded in their covariance matrices as in Eqs. \eqref{single_squeezing_covariance matrix}, \eqref{c_pm_expression} and \eqref{z_expression}. Concretely, in the main text we have used a squeezing angle encoding setting $r_{k} = 1$ and $s_{k} = 3\pi s_{k}/4$. The input could also be encoded in the squeezing strength $r_{k}$, yielding a different degree distribution \cite{Nokkala2021,Mujal_nonlinearity}. In this appendix we focus our attention on angle encodings with $r_{k} = 1$ and $\phi_{k} = \beta s_{k}$, where $\beta > 0$. In Eqs. \eqref{c_pm_expression} and \eqref{z_expression} the input nonlinearity comes from the functions $\sin(\phi_{k})$ and $\cos(\phi_{k})$. We now write the Taylor expansion of these functions as
\begin{align}
    \sin(\beta s_{k}) &= \beta s_{k} + \frac{\beta^{3} s_{k}^{3}}{6} + \mathcal{O}(\beta^{5}) \\
    \cos(\beta s_{k}) &= 1 - \frac{\beta^{2}s_{k}^{2}}{2} + \mathcal{O}(\beta^{4}) \ .
\end{align}
And so we actually see that the smaller the value of $\beta$, the closer $c_{\pm}(s_{k})$ ($z(s_{k})$) are from being linear functions of $s_{k}$ and $s_{k}^{3}$ ($s_{k}^{2}$). So for smaller values of $\beta$ the linear and low nonlinear contributions to the IPC will be greater. On the other hand, as we increase the value of $\beta$, the high nonlinear terms will be more relevant. This is visualized in Fig. \ref{figure-encoding}.

\section{Ensemble averages and ideal case limit} \label{appendix-D}
In this section, we derive the expression for the covariance matrix of the output signal in Eq. \eqref{output-r} for an infinite ensemble of pulses, %
so that we get rid of any statistical error. It is relevant to remark that, as we are not considering any filter, every pulse in the ensemble is conditioned to a measurement outcome history throughout a certain number of round trips. As we will see, by averaging over an ensemble of conditioned realizations with no filters, the unconditional state arises. As it was already commented in section \ref{section-II}, this is an already known feature of conditional Gaussian states under general-dyne monitoring, \cite{conditional_unconditional_gaussian}.
\\
\\
To denote the ensemble pulses we add a label $m$, which ranges from 1 to $M$ (the total number of pulses in the ensemble). In that manner, we take Eq. \eqref{output-r} and add the ensemble label, so the measurement outcome of the $m$-th pulse is
\begin{equation} \label{outcome-r-ensemble}
    \mathbf{r}_{\text{out}}^{(k,m)} = \mathbf{r}_{\text{HD}}^{(k,m)} + \sqrt{\sigma_{\text{HD}}^{(k)} + \sigma_{\text{m}}} \mathbf{u}^{(k,m)} \ .
\end{equation}
In figure \ref{figure-1}b the outcome measurement of the first term of $\mathbf{r}_{\text{out}}^{(k,m)}$ for some number of pulses in the ensemble was shown. As we discussed at the end of subsection \ref{gaussian-measurements-appendix}, for homodyne detection of the $x$-quadratures the readout vector has $N$ terms (instead of the $2N$ terms that $\mathbf{r}_{\text{out}}^{(k,m)}$ in a more general scenario). For this section, we arrive at the final expressions in the general case and, then, take the homodyne limit. In section \ref{section-II}, we introduced the ensemble averages as 
\begin{equation}
\left \langle A^{(k)} \right \rangle_{M} = \frac{1}{M} \sum_{m=1}^{M} A^{(k,m)}
\end{equation}
for any generic measured observable, $A$. We use the following notation to denote the average limit of an infinite number of ensemble realization: $\left \langle A^{(k)} \right \rangle \equiv \lim_{M \to \infty} \left \langle A^{(k)} \right \rangle_{M}$. Having this defined, we consider the expected value of the covariance of the output signal from Eq. \eqref{outcome-r-ensemble} and obtain
\begin{equation} \label{second-moments-early}
\begin{aligned}
    \left \langle \mathbf{r}_{\text{out}}^{(k)} \mathbf{r}_{\text{out}}^{(k) \top} \right\rangle = \left\langle\mathbf{r}_{\text{HD}}^{(k)} \mathbf{r}_{\text{HD}}^{(k) \top}\right\rangle + \sigma_{\text{HD}}^{(k)} + \sigma_{\text{m}} \ ,
\end{aligned}
\end{equation}
in which we have taken advantage of the fact that the terms of every vector $\mathbf{u}^{(k,m)}$ are taken from a one-dimensional normal distribution with zero mean and variance equal to one. If we substitute $\mathbf{r}_{\text{HD}}^{(k,m)}$ with its expression from Eq. \eqref{PD-desarrollada}, perform the average over infinite pulses, and also substitute $\sigma_{\text{HD}}^{(k)}$ from its expression in Eq. \eqref{sigma-PD-long}, we can derive the resulting expression
\begin{equation} \label{out-BA-vanish}
\begin{aligned}
    \left \langle \mathbf{r}_{\text{out}}^{(k)} \mathbf{r}_{\text{out}}^{(k) \top} \right\rangle &= R S_{2} \sigma_{\text{anc}}^{(k)} S_{2}^{\top} + \sigma_{\text{m}} \\
    &+ T^{2}S_{2}\left[\sum_{d=1}^{k-1} R^{d-1} S_{1}^{d} \sigma_{\text{anc}}^{(k-d)} \left(S_{1}^{d}\right)^{\top}\right]S_{2}^{\top} \ .
\end{aligned}
\end{equation}
We already note that these second moments resemble the unconditional dynamics, as there are no terms depending on $\sigma_{\text{corr}}^{(k-d)} \left[ \sigma_{\text{HD}}^{(k-d)} + \sigma_{\text{m}} \right]^{-1} \sigma_{\text{corr}}^{(k-d) \top}$, which determined the conditional evolution of the feedback pulses, as seen in Eqs. \eqref{BA-r-k+1} and \eqref{BA-sigma-k+1}.
Although the pulses that reach the detector are not, in general, vacuum states (due to the conditional displacements), we can easily see that $\left \langle \mathbf{r}_{\text{out}}^{(k)} \right\rangle = \mathbf{0}$ (back to the unconditional scenario). In that case, the second moments expression that we derived in Eq. \eqref{out-BA-vanish} is equal to the covariance matrix of the output signal, that is, $\left \langle \mathbf{r}_{\text{out}}^{(k)} \mathbf{r}_{\text{out}}^{(k) \top} \right\rangle = \text{cov} \left( \mathbf{r}_{\text{out}}^{(k)} \right)$. We rename $\text{cov} \left( \mathbf{r}_{\text{out}}^{(k)} \right)$ to $\sigma_{\text{ideal}}^{(k)}$, yielding
\begin{equation} \label{ideal-decomposition-appendix}
    \sigma_{\text{ideal}}^{(k)} = \sum_{d=0}^{\infty} \gamma_{d}^{(k)} + \sigma_{\text{m}} \ ,
\end{equation}
in which each $\gamma_{d}^{(k)}$ is the term including $\sigma_{\text{anc}}^{(k-d)}$ from Eq. \eqref{out-BA-vanish}. Expression \eqref{ideal-decomposition-appendix} is similar to Eq. \eqref{ideal-decomposition}. The only difference lies in the fact that in Eq. \eqref{ideal-decomposition} the covariance corresponds to the position quadratures block as we have considered homodyne detection of the $x$-quadratures to obtain $\sigma_{\text{ideal}}^{(k)}$. This is equivalent to tracing out the $p$-quadrature degrees of freedom from every term in eq. \eqref{ideal-decomposition-appendix}, taking into account that $\text{Tr}_{\mathbf{p}}\left(\sigma_{\text{m}} \right) = 0$ (for this measurement scheme). This yields the expression
\begin{equation} \label{gamma-homodyne-eqs}
\begin{aligned}
    \gamma_{0}^{(k)} &= R \text{Tr}_{\mathbf{p}} \left\{S_{2} \sigma_{\text{anc}}^{(k)} S_{2}^{\top} \right\} \\
    \gamma_{d}^{(k)} &= (1-R)^{2}R^{d-1}\text{Tr}_{\mathbf{p}} \left\{S_{2} S_{1}^{d} \sigma_{\text{anc}}^{(k-d)} \left(S_{1}^{d}\right)^{\top}S_{2}^{\top}\right\} \\
    &\text{for} \ d \geq 1 \ ,
\end{aligned}
\end{equation}
in the case of homodyne detection of the $x$-quadratures. We have replaced the transmissivity $T$ by $1 - R$ to make the relation between Eqs. \eqref{gamma-homodyne-eqs} and \eqref{eq-gamma-decay} more evident. It can be noted from Eq. \eqref{gamma-homodyne-eqs} that the dependence on $d$ is not only found on $R^{d-1}$, but also inside the trace (in the terms $S_{1}^{d}$). However, we observe numerically that averaging among different Hamiltonians and different input strings removes every dependency on $d$ apart from the one in $R^{d-1}$.

\section{Fading memory condition and echo state property} \label{RC-conditions-proof}
In this section, we will briefly state how both the fading memory condition and the echo state property \cite{Konkoli2017,RCbook}, which are necessary conditions for a functional RC platform, are fulfilled by our platform. We begin from the theorem formulated in \cite{Nokkala2021} for linear Gaussian dynamics. We have a linear Gaussian system whose quadrature operators evolve at the $k$-th time step such as
\begin{equation}
    \hat{\mathbf{r}}_{\text{R}}^{(k+1)} = A \hat{\mathbf{r}}_{\text{R}}^{(k)} + B \hat{\mathbf{r}}_{\text{anc}}^{(k)} \ ,
\end{equation}
where $\hat{\mathbf{r}}_{\text{R}}^{(k)}$ is the quadrature operator vector of the reservoir and $\hat{\mathbf{r}}_{\text{anc}}^{(k)}$ the one of the ancilla input, both at time step $k$. Then both the echo state property and the fading memory condition are fulfilled if $\rho\left[A\right] < 1$, being $\rho\left[ \bullet \right]$ the \textit{spectral radius} of a matrix. In our case, the symplectic matrix determining the dynamics of our platform is the one from Eq. \eqref{S-prime-eq}, which can be explicitly written in the form
\begin{equation}
    S'(\Delta t) = \left(
    \begin{array}{cc}
        \sqrt{R}S_{1}(\Delta t) & -\sqrt{1-R} S_{1}(\Delta t) \\
        \sqrt{1-R} S_{2}(\Delta t) & \sqrt{R} S_{2}(\Delta t)
    \end{array}
    \right) \ .
\end{equation}
In our platform, it can be clearly seen how $A \equiv \sqrt{R} S_{1}(\Delta t)$, so we must have $\rho\left[\sqrt{R}S_{1}(\Delta t)\right] = \sqrt{R} \rho\left[S_{1}(\Delta t)\right] < 1$. It can be shown that $\rho\left[S_{1}(\Delta t)\right] = 1$. The proof is carried out as follows: from the time evolution of the quadrature operators in the Heisenberg picture under a quadratic Hamiltonian (as on Eq. \eqref{hamiltonian}) it can be shown that the resulting symplectic transformation can be written as $S_{1}(\Delta t) = \exp\left( \Omega H_{1} \Delta t \right)$, where $H_{1}$ is a $2N \times 2N$ symmetric matrix and $\Omega = \bigoplus_{i=1}^{N} \left(\begin{array}{cc}
    0 & 1 \\
    -1 & 0
\end{array}\right)$. We have only considered stable Hamiltonians in our simulations, so $H_{1} > 0$ (positive definite). In this scenario, it can be shown that the eigenvalues of $\Omega H_{1}$ are purely imaginary \cite{serafini2017}, which means that $\rho\left[ \exp\left( \Omega H_{1} \Delta t \right) \right] = 1$. We thus have that in our platform $\rho(A) = \sqrt{R}$, which is always $< 1$, and thus both the fading memory condition and the echo state property are fulfilled.

\section{Resolution analysis}
In this section, we will analyze in detail the main mathematical relations concerning the SNR of the delayed terms, $\gamma_{d}^{(k)}$. We will thus show how we have obtained the main theoretical results from section \ref{finite-pulses-case}. We start from the delayed SNR matrix, $\left|\gamma_{d}^{(k)}/\xi_{M}^{(k)}\right|$, that was introduced in section \ref{finite-pulses-case}. If we average the SNR among realizations of the input story and the Hamiltonians, we observe that
\begin{equation} \label{SNR-gamma-d_gamma-1}%
    \left\langle\left|\frac{\gamma_{d}^{(k)}}{ \xi_{M}^{(k)} } \right| \right\rangle_{\mathbf{s},\hat{H}}  = R^{d-1} \left\langle\left|\frac{\gamma_{1}^{(k)}}{ \xi_{M}^{(k)} } \right| \right\rangle_{\mathbf{s},\hat{H}} \ .
\end{equation}
Eq. \eqref{SNR-gamma-d_gamma-1} is a key numerical observation and will be important in the following arguments.
For the rest of the section, the subindex $\mathbf{s},\hat{H}$ and the label $(k)$ are removed for simplicity.
That is, when averaging over realizations the magnitude of the ideal delayed terms only differs on a factor equal to $R^{d-1}$. In Eq. \eqref{SNR-gamma-d_gamma-1} the $\gamma_{1}$ term can be further expanded as
\begin{equation} \label{SNR-gamma-1}%
    \left\langle\left|\frac{\gamma_{1}}{ \xi_{M} } \right| \right\rangle  = \sqrt{M} (1-R)^{2} \mathcal{C}_{\text{SNR}} \ .
\end{equation}
In Eq. \eqref{SNR-gamma-1} the term $\sqrt{M}$ comes from the dependency of $M$ of the noise, while the term $(1-R)^{2}$ denotes the transmissivity dependence from Eq. \eqref{eq-gamma-decay} that has been already commented. The term $\mathcal{C}_{\text{SNR}}$ is neither a function of $M$, $R$ nor $d$, but can be dependent on all the other parameters (including $N$).
From Eqs. \eqref{SNR-gamma-d_gamma-1} and \eqref{SNR-gamma-1} we are going to derive the main theoretical equations regarding the SNR.

\subsection{Derivation of Eq. \eqref{SNR-m-scaling}} \label{App-F1}

In this subsection, we derive an expression that accounts for how much we have to scale the number of measurements $M$ to be able to resolve $\Delta d$ steps in the past. We start by considering that we can already, with $M$ measurements, properly resolve inputs up to a certain delay $d$. That is, the SNR of $\gamma_{d}^{(k)}$ for a statistical noise $\xi_{M}^{(k)}$ is high enough. We now impose the SNR of $\gamma_{d+\Delta d}$ to be equal to the one of $\gamma_{d}^{(k)}$. This would require a higher number of measurements, which we call $M'$. This equality can be written as follows
\begin{equation} \label{SNR-equality}
    \left\langle\left|\frac{\gamma_{d}}{ \xi_{M} } \right|\right\rangle = \left\langle\left|\frac{\gamma_{d+\Delta d}}{ \xi_{M'} } \right|\right\rangle \ .
\end{equation}
If we now substitute from Eq. \eqref{SNR-gamma-d_gamma-1} and \eqref{SNR-gamma-1}, we get
\begin{equation}
    \sqrt{M} R^{d} = \sqrt{M^{\prime}} R^{d+\Delta d} \ .
\end{equation}
By reordering and simplifying the equation above we end up with the following result
\begin{equation} \label{SNR-m-scaling-appendix}
    \frac{M'}{M} = R^{-2\Delta d} \ ,
\end{equation}
which is the one shown in Eq. \eqref{SNR-m-scaling}. We can isolate the term $\Delta d$ from Eq. \eqref{SNR-m-scaling-appendix} to yield the following result
\begin{equation} \label{Delta_d_eq}
    \Delta d = \log_{R} \sqrt{\frac{M}{M'}} \ .
\end{equation}
This relation relates the measurement increase, $M'/M$, with the resolution enhancement, $\Delta d$. We note that a linear increase in the delay resolution yields an exponential increase in the number of measurements.
\subsection{SNR main equations} \label{App-F2}
The SNR lines shown in dBs in Fig. \ref{figure-3}b and \ref{figure-3}c have a simple mathematical representation. From Eq. \eqref{SNR-gamma-d_gamma-1} we can infer that
\begin{equation} \label{log-SNR-linear}
\begin{aligned}
 10 \log_{10}\left\langle\left|\frac{\gamma_{d}}{ \xi_{M} } \right|\right\rangle &= 10 \log_{10}\left\langle\left|\frac{\gamma_{1}}{ \xi_{M} } \right|\right\rangle \\
 &+ 10 (d-1) \log_{10} R\ ,
 \end{aligned}
\end{equation}
which is just a linear equation of $d$. In visual representations, such as in Fig. \ref{figure-3}b and \ref{figure-3}c, the slope corresponds to $10\log_{10} R$ and the height to $10 \log_{10} \langle| \gamma_{1} \xi_{M}^{-1} |\rangle$. Using Eq. \eqref{SNR-gamma-1}, we can further expand the SNR of $\gamma_{1}^{(k)}$ inside the logarithm as
\begin{equation} \label{height-equation}
\begin{aligned}
    10 \log_{10}\left\langle\left|\frac{\gamma_{1}}{ \xi_{M} } \right| \right\rangle  &= 10 \left[ \frac{1}{2}\log_{10} M + 2 \log_{10} (1-R) \right. \\
    &+ \left. \log_{10}\mathcal{C}_{\text{SNR}} \right] \ .
\end{aligned}
\end{equation}
The SNR thus has a simple behavior when changing the parameters $M$ and $R$. For instance, changing the number of measurements from $M$ to $M'$ yields a difference on the SNR of $\gamma_{1}^{(k)}$ of
\begin{equation} \label{Delta-h-m}
\begin{aligned}
    \left\langle\left|\frac{\gamma_{1}}{\xi_{M'}}\right|\right\rangle - \left\langle\left|\frac{\gamma_{1}}{\xi_{M}}\right|\right\rangle = 5\log_{10}\left(\frac{M^{\prime}}{M}\right) \ .
\end{aligned}
\end{equation}
For the case of changing the reflectivity from $R$ to $R'$, the difference on the $\gamma_{1}^{(k)}$ term is given by
\begin{equation} \label{Delta-h-R}
    \left\langle\left|\frac{\gamma_{1}(R')}{\xi_{M}}\right|\right\rangle - \left\langle\left|\frac{\gamma_{1}(R)}{\xi_{M}}\right|\right\rangle = 20\log_{10}\left(\frac{1-R^{\prime}}{1-R}\right) \ .
\end{equation}
The term multiplying $(d-1)$ (slope) also changes to $10\log_{10} R'$.
\subsection{Derivation of Eq. \eqref{R-m-functions-scaling}} \label{App-F3}
\begin{figure}[h!]
    \centering
    \includegraphics[width=\linewidth]{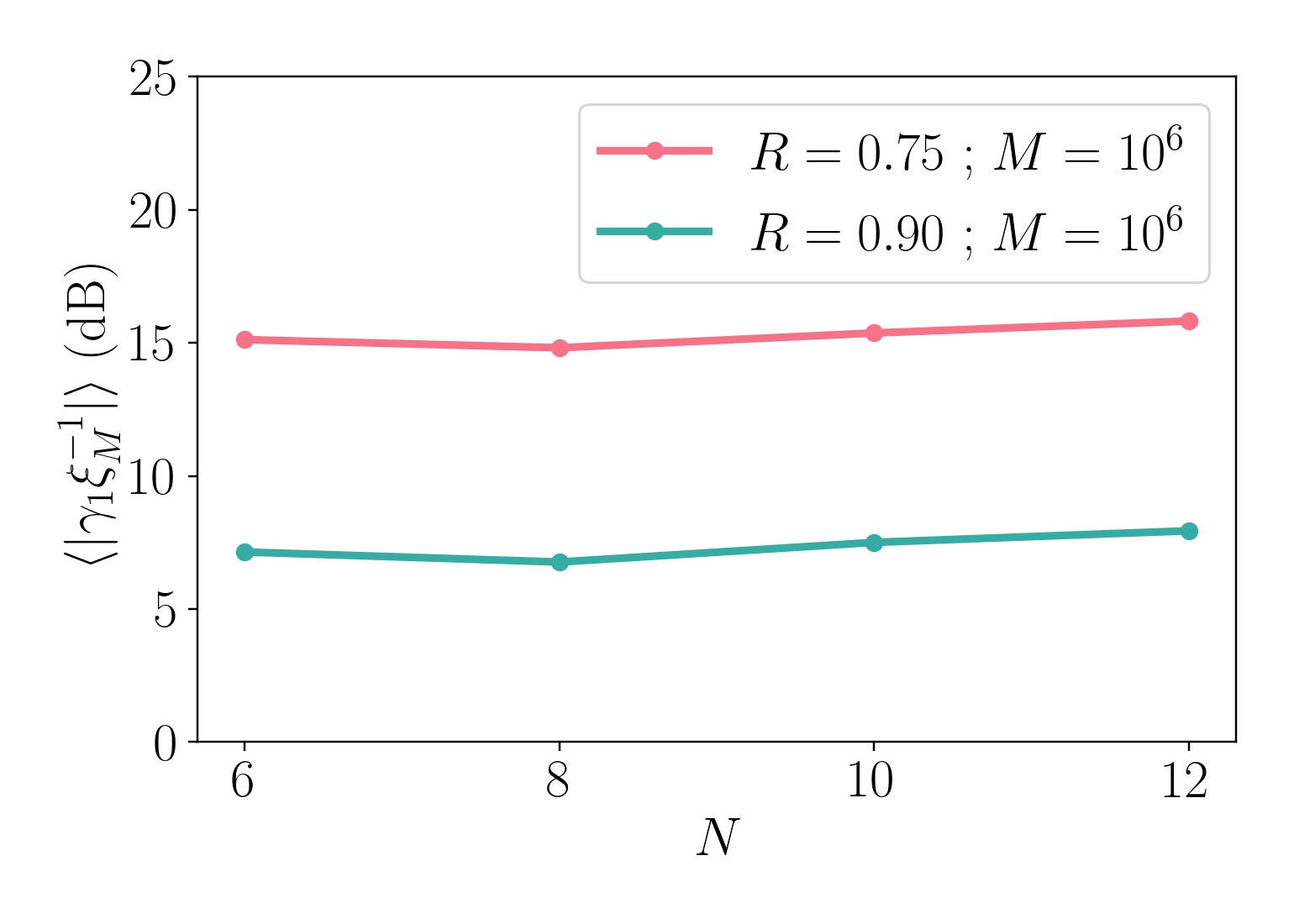}
    \caption{SNR of $\gamma_{1}^{(k)}$ as a function of $N$. The different lines correspond to a different value of $R$, while $M = 10^{6}$ for both.}
    \label{figure-5}
\end{figure}

In this subsection, we now consider a dependency with the size of the system, $N$. We aim to get an equation that allows us to improve the delay resolution quadratically without an exponential scaling of $M$. We argued that only the SNR of $\gamma_{1}^{(k)}$ has a dependency on $N$ through the term $\mathcal{C}_{\text{SNR}}$ in Eq. \eqref{SNR-gamma-1}. In the size intervals we have considered for this article (up to $N = 12$), $\langle| \gamma_{1} \xi_{M}^{-1} |\rangle$ remained practically constant with $N$, as it is shown in Fig. \ref{figure-5}. We have thus taken the approximation in which we consider no dependency on $N$.
\\
\\
The strategy to get a quadratic increase in resolution as we increase $N$, which ensures that the normalized IPC at least remains constant, is conceived as follows. We know that for higher values of $R$ it is less costly to resolve further terms into the past. We are thus interested in increasing the value of $R$ with $N$. We consider the following function
\begin{equation} \label{R-N-dependency}
    R(N) = 1 - \frac{\mathcal{C}}{N^{2}} \ ,
\end{equation}
in which $\mathcal{C}$ is an arbitrary constant. We recall the SNR equality from Eq. \eqref{SNR-equality} that we used to obtain Eq. \eqref{SNR-m-scaling-appendix}, taking into account that now neither $R$ nor $M$ remain constant. We take the delay so that, for a fixed SNR, $d(N) = \alpha N^{2}$, where $\alpha$ is an arbitrary constant. This will ensure that the resolution scales quadratically with $N$. We can rewrite the equality in Eq. \eqref{SNR-equality} as
\begin{equation}
    \sqrt{M} (1-R)^{2} R^{d-1} = \sqrt{M'} (1-R')^{2} (R')^{d'-1} \ ,
\end{equation}
where $M$, $R$ and $d$ are functions of $N$ and $M'$, $R'$ and $d'$ are functions of $N'$. The equality above can be reordered as
\begin{equation} \label{M'/M-scaling}
    \sqrt{\frac{M'}{M}} = \left( \frac{1-R}{1-R'} \right)^{2} \frac{R^{d-1}}{(R')^{d'-1}} \ .
\end{equation}
We now turn our attention to the last term regarding the $R^{d-1}$ and $(R')^{d'-1}$. If we substitute the chosen functions $R(N)$ and $d(N)$ in the term $R^{d-1}$, we can perform a series expansion of $N^{-1}$, yielding
\begin{equation}
\begin{aligned}
    R(N)^{d(N)-1} &= \left( 1 - \frac{\mathcal{C}}{N^{2}} \right)^{\alpha N^{2}-1} \\
    &= e^{-\mathcal{C}\alpha} + \mathcal{O}\left(N^{-2}\right) \ .
\end{aligned}
\end{equation}
Thus, the fraction of $R^{d-1}/(R')^{d'-1} \simeq 1$. With this, Eq. \eqref{M'/M-scaling} can be rewritten as
\begin{equation} \label{M'/M-definitive}
    \frac{M'}{M} \simeq \left( \frac{N'}{N} \right)^{8} \ ,
\end{equation}
We end up with $M(N) \propto N^{8}$, which is the polynomial scaling condition from Eq. \eqref{R-m-functions-scaling}.

\newpage
\begin{thebibliography}{73}%
\makeatletter
\providecommand \@ifxundefined [1]{%
 \@ifx{#1\undefined}
}%
\providecommand \@ifnum [1]{%
 \ifnum #1\expandafter \@firstoftwo
 \else \expandafter \@secondoftwo
 \fi
}%
\providecommand \@ifx [1]{%
 \ifx #1\expandafter \@firstoftwo
 \else \expandafter \@secondoftwo
 \fi
}%
\providecommand \natexlab [1]{#1}%
\providecommand \enquote  [1]{``#1''}%
\providecommand \bibnamefont  [1]{#1}%
\providecommand \bibfnamefont [1]{#1}%
\providecommand \citenamefont [1]{#1}%
\providecommand \href@noop [0]{\@secondoftwo}%
\providecommand \href [0]{\begingroup \@sanitize@url \@href}%
\providecommand \@href[1]{\@@startlink{#1}\@@href}%
\providecommand \@@href[1]{\endgroup#1\@@endlink}%
\providecommand \@sanitize@url [0]{\catcode `\\12\catcode `\$12\catcode
  `\&12\catcode `\#12\catcode `\^12\catcode `\_12\catcode `\%12\relax}%
\providecommand \@@startlink[1]{}%
\providecommand \@@endlink[0]{}%
\providecommand \url  [0]{\begingroup\@sanitize@url \@url }%
\providecommand \@url [1]{\endgroup\@href {#1}{\urlprefix }}%
\providecommand \urlprefix  [0]{URL }%
\providecommand \Eprint [0]{\href }%
\providecommand \doibase [0]{https://doi.org/}%
\providecommand \selectlanguage [0]{\@gobble}%
\providecommand \bibinfo  [0]{\@secondoftwo}%
\providecommand \bibfield  [0]{\@secondoftwo}%
\providecommand \translation [1]{[#1]}%
\providecommand \BibitemOpen [0]{}%
\providecommand \bibitemStop [0]{}%
\providecommand \bibitemNoStop [0]{.\EOS\space}%
\providecommand \EOS [0]{\spacefactor3000\relax}%
\providecommand \BibitemShut  [1]{\csname bibitem#1\endcsname}%
\let\auto@bib@innerbib\@empty
\bibitem [{\citenamefont {Slussarenko}\ and\ \citenamefont
  {Pryde}(2019)}]{doi:10.1063/1.5115814}%
  \BibitemOpen
  \bibfield  {author} {\bibinfo {author} {\bibfnamefont {S.}~\bibnamefont
  {Slussarenko}}\ and\ \bibinfo {author} {\bibfnamefont {G.~J.}\ \bibnamefont
  {Pryde}},\ }\bibfield  {title} {\bibinfo {title} {Photonic quantum
  information processing: A concise review},\ }\href
  {https://doi.org/10.1063/1.5115814} {\bibfield  {journal} {\bibinfo
  {journal} {Applied Physics Reviews}\ }\textbf {\bibinfo {volume} {6}},\
  \bibinfo {pages} {041303} (\bibinfo {year} {2019})},\ \Eprint
  {https://arxiv.org/abs/https://doi.org/10.1063/1.5115814}
  {https://doi.org/10.1063/1.5115814} \BibitemShut {NoStop}%
\bibitem [{\citenamefont {Takeda}\ and\ \citenamefont
  {Furusawa}(2019)}]{Takeda2019}%
  \BibitemOpen
  \bibfield  {author} {\bibinfo {author} {\bibfnamefont {S.}~\bibnamefont
  {Takeda}}\ and\ \bibinfo {author} {\bibfnamefont {A.}~\bibnamefont
  {Furusawa}},\ }\bibfield  {title} {\bibinfo {title} {Toward large-scale
  fault-tolerant universal photonic quantum computing},\ }\href
  {https://doi.org/10.1063/1.5100160} {\bibfield  {journal} {\bibinfo
  {journal} {APL Photonics}\ }\textbf {\bibinfo {volume} {4}},\ \bibinfo
  {pages} {060902} (\bibinfo {year} {2019})},\ \Eprint
  {https://arxiv.org/abs/https://doi.org/10.1063/1.5100160}
  {https://doi.org/10.1063/1.5100160} \BibitemShut {NoStop}%
\bibitem [{\citenamefont {Wang}\ \emph {et~al.}(2019)\citenamefont {Wang},
  \citenamefont {Qin}, \citenamefont {Ding}, \citenamefont {Chen},
  \citenamefont {Chen}, \citenamefont {You}, \citenamefont {He}, \citenamefont
  {Jiang}, \citenamefont {You}, \citenamefont {Wang}, \citenamefont
  {Schneider}, \citenamefont {Renema}, \citenamefont {H{\"{o}}fling},
  \citenamefont {Lu},\ and\ \citenamefont {Pan}}]{Wang2019BosonSpace}%
  \BibitemOpen
  \bibfield  {author} {\bibinfo {author} {\bibfnamefont {H.}~\bibnamefont
  {Wang}}, \bibinfo {author} {\bibfnamefont {J.}~\bibnamefont {Qin}}, \bibinfo
  {author} {\bibfnamefont {X.}~\bibnamefont {Ding}}, \bibinfo {author}
  {\bibfnamefont {M.~C.}\ \bibnamefont {Chen}}, \bibinfo {author}
  {\bibfnamefont {S.}~\bibnamefont {Chen}}, \bibinfo {author} {\bibfnamefont
  {X.}~\bibnamefont {You}}, \bibinfo {author} {\bibfnamefont {Y.~M.}\
  \bibnamefont {He}}, \bibinfo {author} {\bibfnamefont {X.}~\bibnamefont
  {Jiang}}, \bibinfo {author} {\bibfnamefont {L.}~\bibnamefont {You}}, \bibinfo
  {author} {\bibfnamefont {Z.}~\bibnamefont {Wang}}, \bibinfo {author}
  {\bibfnamefont {C.}~\bibnamefont {Schneider}}, \bibinfo {author}
  {\bibfnamefont {J.~J.}\ \bibnamefont {Renema}}, \bibinfo {author}
  {\bibfnamefont {S.}~\bibnamefont {H{\"{o}}fling}}, \bibinfo {author}
  {\bibfnamefont {C.~Y.}\ \bibnamefont {Lu}},\ and\ \bibinfo {author}
  {\bibfnamefont {J.~W.}\ \bibnamefont {Pan}},\ }\bibfield  {title} {\bibinfo
  {title} {{Boson Sampling with 20 Input Photons and a 60-Mode Interferometer
  in a 1014 -Dimensional Hilbert Space}},\ }\href
  {https://doi.org/10.1103/PhysRevLett.123.250503} {\bibfield  {journal}
  {\bibinfo  {journal} {Physical Review Letters}\ }\textbf {\bibinfo {volume}
  {123}},\ \bibinfo {pages} {250503} (\bibinfo {year} {2019})}\BibitemShut
  {NoStop}%
\bibitem [{\citenamefont {Zhong}\ \emph {et~al.}(2020)\citenamefont {Zhong},
  \citenamefont {Wang}, \citenamefont {Deng}, \citenamefont {Chen},
  \citenamefont {Peng}, \citenamefont {Luo}, \citenamefont {Qin}, \citenamefont
  {Wu}, \citenamefont {Ding}, \citenamefont {Hu}, \citenamefont {Hu},
  \citenamefont {Yang}, \citenamefont {Zhang}, \citenamefont {Li},
  \citenamefont {Li}, \citenamefont {Jiang}, \citenamefont {Gan}, \citenamefont
  {Yang}, \citenamefont {You}, \citenamefont {Wang}, \citenamefont {Li},
  \citenamefont {Liu}, \citenamefont {Lu},\ and\ \citenamefont
  {Pan}}]{Zhong2020QuantumPhotons}%
  \BibitemOpen
  \bibfield  {author} {\bibinfo {author} {\bibfnamefont {H.~S.}\ \bibnamefont
  {Zhong}}, \bibinfo {author} {\bibfnamefont {H.}~\bibnamefont {Wang}},
  \bibinfo {author} {\bibfnamefont {Y.~H.}\ \bibnamefont {Deng}}, \bibinfo
  {author} {\bibfnamefont {M.~C.}\ \bibnamefont {Chen}}, \bibinfo {author}
  {\bibfnamefont {L.~C.}\ \bibnamefont {Peng}}, \bibinfo {author}
  {\bibfnamefont {Y.~H.}\ \bibnamefont {Luo}}, \bibinfo {author} {\bibfnamefont
  {J.}~\bibnamefont {Qin}}, \bibinfo {author} {\bibfnamefont {D.}~\bibnamefont
  {Wu}}, \bibinfo {author} {\bibfnamefont {X.}~\bibnamefont {Ding}}, \bibinfo
  {author} {\bibfnamefont {Y.}~\bibnamefont {Hu}}, \bibinfo {author}
  {\bibfnamefont {P.}~\bibnamefont {Hu}}, \bibinfo {author} {\bibfnamefont
  {X.~Y.}\ \bibnamefont {Yang}}, \bibinfo {author} {\bibfnamefont {W.~J.}\
  \bibnamefont {Zhang}}, \bibinfo {author} {\bibfnamefont {H.}~\bibnamefont
  {Li}}, \bibinfo {author} {\bibfnamefont {Y.}~\bibnamefont {Li}}, \bibinfo
  {author} {\bibfnamefont {X.}~\bibnamefont {Jiang}}, \bibinfo {author}
  {\bibfnamefont {L.}~\bibnamefont {Gan}}, \bibinfo {author} {\bibfnamefont
  {G.}~\bibnamefont {Yang}}, \bibinfo {author} {\bibfnamefont {L.}~\bibnamefont
  {You}}, \bibinfo {author} {\bibfnamefont {Z.}~\bibnamefont {Wang}}, \bibinfo
  {author} {\bibfnamefont {L.}~\bibnamefont {Li}}, \bibinfo {author}
  {\bibfnamefont {N.~L.}\ \bibnamefont {Liu}}, \bibinfo {author} {\bibfnamefont
  {C.~Y.}\ \bibnamefont {Lu}},\ and\ \bibinfo {author} {\bibfnamefont {J.~W.}\
  \bibnamefont {Pan}},\ }\bibfield  {title} {\bibinfo {title} {{Quantum
  computational advantage using photons}},\ }\href
  {https://doi.org/10.1126/science.abe8770} {\bibfield  {journal} {\bibinfo
  {journal} {Science}\ }\textbf {\bibinfo {volume} {370}},\ \bibinfo {pages}
  {1460} (\bibinfo {year} {2020})}\BibitemShut {NoStop}%
\bibitem [{\citenamefont {Yoshikawa}\ \emph {et~al.}(2016)\citenamefont
  {Yoshikawa}, \citenamefont {Yokoyama}, \citenamefont {Kaji}, \citenamefont
  {Sornphiphatphong}, \citenamefont {Shiozawa}, \citenamefont {Makino},\ and\
  \citenamefont {Furusawa}}]{Yoshikawa2016}%
  \BibitemOpen
  \bibfield  {author} {\bibinfo {author} {\bibfnamefont {J.-i.}\ \bibnamefont
  {Yoshikawa}}, \bibinfo {author} {\bibfnamefont {S.}~\bibnamefont {Yokoyama}},
  \bibinfo {author} {\bibfnamefont {T.}~\bibnamefont {Kaji}}, \bibinfo {author}
  {\bibfnamefont {C.}~\bibnamefont {Sornphiphatphong}}, \bibinfo {author}
  {\bibfnamefont {Y.}~\bibnamefont {Shiozawa}}, \bibinfo {author}
  {\bibfnamefont {K.}~\bibnamefont {Makino}},\ and\ \bibinfo {author}
  {\bibfnamefont {A.}~\bibnamefont {Furusawa}},\ }\bibfield  {title} {\bibinfo
  {title} {Invited article: Generation of one-million-mode continuous-variable
  cluster state by unlimited time-domain multiplexing},\ }\href
  {https://doi.org/10.1063/1.4962732} {\bibfield  {journal} {\bibinfo
  {journal} {APL Photonics}\ }\textbf {\bibinfo {volume} {1}},\ \bibinfo
  {pages} {060801} (\bibinfo {year} {2016})},\ \Eprint
  {https://arxiv.org/abs/https://doi.org/10.1063/1.4962732}
  {https://doi.org/10.1063/1.4962732} \BibitemShut {NoStop}%
\bibitem [{\citenamefont {Cai}\ \emph {et~al.}(2017)\citenamefont {Cai},
  \citenamefont {Roslund}, \citenamefont {Ferrini}, \citenamefont {Arzani},
  \citenamefont {Xu}, \citenamefont {Fabre},\ and\ \citenamefont
  {Treps}}]{Cai2017b}%
  \BibitemOpen
  \bibfield  {author} {\bibinfo {author} {\bibfnamefont {Y.}~\bibnamefont
  {Cai}}, \bibinfo {author} {\bibfnamefont {J.}~\bibnamefont {Roslund}},
  \bibinfo {author} {\bibfnamefont {G.}~\bibnamefont {Ferrini}}, \bibinfo
  {author} {\bibfnamefont {F.}~\bibnamefont {Arzani}}, \bibinfo {author}
  {\bibfnamefont {X.}~\bibnamefont {Xu}}, \bibinfo {author} {\bibfnamefont
  {C.}~\bibnamefont {Fabre}},\ and\ \bibinfo {author} {\bibfnamefont
  {N.}~\bibnamefont {Treps}},\ }\bibfield  {title} {\bibinfo {title} {Multimode
  entanglement in reconfigurable graph states using optical frequency combs},\
  }\href {https://doi.org/10.1038/ncomms15645} {\bibfield  {journal} {\bibinfo
  {journal} {Nature Communications}\ }\textbf {\bibinfo {volume} {8}},\
  \bibinfo {pages} {15645} (\bibinfo {year} {2017})}\BibitemShut {NoStop}%
\bibitem [{\citenamefont {Peruzzo}\ \emph {et~al.}(2014)\citenamefont
  {Peruzzo}, \citenamefont {McClean}, \citenamefont {Shadbolt}, \citenamefont
  {Yung}, \citenamefont {Zhou}, \citenamefont {Love}, \citenamefont
  {Aspuru-Guzik},\ and\ \citenamefont {O'Brien}}]{Peruzzo2014}%
  \BibitemOpen
  \bibfield  {author} {\bibinfo {author} {\bibfnamefont {A.}~\bibnamefont
  {Peruzzo}}, \bibinfo {author} {\bibfnamefont {J.}~\bibnamefont {McClean}},
  \bibinfo {author} {\bibfnamefont {P.}~\bibnamefont {Shadbolt}}, \bibinfo
  {author} {\bibfnamefont {M.-H.}\ \bibnamefont {Yung}}, \bibinfo {author}
  {\bibfnamefont {X.-Q.}\ \bibnamefont {Zhou}}, \bibinfo {author}
  {\bibfnamefont {P.~J.}\ \bibnamefont {Love}}, \bibinfo {author}
  {\bibfnamefont {A.}~\bibnamefont {Aspuru-Guzik}},\ and\ \bibinfo {author}
  {\bibfnamefont {J.~L.}\ \bibnamefont {O'Brien}},\ }\bibfield  {title}
  {\bibinfo {title} {A variational eigenvalue solver on a photonic quantum
  processor},\ }\href {https://doi.org/10.1038/ncomms5213} {\bibfield
  {journal} {\bibinfo  {journal} {Nature Communications}\ }\textbf {\bibinfo
  {volume} {5}},\ \bibinfo {pages} {4213} (\bibinfo {year} {2014})}\BibitemShut
  {NoStop}%
\bibitem [{\citenamefont {McMahon}\ \emph {et~al.}(2016)\citenamefont
  {McMahon}, \citenamefont {Marandi}, \citenamefont {Haribara}, \citenamefont
  {Hamerly}, \citenamefont {Langrock}, \citenamefont {Tamate}, \citenamefont
  {Inagaki}, \citenamefont {Takesue}, \citenamefont {Utsunomiya}, \citenamefont
  {Aihara}, \citenamefont {Byer}, \citenamefont {Fejer}, \citenamefont
  {Mabuchi},\ and\ \citenamefont {Yamamoto}}]{McMahon2016}%
  \BibitemOpen
  \bibfield  {author} {\bibinfo {author} {\bibfnamefont {P.~L.}\ \bibnamefont
  {McMahon}}, \bibinfo {author} {\bibfnamefont {A.}~\bibnamefont {Marandi}},
  \bibinfo {author} {\bibfnamefont {Y.}~\bibnamefont {Haribara}}, \bibinfo
  {author} {\bibfnamefont {R.}~\bibnamefont {Hamerly}}, \bibinfo {author}
  {\bibfnamefont {C.}~\bibnamefont {Langrock}}, \bibinfo {author}
  {\bibfnamefont {S.}~\bibnamefont {Tamate}}, \bibinfo {author} {\bibfnamefont
  {T.}~\bibnamefont {Inagaki}}, \bibinfo {author} {\bibfnamefont
  {H.}~\bibnamefont {Takesue}}, \bibinfo {author} {\bibfnamefont
  {S.}~\bibnamefont {Utsunomiya}}, \bibinfo {author} {\bibfnamefont
  {K.}~\bibnamefont {Aihara}}, \bibinfo {author} {\bibfnamefont {R.~L.}\
  \bibnamefont {Byer}}, \bibinfo {author} {\bibfnamefont {M.~M.}\ \bibnamefont
  {Fejer}}, \bibinfo {author} {\bibfnamefont {H.}~\bibnamefont {Mabuchi}},\
  and\ \bibinfo {author} {\bibfnamefont {Y.}~\bibnamefont {Yamamoto}},\
  }\bibfield  {title} {\bibinfo {title} {A fully programmable 100-spin coherent
  ising machine with all-to-all connections},\ }\href
  {https://doi.org/10.1126/science.aah5178} {\bibfield  {journal} {\bibinfo
  {journal} {Science}\ }\textbf {\bibinfo {volume} {354}},\ \bibinfo {pages}
  {614} (\bibinfo {year} {2016})},\ \Eprint
  {https://arxiv.org/abs/https://www.science.org/doi/pdf/10.1126/science.aah5178}
  {https://www.science.org/doi/pdf/10.1126/science.aah5178} \BibitemShut
  {NoStop}%
\bibitem [{\citenamefont {Honjo}\ \emph {et~al.}(2021)\citenamefont {Honjo},
  \citenamefont {Sonobe}, \citenamefont {Inaba}, \citenamefont {Inagaki},
  \citenamefont {Ikuta}, \citenamefont {Yamada}, \citenamefont {Kazama},
  \citenamefont {Enbutsu}, \citenamefont {Umeki}, \citenamefont {Kasahara},
  \citenamefont {ichi Kawarabayashi},\ and\ \citenamefont
  {Takesue}}]{Honjo2021}%
  \BibitemOpen
  \bibfield  {author} {\bibinfo {author} {\bibfnamefont {T.}~\bibnamefont
  {Honjo}}, \bibinfo {author} {\bibfnamefont {T.}~\bibnamefont {Sonobe}},
  \bibinfo {author} {\bibfnamefont {K.}~\bibnamefont {Inaba}}, \bibinfo
  {author} {\bibfnamefont {T.}~\bibnamefont {Inagaki}}, \bibinfo {author}
  {\bibfnamefont {T.}~\bibnamefont {Ikuta}}, \bibinfo {author} {\bibfnamefont
  {Y.}~\bibnamefont {Yamada}}, \bibinfo {author} {\bibfnamefont
  {T.}~\bibnamefont {Kazama}}, \bibinfo {author} {\bibfnamefont
  {K.}~\bibnamefont {Enbutsu}}, \bibinfo {author} {\bibfnamefont
  {T.}~\bibnamefont {Umeki}}, \bibinfo {author} {\bibfnamefont
  {R.}~\bibnamefont {Kasahara}}, \bibinfo {author} {\bibfnamefont
  {K.}~\bibnamefont {ichi Kawarabayashi}},\ and\ \bibinfo {author}
  {\bibfnamefont {H.}~\bibnamefont {Takesue}},\ }\bibfield  {title} {\bibinfo
  {title} {100,000-spin coherent ising machine},\ }\href
  {https://doi.org/10.1126/sciadv.abh0952} {\bibfield  {journal} {\bibinfo
  {journal} {Science Advances}\ }\textbf {\bibinfo {volume} {7}},\ \bibinfo
  {pages} {eabh0952} (\bibinfo {year} {2021})},\ \Eprint
  {https://arxiv.org/abs/https://www.science.org/doi/pdf/10.1126/sciadv.abh0952}
  {https://www.science.org/doi/pdf/10.1126/sciadv.abh0952} \BibitemShut
  {NoStop}%
\bibitem [{\citenamefont {Pierangeli}\ \emph {et~al.}(2019)\citenamefont
  {Pierangeli}, \citenamefont {Marcucci},\ and\ \citenamefont
  {Conti}}]{pierangeli2019large}%
  \BibitemOpen
  \bibfield  {author} {\bibinfo {author} {\bibfnamefont {D.}~\bibnamefont
  {Pierangeli}}, \bibinfo {author} {\bibfnamefont {G.}~\bibnamefont
  {Marcucci}},\ and\ \bibinfo {author} {\bibfnamefont {C.}~\bibnamefont
  {Conti}},\ }\bibfield  {title} {\bibinfo {title} {Large-scale photonic ising
  machine by spatial light modulation},\ }\href
  {https://doi.org/10.1103/PhysRevLett.122.213902} {\bibfield  {journal}
  {\bibinfo  {journal} {Phys. Rev. Lett.}\ }\textbf {\bibinfo {volume} {122}},\
  \bibinfo {pages} {213902} (\bibinfo {year} {2019})}\BibitemShut {NoStop}%
\bibitem [{\citenamefont {B{\"o}hm}\ \emph {et~al.}(2019)\citenamefont
  {B{\"o}hm}, \citenamefont {Verschaffelt},\ and\ \citenamefont {Van~der
  Sande}}]{bohm2019poor}%
  \BibitemOpen
  \bibfield  {author} {\bibinfo {author} {\bibfnamefont {F.}~\bibnamefont
  {B{\"o}hm}}, \bibinfo {author} {\bibfnamefont {G.}~\bibnamefont
  {Verschaffelt}},\ and\ \bibinfo {author} {\bibfnamefont {G.}~\bibnamefont
  {Van~der Sande}},\ }\bibfield  {title} {\bibinfo {title} {A poor man’s
  coherent {Ising} machine based on opto-electronic feedback systems for
  solving optimization problems},\ }\href@noop {} {\bibfield  {journal}
  {\bibinfo  {journal} {Nature communications}\ }\textbf {\bibinfo {volume}
  {10}},\ \bibinfo {pages} {3538} (\bibinfo {year} {2019})}\BibitemShut
  {NoStop}%
\bibitem [{\citenamefont {Wyffels}\ and\ \citenamefont
  {Schrauwen}(2010)}]{WYFFELS20101958}%
  \BibitemOpen
  \bibfield  {author} {\bibinfo {author} {\bibfnamefont {F.}~\bibnamefont
  {Wyffels}}\ and\ \bibinfo {author} {\bibfnamefont {B.}~\bibnamefont
  {Schrauwen}},\ }\bibfield  {title} {\bibinfo {title} {A comparative study of
  reservoir computing strategies for monthly time series prediction},\ }\href
  {https://doi.org/https://doi.org/10.1016/j.neucom.2010.01.016} {\bibfield
  {journal} {\bibinfo  {journal} {Neurocomputing}\ }\textbf {\bibinfo {volume}
  {73}},\ \bibinfo {pages} {1958} (\bibinfo {year} {2010})},\ \bibinfo {note}
  {subspace Learning / Selected papers from the European Symposium on Time
  Series Prediction}\BibitemShut {NoStop}%
\bibitem [{\citenamefont {Lin}\ \emph {et~al.}(2009)\citenamefont {Lin},
  \citenamefont {Yang},\ and\ \citenamefont {Song}}]{LIN20097313}%
  \BibitemOpen
  \bibfield  {author} {\bibinfo {author} {\bibfnamefont {X.}~\bibnamefont
  {Lin}}, \bibinfo {author} {\bibfnamefont {Z.}~\bibnamefont {Yang}},\ and\
  \bibinfo {author} {\bibfnamefont {Y.}~\bibnamefont {Song}},\ }\bibfield
  {title} {\bibinfo {title} {Short-term stock price prediction based on echo
  state networks},\ }\href
  {https://doi.org/https://doi.org/10.1016/j.eswa.2008.09.049} {\bibfield
  {journal} {\bibinfo  {journal} {Expert Systems with Applications}\ }\textbf
  {\bibinfo {volume} {36}},\ \bibinfo {pages} {7313} (\bibinfo {year}
  {2009})}\BibitemShut {NoStop}%
\bibitem [{\citenamefont {Ilies}\ \emph {et~al.}(2007)\citenamefont {Ilies},
  \citenamefont {Jaeger}, \citenamefont {Kosuchinas},\ and\ \citenamefont
  {Rincon}}]{Ilies07steppingforward}%
  \BibitemOpen
  \bibfield  {author} {\bibinfo {author} {\bibfnamefont {I.}~\bibnamefont
  {Ilies}}, \bibinfo {author} {\bibfnamefont {H.}~\bibnamefont {Jaeger}},
  \bibinfo {author} {\bibfnamefont {O.}~\bibnamefont {Kosuchinas}},\ and\
  \bibinfo {author} {\bibfnamefont {M.}~\bibnamefont {Rincon}},\ }\href@noop {}
  {\emph {\bibinfo {title} {Stepping forward through echoes of the past:
  forecasting with Echo State Networks}}},\ \bibinfo {type} {Tech. Rep.}\
  (\bibinfo {year} {2007})\BibitemShut {NoStop}%
\bibitem [{\citenamefont {Nakajima}\ and\ \citenamefont
  {Fischer}(2021)}]{RCbook}%
  \BibitemOpen
  \bibfield  {author} {\bibinfo {author} {\bibfnamefont {K.}~\bibnamefont
  {Nakajima}}\ and\ \bibinfo {author} {\bibfnamefont {I.}~\bibnamefont
  {Fischer}},\ }\href@noop {} {\emph {\bibinfo {title} {Reservoir Computing:
  Theory, Physical Implementations, and Applications}}}\ (\bibinfo  {publisher}
  {Springer},\ \bibinfo {year} {2021})\BibitemShut {NoStop}%
\bibitem [{\citenamefont {Coulibaly}(2010)}]{COULIBALY201076}%
  \BibitemOpen
  \bibfield  {author} {\bibinfo {author} {\bibfnamefont {P.}~\bibnamefont
  {Coulibaly}},\ }\bibfield  {title} {\bibinfo {title} {Reservoir computing
  approach to great lakes water level forecasting},\ }\href
  {https://doi.org/https://doi.org/10.1016/j.jhydrol.2009.11.027} {\bibfield
  {journal} {\bibinfo  {journal} {Journal of Hydrology}\ }\textbf {\bibinfo
  {volume} {381}},\ \bibinfo {pages} {76} (\bibinfo {year} {2010})}\BibitemShut
  {NoStop}%
\bibitem [{\citenamefont {Triefenbach}\ \emph {et~al.}(2010)\citenamefont
  {Triefenbach}, \citenamefont {Jalalvand}, \citenamefont {Schrauwen},\ and\
  \citenamefont {Martens}}]{NIPS2010_2ca65f58}%
  \BibitemOpen
  \bibfield  {author} {\bibinfo {author} {\bibfnamefont {F.}~\bibnamefont
  {Triefenbach}}, \bibinfo {author} {\bibfnamefont {A.}~\bibnamefont
  {Jalalvand}}, \bibinfo {author} {\bibfnamefont {B.}~\bibnamefont
  {Schrauwen}},\ and\ \bibinfo {author} {\bibfnamefont {J.-p.}\ \bibnamefont
  {Martens}},\ }\bibfield  {title} {\bibinfo {title} {Phoneme recognition with
  large hierarchical reservoirs},\ }in\ \href
  {https://proceedings.neurips.cc/paper/2010/file/2ca65f58e35d9ad45bf7f3ae5cfd08f1-Paper.pdf}
  {\emph {\bibinfo {booktitle} {Advances in Neural Information Processing
  Systems}}},\ Vol.~\bibinfo {volume} {23},\ \bibinfo {editor} {edited by\
  \bibinfo {editor} {\bibfnamefont {J.}~\bibnamefont {Lafferty}}, \bibinfo
  {editor} {\bibfnamefont {C.}~\bibnamefont {Williams}}, \bibinfo {editor}
  {\bibfnamefont {J.}~\bibnamefont {Shawe-Taylor}}, \bibinfo {editor}
  {\bibfnamefont {R.}~\bibnamefont {Zemel}},\ and\ \bibinfo {editor}
  {\bibfnamefont {A.}~\bibnamefont {Culotta}}}\ (\bibinfo  {publisher} {Curran
  Associates, Inc.},\ \bibinfo {year} {2010})\BibitemShut {NoStop}%
\bibitem [{\citenamefont {Wang}\ \emph {et~al.}(2016)\citenamefont {Wang},
  \citenamefont {Wang},\ and\ \citenamefont {Liu}}]{WANG2016237}%
  \BibitemOpen
  \bibfield  {author} {\bibinfo {author} {\bibfnamefont {L.}~\bibnamefont
  {Wang}}, \bibinfo {author} {\bibfnamefont {Z.}~\bibnamefont {Wang}},\ and\
  \bibinfo {author} {\bibfnamefont {S.}~\bibnamefont {Liu}},\ }\bibfield
  {title} {\bibinfo {title} {An effective multivariate time series
  classification approach using echo state network and adaptive differential
  evolution algorithm},\ }\href
  {https://doi.org/https://doi.org/10.1016/j.eswa.2015.08.055} {\bibfield
  {journal} {\bibinfo  {journal} {Expert Systems with Applications}\ }\textbf
  {\bibinfo {volume} {43}},\ \bibinfo {pages} {237} (\bibinfo {year}
  {2016})}\BibitemShut {NoStop}%
\bibitem [{\citenamefont {Verstraeten}\ \emph {et~al.}(2007)\citenamefont
  {Verstraeten}, \citenamefont {Schrauwen}, \citenamefont {D’Haene},\ and\
  \citenamefont {Stroobandt}}]{VERSTRAETEN2007391}%
  \BibitemOpen
  \bibfield  {author} {\bibinfo {author} {\bibfnamefont {D.}~\bibnamefont
  {Verstraeten}}, \bibinfo {author} {\bibfnamefont {B.}~\bibnamefont
  {Schrauwen}}, \bibinfo {author} {\bibfnamefont {M.}~\bibnamefont
  {D’Haene}},\ and\ \bibinfo {author} {\bibfnamefont {D.}~\bibnamefont
  {Stroobandt}},\ }\bibfield  {title} {\bibinfo {title} {An experimental
  unification of reservoir computing methods},\ }\href
  {https://doi.org/https://doi.org/10.1016/j.neunet.2007.04.003} {\bibfield
  {journal} {\bibinfo  {journal} {Neural Networks}\ }\textbf {\bibinfo {volume}
  {20}},\ \bibinfo {pages} {391} (\bibinfo {year} {2007})},\ \bibinfo {note}
  {echo State Networks and Liquid State Machines}\BibitemShut {NoStop}%
\bibitem [{\citenamefont {Tanaka}\ \emph {et~al.}(2019)\citenamefont {Tanaka},
  \citenamefont {Yamane}, \citenamefont {Héroux}, \citenamefont {Nakane},
  \citenamefont {Kanazawa}, \citenamefont {Takeda}, \citenamefont {Numata},
  \citenamefont {Nakano},\ and\ \citenamefont {Hirose}}]{tanaka2019}%
  \BibitemOpen
  \bibfield  {author} {\bibinfo {author} {\bibfnamefont {G.}~\bibnamefont
  {Tanaka}}, \bibinfo {author} {\bibfnamefont {T.}~\bibnamefont {Yamane}},
  \bibinfo {author} {\bibfnamefont {J.~B.}\ \bibnamefont {Héroux}}, \bibinfo
  {author} {\bibfnamefont {R.}~\bibnamefont {Nakane}}, \bibinfo {author}
  {\bibfnamefont {N.}~\bibnamefont {Kanazawa}}, \bibinfo {author}
  {\bibfnamefont {S.}~\bibnamefont {Takeda}}, \bibinfo {author} {\bibfnamefont
  {H.}~\bibnamefont {Numata}}, \bibinfo {author} {\bibfnamefont
  {D.}~\bibnamefont {Nakano}},\ and\ \bibinfo {author} {\bibfnamefont
  {A.}~\bibnamefont {Hirose}},\ }\bibfield  {title} {\bibinfo {title} {Recent
  advances in physical reservoir computing: A review},\ }\href
  {https://doi.org/https://doi.org/10.1016/j.neunet.2019.03.005} {\bibfield
  {journal} {\bibinfo  {journal} {Neural Networks}\ }\textbf {\bibinfo {volume}
  {115}},\ \bibinfo {pages} {100} (\bibinfo {year} {2019})}\BibitemShut
  {NoStop}%
\bibitem [{\citenamefont {Brunner}\ \emph {et~al.}(2013)\citenamefont
  {Brunner}, \citenamefont {Soriano}, \citenamefont {Mirasso},\ and\
  \citenamefont {Fischer}}]{brunner2013parallel}%
  \BibitemOpen
  \bibfield  {author} {\bibinfo {author} {\bibfnamefont {D.}~\bibnamefont
  {Brunner}}, \bibinfo {author} {\bibfnamefont {M.~C.}\ \bibnamefont
  {Soriano}}, \bibinfo {author} {\bibfnamefont {C.~R.}\ \bibnamefont
  {Mirasso}},\ and\ \bibinfo {author} {\bibfnamefont {I.}~\bibnamefont
  {Fischer}},\ }\bibfield  {title} {\bibinfo {title} {Parallel photonic
  information processing at gigabyte per second data rates using transient
  states},\ }\href@noop {} {\bibfield  {journal} {\bibinfo  {journal} {Nature
  communications}\ }\textbf {\bibinfo {volume} {4}},\ \bibinfo {pages} {1364}
  (\bibinfo {year} {2013})}\BibitemShut {NoStop}%
\bibitem [{\citenamefont {Vandoorne}\ \emph {et~al.}(2014)\citenamefont
  {Vandoorne}, \citenamefont {Mechet}, \citenamefont {Van~Vaerenbergh},
  \citenamefont {Fiers}, \citenamefont {Morthier}, \citenamefont {Verstraeten},
  \citenamefont {Schrauwen}, \citenamefont {Dambre},\ and\ \citenamefont
  {Bienstman}}]{vandoorne2014experimental}%
  \BibitemOpen
  \bibfield  {author} {\bibinfo {author} {\bibfnamefont {K.}~\bibnamefont
  {Vandoorne}}, \bibinfo {author} {\bibfnamefont {P.}~\bibnamefont {Mechet}},
  \bibinfo {author} {\bibfnamefont {T.}~\bibnamefont {Van~Vaerenbergh}},
  \bibinfo {author} {\bibfnamefont {M.}~\bibnamefont {Fiers}}, \bibinfo
  {author} {\bibfnamefont {G.}~\bibnamefont {Morthier}}, \bibinfo {author}
  {\bibfnamefont {D.}~\bibnamefont {Verstraeten}}, \bibinfo {author}
  {\bibfnamefont {B.}~\bibnamefont {Schrauwen}}, \bibinfo {author}
  {\bibfnamefont {J.}~\bibnamefont {Dambre}},\ and\ \bibinfo {author}
  {\bibfnamefont {P.}~\bibnamefont {Bienstman}},\ }\bibfield  {title} {\bibinfo
  {title} {Experimental demonstration of reservoir computing on a silicon
  photonics chip},\ }\href@noop {} {\bibfield  {journal} {\bibinfo  {journal}
  {Nature communications}\ }\textbf {\bibinfo {volume} {5}},\ \bibinfo {pages}
  {3541} (\bibinfo {year} {2014})}\BibitemShut {NoStop}%
\bibitem [{\citenamefont {Larger}\ \emph {et~al.}(2017)\citenamefont {Larger},
  \citenamefont {Bayl{\'o}n-Fuentes}, \citenamefont {Martinenghi},
  \citenamefont {Udaltsov}, \citenamefont {Chembo},\ and\ \citenamefont
  {Jacquot}}]{larger2017high}%
  \BibitemOpen
  \bibfield  {author} {\bibinfo {author} {\bibfnamefont {L.}~\bibnamefont
  {Larger}}, \bibinfo {author} {\bibfnamefont {A.}~\bibnamefont
  {Bayl{\'o}n-Fuentes}}, \bibinfo {author} {\bibfnamefont {R.}~\bibnamefont
  {Martinenghi}}, \bibinfo {author} {\bibfnamefont {V.~S.}\ \bibnamefont
  {Udaltsov}}, \bibinfo {author} {\bibfnamefont {Y.~K.}\ \bibnamefont
  {Chembo}},\ and\ \bibinfo {author} {\bibfnamefont {M.}~\bibnamefont
  {Jacquot}},\ }\bibfield  {title} {\bibinfo {title} {High-speed photonic
  reservoir computing using a time-delay-based architecture: Million words per
  second classification},\ }\href@noop {} {\bibfield  {journal} {\bibinfo
  {journal} {Physical Review X}\ }\textbf {\bibinfo {volume} {7}},\ \bibinfo
  {pages} {011015} (\bibinfo {year} {2017})}\BibitemShut {NoStop}%
\bibitem [{\citenamefont {Van Der~Sande}\ \emph {et~al.}(2017)\citenamefont
  {Van Der~Sande}, \citenamefont {Brunner},\ and\ \citenamefont
  {Soriano}}]{VanDerSande2017}%
  \BibitemOpen
  \bibfield  {author} {\bibinfo {author} {\bibfnamefont {G.}~\bibnamefont {Van
  Der~Sande}}, \bibinfo {author} {\bibfnamefont {D.}~\bibnamefont {Brunner}},\
  and\ \bibinfo {author} {\bibfnamefont {M.~C.}\ \bibnamefont {Soriano}},\
  }\bibfield  {title} {\bibinfo {title} {{Advances in photonic reservoir
  computing}},\ }\href {https://doi.org/10.1515/nanoph-2016-0132} {\bibfield
  {journal} {\bibinfo  {journal} {Nanophotonics}\ }\textbf {\bibinfo {volume}
  {6}},\ \bibinfo {pages} {561} (\bibinfo {year} {2017})}\BibitemShut {NoStop}%
\bibitem [{\citenamefont {Mujal}\ \emph
  {et~al.}(2021{\natexlab{a}})\citenamefont {Mujal}, \citenamefont
  {Mart{\'\i}nez-Pe\~{n}a}, \citenamefont {Nokkala}, \citenamefont
  {Garc{\'\i}a-Beni}, \citenamefont {Giorgi}, \citenamefont {Soriano},\ and\
  \citenamefont {Zambrini}}]{Mujal2021}%
  \BibitemOpen
  \bibfield  {author} {\bibinfo {author} {\bibfnamefont {P.}~\bibnamefont
  {Mujal}}, \bibinfo {author} {\bibfnamefont {R.}~\bibnamefont
  {Mart{\'\i}nez-Pe\~{n}a}}, \bibinfo {author} {\bibfnamefont {J.}~\bibnamefont
  {Nokkala}}, \bibinfo {author} {\bibfnamefont {J.}~\bibnamefont
  {Garc{\'\i}a-Beni}}, \bibinfo {author} {\bibfnamefont {G.~L.}\ \bibnamefont
  {Giorgi}}, \bibinfo {author} {\bibfnamefont {M.~C.}\ \bibnamefont
  {Soriano}},\ and\ \bibinfo {author} {\bibfnamefont {R.}~\bibnamefont
  {Zambrini}},\ }\bibfield  {title} {\bibinfo {title} {Opportunities in quantum
  reservoir computing and extreme learning machines},\ }\href
  {https://doi.org/https://doi.org/10.1002/qute.202100027} {\bibfield
  {journal} {\bibinfo  {journal} {Advanced Quantum Technologies}\ }\textbf
  {\bibinfo {volume} {4}},\ \bibinfo {pages} {2100027} (\bibinfo {year}
  {2021}{\natexlab{a}})}\BibitemShut {NoStop}%
\bibitem [{\citenamefont {Ghosh}\ \emph {et~al.}(2021)\citenamefont {Ghosh},
  \citenamefont {Nakajima}, \citenamefont {Krisnanda}, \citenamefont {Fujii},\
  and\ \citenamefont {Liew}}]{ghosh-RC-review}%
  \BibitemOpen
  \bibfield  {author} {\bibinfo {author} {\bibfnamefont {S.}~\bibnamefont
  {Ghosh}}, \bibinfo {author} {\bibfnamefont {K.}~\bibnamefont {Nakajima}},
  \bibinfo {author} {\bibfnamefont {T.}~\bibnamefont {Krisnanda}}, \bibinfo
  {author} {\bibfnamefont {K.}~\bibnamefont {Fujii}},\ and\ \bibinfo {author}
  {\bibfnamefont {T.~C.~H.}\ \bibnamefont {Liew}},\ }\bibfield  {title}
  {\bibinfo {title} {Quantum neuromorphic computing with reservoir computing
  networks},\ }\href {https://doi.org/https://doi.org/10.1002/qute.202100053}
  {\bibfield  {journal} {\bibinfo  {journal} {Advanced Quantum Technologies}\
  }\textbf {\bibinfo {volume} {4}},\ \bibinfo {pages} {2100053} (\bibinfo
  {year} {2021})}\BibitemShut {NoStop}%
\bibitem [{\citenamefont {Markovi{\'{c}}}\ \emph {et~al.}(2020)\citenamefont
  {Markovi{\'{c}}}, \citenamefont {Mizrahi}, \citenamefont {Querlioz},\ and\
  \citenamefont {Grollier}}]{neurom_review_grollier}%
  \BibitemOpen
  \bibfield  {author} {\bibinfo {author} {\bibfnamefont {D.}~\bibnamefont
  {Markovi{\'{c}}}}, \bibinfo {author} {\bibfnamefont {A.}~\bibnamefont
  {Mizrahi}}, \bibinfo {author} {\bibfnamefont {D.}~\bibnamefont {Querlioz}},\
  and\ \bibinfo {author} {\bibfnamefont {J.}~\bibnamefont {Grollier}},\
  }\bibfield  {title} {\bibinfo {title} {Physics for neuromorphic computing},\
  }\href {https://doi.org/10.1038/s42254-020-0208-2} {\bibfield  {journal}
  {\bibinfo  {journal} {Nature Reviews Physics}\ }\textbf {\bibinfo {volume}
  {2}},\ \bibinfo {pages} {499} (\bibinfo {year} {2020})}\BibitemShut {NoStop}%
\bibitem [{\citenamefont {Mujal}\ \emph {et~al.}(2022)\citenamefont {Mujal},
  \citenamefont {Martínez-Peña}, \citenamefont {Giorgi}, \citenamefont
  {Soriano},\ and\ \citenamefont {Zambrini}}]{time-series-QRC-measurements}%
  \BibitemOpen
  \bibfield  {author} {\bibinfo {author} {\bibfnamefont {P.}~\bibnamefont
  {Mujal}}, \bibinfo {author} {\bibfnamefont {R.}~\bibnamefont
  {Martínez-Peña}}, \bibinfo {author} {\bibfnamefont {G.~L.}\ \bibnamefont
  {Giorgi}}, \bibinfo {author} {\bibfnamefont {M.~C.}\ \bibnamefont
  {Soriano}},\ and\ \bibinfo {author} {\bibfnamefont {R.}~\bibnamefont
  {Zambrini}},\ }\href {https://doi.org/10.48550/ARXIV.2205.06809} {\bibinfo
  {title} {Time series quantum reservoir computing with weak and projective
  measurements}} (\bibinfo {year} {2022}),\ \Eprint
  {https://arxiv.org/abs/2205.06809} {arXiv:2205.06809 [quant-ph]} \BibitemShut
  {NoStop}%
\bibitem [{\citenamefont {Chen}\ \emph {et~al.}(2020)\citenamefont {Chen},
  \citenamefont {Nurdin},\ and\ \citenamefont
  {Yamamoto}}]{PhysRevApplied.14.024065}%
  \BibitemOpen
  \bibfield  {author} {\bibinfo {author} {\bibfnamefont {J.}~\bibnamefont
  {Chen}}, \bibinfo {author} {\bibfnamefont {H.~I.}\ \bibnamefont {Nurdin}},\
  and\ \bibinfo {author} {\bibfnamefont {N.}~\bibnamefont {Yamamoto}},\
  }\bibfield  {title} {\bibinfo {title} {Temporal information processing on
  noisy quantum computers},\ }\href
  {https://doi.org/10.1103/PhysRevApplied.14.024065} {\bibfield  {journal}
  {\bibinfo  {journal} {Phys. Rev. Applied}\ }\textbf {\bibinfo {volume}
  {14}},\ \bibinfo {pages} {024065} (\bibinfo {year} {2020})}\BibitemShut
  {NoStop}%
\bibitem [{\citenamefont {Nokkala}\ \emph {et~al.}(2021)\citenamefont
  {Nokkala}, \citenamefont {Mart{\'i}nez-Pe{\~{n}}a}, \citenamefont {Giorgi},
  \citenamefont {Parigi}, \citenamefont {Soriano},\ and\ \citenamefont
  {Zambrini}}]{Nokkala2021}%
  \BibitemOpen
  \bibfield  {author} {\bibinfo {author} {\bibfnamefont {J.}~\bibnamefont
  {Nokkala}}, \bibinfo {author} {\bibfnamefont {R.}~\bibnamefont
  {Mart{\'i}nez-Pe{\~{n}}a}}, \bibinfo {author} {\bibfnamefont {G.~L.}\
  \bibnamefont {Giorgi}}, \bibinfo {author} {\bibfnamefont {V.}~\bibnamefont
  {Parigi}}, \bibinfo {author} {\bibfnamefont {M.~C.}\ \bibnamefont
  {Soriano}},\ and\ \bibinfo {author} {\bibfnamefont {R.}~\bibnamefont
  {Zambrini}},\ }\bibfield  {title} {\bibinfo {title} {Gaussian states of
  continuous-variable quantum systems provide universal and versatile reservoir
  computing},\ }\href {https://doi.org/10.1038/s42005-021-00556-w} {\bibfield
  {journal} {\bibinfo  {journal} {Communications Physics}\ }\textbf {\bibinfo
  {volume} {4}},\ \bibinfo {pages} {53} (\bibinfo {year} {2021})}\BibitemShut
  {NoStop}%
\bibitem [{\citenamefont {Medeiros~de Ara\'ujo}\ \emph
  {et~al.}(2014)\citenamefont {Medeiros~de Ara\'ujo}, \citenamefont {Roslund},
  \citenamefont {Cai}, \citenamefont {Ferrini}, \citenamefont {Fabre},\ and\
  \citenamefont {Treps}}]{Araujo2014}%
  \BibitemOpen
  \bibfield  {author} {\bibinfo {author} {\bibfnamefont {R.}~\bibnamefont
  {Medeiros~de Ara\'ujo}}, \bibinfo {author} {\bibfnamefont {J.}~\bibnamefont
  {Roslund}}, \bibinfo {author} {\bibfnamefont {Y.}~\bibnamefont {Cai}},
  \bibinfo {author} {\bibfnamefont {G.}~\bibnamefont {Ferrini}}, \bibinfo
  {author} {\bibfnamefont {C.}~\bibnamefont {Fabre}},\ and\ \bibinfo {author}
  {\bibfnamefont {N.}~\bibnamefont {Treps}},\ }\bibfield  {title} {\bibinfo
  {title} {Full characterization of a highly multimode entangled state embedded
  in an optical frequency comb using pulse shaping},\ }\href
  {https://doi.org/10.1103/PhysRevA.89.053828} {\bibfield  {journal} {\bibinfo
  {journal} {Phys. Rev. A}\ }\textbf {\bibinfo {volume} {89}},\ \bibinfo
  {pages} {053828} (\bibinfo {year} {2014})}\BibitemShut {NoStop}%
\bibitem [{\citenamefont {Roslund}\ \emph {et~al.}(2014)\citenamefont
  {Roslund}, \citenamefont {de~Ara{\'u}jo}, \citenamefont {Jiang},
  \citenamefont {Fabre},\ and\ \citenamefont {Treps}}]{Roslund2014}%
  \BibitemOpen
  \bibfield  {author} {\bibinfo {author} {\bibfnamefont {J.}~\bibnamefont
  {Roslund}}, \bibinfo {author} {\bibfnamefont {R.~M.}\ \bibnamefont
  {de~Ara{\'u}jo}}, \bibinfo {author} {\bibfnamefont {S.}~\bibnamefont
  {Jiang}}, \bibinfo {author} {\bibfnamefont {C.}~\bibnamefont {Fabre}},\ and\
  \bibinfo {author} {\bibfnamefont {N.}~\bibnamefont {Treps}},\ }\bibfield
  {title} {\bibinfo {title} {Wavelength-multiplexed quantum networks with
  ultrafast frequency combs},\ }\href
  {https://doi.org/10.1038/nphoton.2013.340} {\bibfield  {journal} {\bibinfo
  {journal} {Nature Photonics}\ }\textbf {\bibinfo {volume} {8}},\ \bibinfo
  {pages} {109} (\bibinfo {year} {2014})}\BibitemShut {NoStop}%
\bibitem [{\citenamefont {Nokkala}\ \emph {et~al.}(2018)\citenamefont
  {Nokkala}, \citenamefont {Arzani}, \citenamefont {Galve}, \citenamefont
  {Zambrini}, \citenamefont {Maniscalco}, \citenamefont {Piilo}, \citenamefont
  {Treps},\ and\ \citenamefont {Parigi}}]{NokkalaNJP}%
  \BibitemOpen
  \bibfield  {author} {\bibinfo {author} {\bibfnamefont {J.}~\bibnamefont
  {Nokkala}}, \bibinfo {author} {\bibfnamefont {F.}~\bibnamefont {Arzani}},
  \bibinfo {author} {\bibfnamefont {F.}~\bibnamefont {Galve}}, \bibinfo
  {author} {\bibfnamefont {R.}~\bibnamefont {Zambrini}}, \bibinfo {author}
  {\bibfnamefont {S.}~\bibnamefont {Maniscalco}}, \bibinfo {author}
  {\bibfnamefont {J.}~\bibnamefont {Piilo}}, \bibinfo {author} {\bibfnamefont
  {N.}~\bibnamefont {Treps}},\ and\ \bibinfo {author} {\bibfnamefont
  {V.}~\bibnamefont {Parigi}},\ }\bibfield  {title} {\bibinfo {title}
  {Reconfigurable optical implementation of quantum complex networks},\ }\href
  {https://doi.org/10.1088/1367-2630/aabc77} {\bibfield  {journal} {\bibinfo
  {journal} {New Journal of Physics}\ }\textbf {\bibinfo {volume} {20}},\
  \bibinfo {pages} {053024} (\bibinfo {year} {2018})}\BibitemShut {NoStop}%
\bibitem [{\citenamefont {Fujii}\ and\ \citenamefont
  {Nakajima}(2017)}]{Nakajima2017}%
  \BibitemOpen
  \bibfield  {author} {\bibinfo {author} {\bibfnamefont {K.}~\bibnamefont
  {Fujii}}\ and\ \bibinfo {author} {\bibfnamefont {K.}~\bibnamefont
  {Nakajima}},\ }\bibfield  {title} {\bibinfo {title} {Harnessing
  disordered-ensemble quantum dynamics for machine learning},\ }\href
  {https://doi.org/10.1103/PhysRevApplied.8.024030} {\bibfield  {journal}
  {\bibinfo  {journal} {Phys. Rev. Applied}\ }\textbf {\bibinfo {volume} {8}},\
  \bibinfo {pages} {024030} (\bibinfo {year} {2017})}\BibitemShut {NoStop}%
\bibitem [{\citenamefont {Mart\'{\i}nez-Pe\~na}\ \emph
  {et~al.}(2021)\citenamefont {Mart\'{\i}nez-Pe\~na}, \citenamefont {Giorgi},
  \citenamefont {Nokkala}, \citenamefont {Soriano},\ and\ \citenamefont
  {Zambrini}}]{rodrigo_PRL}%
  \BibitemOpen
  \bibfield  {author} {\bibinfo {author} {\bibfnamefont {R.}~\bibnamefont
  {Mart\'{\i}nez-Pe\~na}}, \bibinfo {author} {\bibfnamefont {G.~L.}\
  \bibnamefont {Giorgi}}, \bibinfo {author} {\bibfnamefont {J.}~\bibnamefont
  {Nokkala}}, \bibinfo {author} {\bibfnamefont {M.~C.}\ \bibnamefont
  {Soriano}},\ and\ \bibinfo {author} {\bibfnamefont {R.}~\bibnamefont
  {Zambrini}},\ }\bibfield  {title} {\bibinfo {title} {Dynamical phase
  transitions in quantum reservoir computing},\ }\href
  {https://doi.org/10.1103/PhysRevLett.127.100502} {\bibfield  {journal}
  {\bibinfo  {journal} {Phys. Rev. Lett.}\ }\textbf {\bibinfo {volume} {127}},\
  \bibinfo {pages} {100502} (\bibinfo {year} {2021})}\BibitemShut {NoStop}%
\bibitem [{\citenamefont {Bravo}\ \emph {et~al.}(2022)\citenamefont {Bravo},
  \citenamefont {Najafi}, \citenamefont {Gao},\ and\ \citenamefont
  {Yelin}}]{Yelin_PRX-Quantum}%
  \BibitemOpen
  \bibfield  {author} {\bibinfo {author} {\bibfnamefont {R.~A.}\ \bibnamefont
  {Bravo}}, \bibinfo {author} {\bibfnamefont {K.}~\bibnamefont {Najafi}},
  \bibinfo {author} {\bibfnamefont {X.}~\bibnamefont {Gao}},\ and\ \bibinfo
  {author} {\bibfnamefont {S.~F.}\ \bibnamefont {Yelin}},\ }\bibfield  {title}
  {\bibinfo {title} {Quantum reservoir computing using arrays of rydberg
  atoms},\ }\href {https://doi.org/10.1103/PRXQuantum.3.030325} {\bibfield
  {journal} {\bibinfo  {journal} {PRX Quantum}\ }\textbf {\bibinfo {volume}
  {3}},\ \bibinfo {pages} {030325} (\bibinfo {year} {2022})}\BibitemShut
  {NoStop}%
\bibitem [{\citenamefont {Nokkala}(2021)}]{johannes}%
  \BibitemOpen
  \bibfield  {author} {\bibinfo {author} {\bibfnamefont {J.}~\bibnamefont
  {Nokkala}},\ }\href {https://doi.org/10.48550/ARXIV.2108.00698} {\bibinfo
  {title} {Online quantum time series processing with random oscillator
  networks}} (\bibinfo {year} {2021}),\ \Eprint
  {https://arxiv.org/abs/2108.00698} {arXiv:2108.00698 [quant-ph]} \BibitemShut
  {NoStop}%
\bibitem [{\citenamefont {Shaked}\ \emph {et~al.}(2018)\citenamefont {Shaked},
  \citenamefont {Michael}, \citenamefont {Vered}, \citenamefont {Bello},
  \citenamefont {Rosenbluh},\ and\ \citenamefont {Pe'er}}]{Shaked2018}%
  \BibitemOpen
  \bibfield  {author} {\bibinfo {author} {\bibfnamefont {Y.}~\bibnamefont
  {Shaked}}, \bibinfo {author} {\bibfnamefont {Y.}~\bibnamefont {Michael}},
  \bibinfo {author} {\bibfnamefont {R.~Z.}\ \bibnamefont {Vered}}, \bibinfo
  {author} {\bibfnamefont {L.}~\bibnamefont {Bello}}, \bibinfo {author}
  {\bibfnamefont {M.}~\bibnamefont {Rosenbluh}},\ and\ \bibinfo {author}
  {\bibfnamefont {A.}~\bibnamefont {Pe'er}},\ }\bibfield  {title} {\bibinfo
  {title} {Lifting the bandwidth limit of optical homodyne measurement with
  broadband parametric amplification},\ }\href
  {https://doi.org/10.1038/s41467-018-03083-5} {\bibfield  {journal} {\bibinfo
  {journal} {Nature Communications}\ }\textbf {\bibinfo {volume} {9}},\
  \bibinfo {pages} {609} (\bibinfo {year} {2018})}\BibitemShut {NoStop}%
\bibitem [{\citenamefont {Takanashi}\ \emph {et~al.}(2020)\citenamefont
  {Takanashi}, \citenamefont {Inoue}, \citenamefont {Kashiwazaki},
  \citenamefont {Kazama}, \citenamefont {Enbutsu}, \citenamefont {Kasahara},
  \citenamefont {Umeki},\ and\ \citenamefont {Furusawa}}]{Takanashi:20}%
  \BibitemOpen
  \bibfield  {author} {\bibinfo {author} {\bibfnamefont {N.}~\bibnamefont
  {Takanashi}}, \bibinfo {author} {\bibfnamefont {A.}~\bibnamefont {Inoue}},
  \bibinfo {author} {\bibfnamefont {T.}~\bibnamefont {Kashiwazaki}}, \bibinfo
  {author} {\bibfnamefont {T.}~\bibnamefont {Kazama}}, \bibinfo {author}
  {\bibfnamefont {K.}~\bibnamefont {Enbutsu}}, \bibinfo {author} {\bibfnamefont
  {R.}~\bibnamefont {Kasahara}}, \bibinfo {author} {\bibfnamefont
  {T.}~\bibnamefont {Umeki}},\ and\ \bibinfo {author} {\bibfnamefont
  {A.}~\bibnamefont {Furusawa}},\ }\bibfield  {title} {\bibinfo {title}
  {All-optical phase-sensitive detection for ultra-fast quantum computation},\
  }\href {https://doi.org/10.1364/OE.405832} {\bibfield  {journal} {\bibinfo
  {journal} {Opt. Express}\ }\textbf {\bibinfo {volume} {28}},\ \bibinfo
  {pages} {34916} (\bibinfo {year} {2020})}\BibitemShut {NoStop}%
\bibitem [{\citenamefont {Wiseman}\ and\ \citenamefont
  {Milburn}(1993{\natexlab{a}})}]{Wiseman_Milburn_quadrature}%
  \BibitemOpen
  \bibfield  {author} {\bibinfo {author} {\bibfnamefont {H.~M.}\ \bibnamefont
  {Wiseman}}\ and\ \bibinfo {author} {\bibfnamefont {G.~J.}\ \bibnamefont
  {Milburn}},\ }\bibfield  {title} {\bibinfo {title} {Quantum theory of
  field-quadrature measurements},\ }\href
  {https://doi.org/10.1103/PhysRevA.47.642} {\bibfield  {journal} {\bibinfo
  {journal} {Phys. Rev. A}\ }\textbf {\bibinfo {volume} {47}},\ \bibinfo
  {pages} {642} (\bibinfo {year} {1993}{\natexlab{a}})}\BibitemShut {NoStop}%
\bibitem [{\citenamefont {Wiseman}\ and\ \citenamefont
  {Milburn}(1993{\natexlab{b}})}]{Wiseman_Milburn_homodyne}%
  \BibitemOpen
  \bibfield  {author} {\bibinfo {author} {\bibfnamefont {H.~M.}\ \bibnamefont
  {Wiseman}}\ and\ \bibinfo {author} {\bibfnamefont {G.~J.}\ \bibnamefont
  {Milburn}},\ }\bibfield  {title} {\bibinfo {title} {Quantum theory of optical
  feedback via homodyne detection},\ }\href
  {https://doi.org/10.1103/PhysRevLett.70.548} {\bibfield  {journal} {\bibinfo
  {journal} {Phys. Rev. Lett.}\ }\textbf {\bibinfo {volume} {70}},\ \bibinfo
  {pages} {548} (\bibinfo {year} {1993}{\natexlab{b}})}\BibitemShut {NoStop}%
\bibitem [{\citenamefont {Kouadou}\ \emph {et~al.}(2022)\citenamefont
  {Kouadou}, \citenamefont {Sansavini}, \citenamefont {Ansquer}, \citenamefont
  {Henaff}, \citenamefont {Treps},\ and\ \citenamefont {Parigi}}]{Kouadou2022}%
  \BibitemOpen
  \bibfield  {author} {\bibinfo {author} {\bibfnamefont {T.}~\bibnamefont
  {Kouadou}}, \bibinfo {author} {\bibfnamefont {F.}~\bibnamefont {Sansavini}},
  \bibinfo {author} {\bibfnamefont {M.}~\bibnamefont {Ansquer}}, \bibinfo
  {author} {\bibfnamefont {J.}~\bibnamefont {Henaff}}, \bibinfo {author}
  {\bibfnamefont {N.}~\bibnamefont {Treps}},\ and\ \bibinfo {author}
  {\bibfnamefont {V.}~\bibnamefont {Parigi}},\ }\href
  {https://doi.org/10.48550/ARXIV.2209.10678} {\bibinfo {title} {Spectrally
  shaped and pulse-by-pulse multiplexed multimode squeezed states of light}}
  (\bibinfo {year} {2022}),\ \Eprint {https://arxiv.org/abs/2209.10678}
  {arXiv:2209.10678 [quant-ph]} \BibitemShut {NoStop}%
\bibitem [{\citenamefont {Madsen}\ \emph {et~al.}(2022)\citenamefont {Madsen},
  \citenamefont {Laudenbach}, \citenamefont {Askarani}, \citenamefont
  {Rortais}, \citenamefont {Vincent}, \citenamefont {Bulmer}, \citenamefont
  {Miatto}, \citenamefont {Neuhaus}, \citenamefont {Helt}, \citenamefont
  {Collins}, \citenamefont {Lita}, \citenamefont {Gerrits}, \citenamefont
  {Nam}, \citenamefont {Vaidya}, \citenamefont {Menotti}, \citenamefont
  {Dhand}, \citenamefont {Vernon}, \citenamefont {Quesada},\ and\ \citenamefont
  {Lavoie}}]{Madsen2022}%
  \BibitemOpen
  \bibfield  {author} {\bibinfo {author} {\bibfnamefont {L.~S.}\ \bibnamefont
  {Madsen}}, \bibinfo {author} {\bibfnamefont {F.}~\bibnamefont {Laudenbach}},
  \bibinfo {author} {\bibfnamefont {M.~F.}\ \bibnamefont {Askarani}}, \bibinfo
  {author} {\bibfnamefont {F.}~\bibnamefont {Rortais}}, \bibinfo {author}
  {\bibfnamefont {T.}~\bibnamefont {Vincent}}, \bibinfo {author} {\bibfnamefont
  {J.~F.~F.}\ \bibnamefont {Bulmer}}, \bibinfo {author} {\bibfnamefont {F.~M.}\
  \bibnamefont {Miatto}}, \bibinfo {author} {\bibfnamefont {L.}~\bibnamefont
  {Neuhaus}}, \bibinfo {author} {\bibfnamefont {L.~G.}\ \bibnamefont {Helt}},
  \bibinfo {author} {\bibfnamefont {M.~J.}\ \bibnamefont {Collins}}, \bibinfo
  {author} {\bibfnamefont {A.~E.}\ \bibnamefont {Lita}}, \bibinfo {author}
  {\bibfnamefont {T.}~\bibnamefont {Gerrits}}, \bibinfo {author} {\bibfnamefont
  {S.~W.}\ \bibnamefont {Nam}}, \bibinfo {author} {\bibfnamefont {V.~D.}\
  \bibnamefont {Vaidya}}, \bibinfo {author} {\bibfnamefont {M.}~\bibnamefont
  {Menotti}}, \bibinfo {author} {\bibfnamefont {I.}~\bibnamefont {Dhand}},
  \bibinfo {author} {\bibfnamefont {Z.}~\bibnamefont {Vernon}}, \bibinfo
  {author} {\bibfnamefont {N.}~\bibnamefont {Quesada}},\ and\ \bibinfo {author}
  {\bibfnamefont {J.}~\bibnamefont {Lavoie}},\ }\bibfield  {title} {\bibinfo
  {title} {Quantum computational advantage with a programmable photonic
  processor},\ }\href {https://doi.org/10.1038/s41586-022-04725-x} {\bibfield
  {journal} {\bibinfo  {journal} {Nature}\ }\textbf {\bibinfo {volume} {606}},\
  \bibinfo {pages} {75} (\bibinfo {year} {2022})}\BibitemShut {NoStop}%
\bibitem [{\citenamefont {Tomoda}\ \emph {et~al.}(2022)\citenamefont {Tomoda},
  \citenamefont {Yoshida}, \citenamefont {Kashiwazaki}, \citenamefont {Umeki},
  \citenamefont {Enomoto},\ and\ \citenamefont {Takeda}}]{Tomoda2022}%
  \BibitemOpen
  \bibfield  {author} {\bibinfo {author} {\bibfnamefont {H.}~\bibnamefont
  {Tomoda}}, \bibinfo {author} {\bibfnamefont {T.}~\bibnamefont {Yoshida}},
  \bibinfo {author} {\bibfnamefont {T.}~\bibnamefont {Kashiwazaki}}, \bibinfo
  {author} {\bibfnamefont {T.}~\bibnamefont {Umeki}}, \bibinfo {author}
  {\bibfnamefont {Y.}~\bibnamefont {Enomoto}},\ and\ \bibinfo {author}
  {\bibfnamefont {S.}~\bibnamefont {Takeda}},\ }\href
  {https://doi.org/10.48550/ARXIV.2209.09458} {\bibinfo {title} {Programmable
  time-multiplexed squeezed light source}} (\bibinfo {year} {2022}),\ \Eprint
  {https://arxiv.org/abs/2209.09458} {arXiv:2209.09458 [quant-ph]} \BibitemShut
  {NoStop}%
\bibitem [{\citenamefont {Chen}\ \emph {et~al.}(2014)\citenamefont {Chen},
  \citenamefont {Menicucci},\ and\ \citenamefont {Pfister}}]{Chen2013}%
  \BibitemOpen
  \bibfield  {author} {\bibinfo {author} {\bibfnamefont {M.}~\bibnamefont
  {Chen}}, \bibinfo {author} {\bibfnamefont {N.~C.}\ \bibnamefont
  {Menicucci}},\ and\ \bibinfo {author} {\bibfnamefont {O.}~\bibnamefont
  {Pfister}},\ }\bibfield  {title} {\bibinfo {title} {Experimental realization
  of multipartite entanglement of 60 modes of a quantum optical frequency
  comb},\ }\href {https://doi.org/10.1103/PhysRevLett.112.120505} {\bibfield
  {journal} {\bibinfo  {journal} {Phys. Rev. Lett.}\ }\textbf {\bibinfo
  {volume} {112}},\ \bibinfo {pages} {120505} (\bibinfo {year}
  {2014})}\BibitemShut {NoStop}%
\bibitem [{\citenamefont {Plick}\ \emph {et~al.}(2018)\citenamefont {Plick},
  \citenamefont {Arzani}, \citenamefont {Treps}, \citenamefont {Diamanti},\
  and\ \citenamefont {Markham}}]{Plick2018}%
  \BibitemOpen
  \bibfield  {author} {\bibinfo {author} {\bibfnamefont {W.~N.}\ \bibnamefont
  {Plick}}, \bibinfo {author} {\bibfnamefont {F.}~\bibnamefont {Arzani}},
  \bibinfo {author} {\bibfnamefont {N.}~\bibnamefont {Treps}}, \bibinfo
  {author} {\bibfnamefont {E.}~\bibnamefont {Diamanti}},\ and\ \bibinfo
  {author} {\bibfnamefont {D.}~\bibnamefont {Markham}},\ }\bibfield  {title}
  {\bibinfo {title} {Violating bell inequalities with entangled optical
  frequency combs and multipixel homodyne detection},\ }\href
  {https://doi.org/10.1103/PhysRevA.98.062101} {\bibfield  {journal} {\bibinfo
  {journal} {Phys. Rev. A}\ }\textbf {\bibinfo {volume} {98}},\ \bibinfo
  {pages} {062101} (\bibinfo {year} {2018})}\BibitemShut {NoStop}%
\bibitem [{\citenamefont {Cai}\ \emph {et~al.}(2021)\citenamefont {Cai},
  \citenamefont {Roslund}, \citenamefont {Thiel}, \citenamefont {Fabre},\ and\
  \citenamefont {Treps}}]{Cai2021}%
  \BibitemOpen
  \bibfield  {author} {\bibinfo {author} {\bibfnamefont {Y.}~\bibnamefont
  {Cai}}, \bibinfo {author} {\bibfnamefont {J.}~\bibnamefont {Roslund}},
  \bibinfo {author} {\bibfnamefont {V.}~\bibnamefont {Thiel}}, \bibinfo
  {author} {\bibfnamefont {C.}~\bibnamefont {Fabre}},\ and\ \bibinfo {author}
  {\bibfnamefont {N.}~\bibnamefont {Treps}},\ }\bibfield  {title} {\bibinfo
  {title} {Quantum enhanced measurement of an optical frequency comb},\ }\href
  {https://doi.org/10.1038/s41534-021-00419-w} {\bibfield  {journal} {\bibinfo
  {journal} {npj Quantum Information}\ }\textbf {\bibinfo {volume} {7}},\
  \bibinfo {pages} {82} (\bibinfo {year} {2021})}\BibitemShut {NoStop}%
\bibitem [{\citenamefont {Larsen}\ \emph {et~al.}(2019)\citenamefont {Larsen},
  \citenamefont {Guo}, \citenamefont {Breum}, \citenamefont
  {Neergaard-Nielsen},\ and\ \citenamefont {Andersen}}]{Larsen2019}%
  \BibitemOpen
  \bibfield  {author} {\bibinfo {author} {\bibfnamefont {M.~V.}\ \bibnamefont
  {Larsen}}, \bibinfo {author} {\bibfnamefont {X.}~\bibnamefont {Guo}},
  \bibinfo {author} {\bibfnamefont {C.~R.}\ \bibnamefont {Breum}}, \bibinfo
  {author} {\bibfnamefont {J.~S.}\ \bibnamefont {Neergaard-Nielsen}},\ and\
  \bibinfo {author} {\bibfnamefont {U.~L.}\ \bibnamefont {Andersen}},\
  }\bibfield  {title} {\bibinfo {title} {Fiber-coupled epr-state generation
  using a single temporally multiplexed squeezed light source},\ }\href
  {https://doi.org/10.1038/s41534-019-0170-y} {\bibfield  {journal} {\bibinfo
  {journal} {npj Quantum Information}\ }\textbf {\bibinfo {volume} {5}},\
  \bibinfo {pages} {46} (\bibinfo {year} {2019})}\BibitemShut {NoStop}%
\bibitem [{\citenamefont {Dambre}\ \emph {et~al.}(2012)\citenamefont {Dambre},
  \citenamefont {Verstraeten}, \citenamefont {Schrauwen},\ and\ \citenamefont
  {Massar}}]{dambre2012information}%
  \BibitemOpen
  \bibfield  {author} {\bibinfo {author} {\bibfnamefont {J.}~\bibnamefont
  {Dambre}}, \bibinfo {author} {\bibfnamefont {D.}~\bibnamefont {Verstraeten}},
  \bibinfo {author} {\bibfnamefont {B.}~\bibnamefont {Schrauwen}},\ and\
  \bibinfo {author} {\bibfnamefont {S.}~\bibnamefont {Massar}},\ }\bibfield
  {title} {\bibinfo {title} {{Information processing capacity of dynamical
  systems}},\ }\href@noop {} {\bibfield  {journal} {\bibinfo  {journal} {Sci.
  Rep.}\ }\textbf {\bibinfo {volume} {2}},\ \bibinfo {pages} {514} (\bibinfo
  {year} {2012})}\BibitemShut {NoStop}%
\bibitem [{\citenamefont {Mart{\'i}nez-Pe{\~{n}}a}\ \emph
  {et~al.}(2020)\citenamefont {Mart{\'i}nez-Pe{\~{n}}a}, \citenamefont
  {Nokkala}, \citenamefont {Giorgi}, \citenamefont {Zambrini},\ and\
  \citenamefont {Soriano}}]{Martinez2020}%
  \BibitemOpen
  \bibfield  {author} {\bibinfo {author} {\bibfnamefont {R.}~\bibnamefont
  {Mart{\'i}nez-Pe{\~{n}}a}}, \bibinfo {author} {\bibfnamefont
  {J.}~\bibnamefont {Nokkala}}, \bibinfo {author} {\bibfnamefont {G.~L.}\
  \bibnamefont {Giorgi}}, \bibinfo {author} {\bibfnamefont {R.}~\bibnamefont
  {Zambrini}},\ and\ \bibinfo {author} {\bibfnamefont {M.~C.}\ \bibnamefont
  {Soriano}},\ }\bibfield  {title} {\bibinfo {title} {Information processing
  capacity of spin-based quantum reservoir computing systems},\ }\bibfield
  {journal} {\bibinfo  {journal} {Cognitive Computation}\ }\href
  {https://doi.org/10.1007/s12559-020-09772-y} {10.1007/s12559-020-09772-y}
  (\bibinfo {year} {2020})\BibitemShut {NoStop}%
\bibitem [{\citenamefont {Vahlbruch}\ \emph {et~al.}(2016)\citenamefont
  {Vahlbruch}, \citenamefont {Mehmet}, \citenamefont {Danzmann},\ and\
  \citenamefont {Schnabel}}]{PhysRevLett.117.110801}%
  \BibitemOpen
  \bibfield  {author} {\bibinfo {author} {\bibfnamefont {H.}~\bibnamefont
  {Vahlbruch}}, \bibinfo {author} {\bibfnamefont {M.}~\bibnamefont {Mehmet}},
  \bibinfo {author} {\bibfnamefont {K.}~\bibnamefont {Danzmann}},\ and\
  \bibinfo {author} {\bibfnamefont {R.}~\bibnamefont {Schnabel}},\ }\bibfield
  {title} {\bibinfo {title} {Detection of 15 db squeezed states of light and
  their application for the absolute calibration of photoelectric quantum
  efficiency},\ }\href {https://doi.org/10.1103/PhysRevLett.117.110801}
  {\bibfield  {journal} {\bibinfo  {journal} {Phys. Rev. Lett.}\ }\textbf
  {\bibinfo {volume} {117}},\ \bibinfo {pages} {110801} (\bibinfo {year}
  {2016})}\BibitemShut {NoStop}%
\bibitem [{\citenamefont {Nokkala}\ \emph {et~al.}(2022)\citenamefont
  {Nokkala}, \citenamefont {Mart{\'\i}nez-Pe\~{n}a}, \citenamefont {Zambrini},\
  and\ \citenamefont {Soriano}}]{9525045}%
  \BibitemOpen
  \bibfield  {author} {\bibinfo {author} {\bibfnamefont {J.}~\bibnamefont
  {Nokkala}}, \bibinfo {author} {\bibfnamefont {R.}~\bibnamefont
  {Mart{\'\i}nez-Pe\~{n}a}}, \bibinfo {author} {\bibfnamefont {R.}~\bibnamefont
  {Zambrini}},\ and\ \bibinfo {author} {\bibfnamefont {M.~C.}\ \bibnamefont
  {Soriano}},\ }\bibfield  {title} {\bibinfo {title} {High-performance
  reservoir computing with fluctuations in linear networks},\ }\href
  {https://doi.org/10.1109/TNNLS.2021.3105695} {\bibfield  {journal} {\bibinfo
  {journal} {IEEE Transactions on Neural Networks and Learning Systems}\
  }\textbf {\bibinfo {volume} {33}},\ \bibinfo {pages} {2664} (\bibinfo {year}
  {2022})}\BibitemShut {NoStop}%
\bibitem [{\citenamefont {Govia}\ \emph {et~al.}(2021)\citenamefont {Govia},
  \citenamefont {Ribeill}, \citenamefont {Rowlands}, \citenamefont {Krovi},\
  and\ \citenamefont {Ohki}}]{PhysRevResearch.3.013077}%
  \BibitemOpen
  \bibfield  {author} {\bibinfo {author} {\bibfnamefont {L.~C.~G.}\
  \bibnamefont {Govia}}, \bibinfo {author} {\bibfnamefont {G.~J.}\ \bibnamefont
  {Ribeill}}, \bibinfo {author} {\bibfnamefont {G.~E.}\ \bibnamefont
  {Rowlands}}, \bibinfo {author} {\bibfnamefont {H.~K.}\ \bibnamefont
  {Krovi}},\ and\ \bibinfo {author} {\bibfnamefont {T.~A.}\ \bibnamefont
  {Ohki}},\ }\bibfield  {title} {\bibinfo {title} {Quantum reservoir computing
  with a single nonlinear oscillator},\ }\href
  {https://doi.org/10.1103/PhysRevResearch.3.013077} {\bibfield  {journal}
  {\bibinfo  {journal} {Phys. Rev. Research}\ }\textbf {\bibinfo {volume}
  {3}},\ \bibinfo {pages} {013077} (\bibinfo {year} {2021})}\BibitemShut
  {NoStop}%
\bibitem [{\citenamefont {Kalfus}\ \emph {et~al.}(2022)\citenamefont {Kalfus},
  \citenamefont {Ribeill}, \citenamefont {Rowlands}, \citenamefont {Krovi},
  \citenamefont {Ohki},\ and\ \citenamefont {Govia}}]{kalfus2022}%
  \BibitemOpen
  \bibfield  {author} {\bibinfo {author} {\bibfnamefont {W.~D.}\ \bibnamefont
  {Kalfus}}, \bibinfo {author} {\bibfnamefont {G.~J.}\ \bibnamefont {Ribeill}},
  \bibinfo {author} {\bibfnamefont {G.~E.}\ \bibnamefont {Rowlands}}, \bibinfo
  {author} {\bibfnamefont {H.~K.}\ \bibnamefont {Krovi}}, \bibinfo {author}
  {\bibfnamefont {T.~A.}\ \bibnamefont {Ohki}},\ and\ \bibinfo {author}
  {\bibfnamefont {L.~C.~G.}\ \bibnamefont {Govia}},\ }\bibfield  {title}
  {\bibinfo {title} {Hilbert space as a computational resource in reservoir
  computing},\ }\href {https://doi.org/10.1103/PhysRevResearch.4.033007}
  {\bibfield  {journal} {\bibinfo  {journal} {Phys. Rev. Research}\ }\textbf
  {\bibinfo {volume} {4}},\ \bibinfo {pages} {033007} (\bibinfo {year}
  {2022})}\BibitemShut {NoStop}%
\bibitem [{\citenamefont {H\"ubner}\ \emph {et~al.}(1989)\citenamefont
  {H\"ubner}, \citenamefont {Abraham},\ and\ \citenamefont {Weiss}}]{santafe1}%
  \BibitemOpen
  \bibfield  {author} {\bibinfo {author} {\bibfnamefont {U.}~\bibnamefont
  {H\"ubner}}, \bibinfo {author} {\bibfnamefont {N.~B.}\ \bibnamefont
  {Abraham}},\ and\ \bibinfo {author} {\bibfnamefont {C.~O.}\ \bibnamefont
  {Weiss}},\ }\bibfield  {title} {\bibinfo {title} {Dimensions and entropies of
  chaotic intensity pulsations in a single-mode far-infrared
  ${\mathrm{nh}}_{3}$ laser},\ }\href
  {https://doi.org/10.1103/PhysRevA.40.6354} {\bibfield  {journal} {\bibinfo
  {journal} {Phys. Rev. A}\ }\textbf {\bibinfo {volume} {40}},\ \bibinfo
  {pages} {6354} (\bibinfo {year} {1989})}\BibitemShut {NoStop}%
\bibitem [{\citenamefont {Weigend}\ and\ \citenamefont
  {Gershenfeld}(1993)}]{santafe2}%
  \BibitemOpen
  \bibfield  {author} {\bibinfo {author} {\bibfnamefont {A.}~\bibnamefont
  {Weigend}}\ and\ \bibinfo {author} {\bibfnamefont {N.}~\bibnamefont
  {Gershenfeld}},\ }\bibfield  {title} {\bibinfo {title} {Results of the time
  series prediction competition at the santa fe institute},\ }in\ \href
  {https://doi.org/10.1109/ICNN.1993.298828} {\emph {\bibinfo {booktitle} {IEEE
  International Conference on Neural Networks}}}\ (\bibinfo {year} {1993})\
  pp.\ \bibinfo {pages} {1786--1793 vol.3}\BibitemShut {NoStop}%
\bibitem [{\citenamefont {Inubushi}\ and\ \citenamefont
  {Yoshimura}(2017)}]{inubushi2017reservoir}%
  \BibitemOpen
  \bibfield  {author} {\bibinfo {author} {\bibfnamefont {M.}~\bibnamefont
  {Inubushi}}\ and\ \bibinfo {author} {\bibfnamefont {K.}~\bibnamefont
  {Yoshimura}},\ }\bibfield  {title} {\bibinfo {title} {Reservoir computing
  beyond memory-nonlinearity trade-off},\ }\href@noop {} {\bibfield  {journal}
  {\bibinfo  {journal} {Scientific reports}\ }\textbf {\bibinfo {volume} {7}},\
  \bibinfo {pages} {10199} (\bibinfo {year} {2017})}\BibitemShut {NoStop}%
\bibitem [{\citenamefont {Harkhoe}\ and\ \citenamefont {Van~der
  Sande}(2019)}]{harkhoe2019delay}%
  \BibitemOpen
  \bibfield  {author} {\bibinfo {author} {\bibfnamefont {K.}~\bibnamefont
  {Harkhoe}}\ and\ \bibinfo {author} {\bibfnamefont {G.}~\bibnamefont {Van~der
  Sande}},\ }\bibfield  {title} {\bibinfo {title} {Delay-based reservoir
  computing using multimode semiconductor lasers: Exploiting the rich carrier
  dynamics},\ }\href@noop {} {\bibfield  {journal} {\bibinfo  {journal} {IEEE
  Journal of Selected Topics in Quantum Electronics}\ }\textbf {\bibinfo
  {volume} {25}},\ \bibinfo {pages} {1502909} (\bibinfo {year}
  {2019})}\BibitemShut {NoStop}%
\bibitem [{\citenamefont {Kumar}\ \emph {et~al.}(2021)\citenamefont {Kumar},
  \citenamefont {Jin}, \citenamefont {Bu}, \citenamefont {Kumar},\ and\
  \citenamefont {Huang}}]{kumar2021efficient}%
  \BibitemOpen
  \bibfield  {author} {\bibinfo {author} {\bibfnamefont {P.}~\bibnamefont
  {Kumar}}, \bibinfo {author} {\bibfnamefont {M.}~\bibnamefont {Jin}}, \bibinfo
  {author} {\bibfnamefont {T.}~\bibnamefont {Bu}}, \bibinfo {author}
  {\bibfnamefont {S.}~\bibnamefont {Kumar}},\ and\ \bibinfo {author}
  {\bibfnamefont {Y.-P.}\ \bibnamefont {Huang}},\ }\bibfield  {title} {\bibinfo
  {title} {Efficient reservoir computing using field programmable gate array
  and electro-optic modulation},\ }\href@noop {} {\bibfield  {journal}
  {\bibinfo  {journal} {OSA Continuum}\ }\textbf {\bibinfo {volume} {4}},\
  \bibinfo {pages} {1086} (\bibinfo {year} {2021})}\BibitemShut {NoStop}%
\bibitem [{\citenamefont {Marzen}(2017)}]{marzen}%
  \BibitemOpen
  \bibfield  {author} {\bibinfo {author} {\bibfnamefont {S.}~\bibnamefont
  {Marzen}},\ }\bibfield  {title} {\bibinfo {title} {Difference between memory
  and prediction in linear recurrent networks},\ }\href
  {https://doi.org/10.1103/PhysRevE.96.032308} {\bibfield  {journal} {\bibinfo
  {journal} {Phys. Rev. E}\ }\textbf {\bibinfo {volume} {96}},\ \bibinfo
  {pages} {032308} (\bibinfo {year} {2017})}\BibitemShut {NoStop}%
\bibitem [{\citenamefont {Brunner}\ \emph {et~al.}(2019)\citenamefont
  {Brunner}, \citenamefont {Soriano},\ and\ \citenamefont {Van~der
  Sande}}]{brunner2019photonic}%
  \BibitemOpen
  \bibfield  {author} {\bibinfo {author} {\bibfnamefont {D.}~\bibnamefont
  {Brunner}}, \bibinfo {author} {\bibfnamefont {M.~C.}\ \bibnamefont
  {Soriano}},\ and\ \bibinfo {author} {\bibfnamefont {G.}~\bibnamefont {Van~der
  Sande}},\ }\href@noop {} {\emph {\bibinfo {title} {Photonic reservoir
  computing}}}\ (\bibinfo  {publisher} {De Gruyter},\ \bibinfo {year}
  {2019})\BibitemShut {NoStop}%
\bibitem [{\citenamefont {Spagnolo}\ \emph {et~al.}(2022)\citenamefont
  {Spagnolo}, \citenamefont {Morris}, \citenamefont {Piacentini}, \citenamefont
  {Antesberger}, \citenamefont {Massa}, \citenamefont {Crespi}, \citenamefont
  {Ceccarelli}, \citenamefont {Osellame},\ and\ \citenamefont
  {Walther}}]{Spagnolo2022}%
  \BibitemOpen
  \bibfield  {author} {\bibinfo {author} {\bibfnamefont {M.}~\bibnamefont
  {Spagnolo}}, \bibinfo {author} {\bibfnamefont {J.}~\bibnamefont {Morris}},
  \bibinfo {author} {\bibfnamefont {S.}~\bibnamefont {Piacentini}}, \bibinfo
  {author} {\bibfnamefont {M.}~\bibnamefont {Antesberger}}, \bibinfo {author}
  {\bibfnamefont {F.}~\bibnamefont {Massa}}, \bibinfo {author} {\bibfnamefont
  {A.}~\bibnamefont {Crespi}}, \bibinfo {author} {\bibfnamefont
  {F.}~\bibnamefont {Ceccarelli}}, \bibinfo {author} {\bibfnamefont
  {R.}~\bibnamefont {Osellame}},\ and\ \bibinfo {author} {\bibfnamefont
  {P.}~\bibnamefont {Walther}},\ }\bibfield  {title} {\bibinfo {title}
  {Experimental photonic quantum memristor},\ }\href
  {https://doi.org/10.1038/s41566-022-00973-5} {\bibfield  {journal} {\bibinfo
  {journal} {Nature Photonics}\ }\textbf {\bibinfo {volume} {16}},\ \bibinfo
  {pages} {318} (\bibinfo {year} {2022})}\BibitemShut {NoStop}%
\bibitem [{\citenamefont {Khan}\ \emph {et~al.}(2021)\citenamefont {Khan},
  \citenamefont {Hu}, \citenamefont {Angelatos},\ and\ \citenamefont
  {Türeci}}]{Khan2021a}%
  \BibitemOpen
  \bibfield  {author} {\bibinfo {author} {\bibfnamefont {S.~A.}\ \bibnamefont
  {Khan}}, \bibinfo {author} {\bibfnamefont {F.}~\bibnamefont {Hu}}, \bibinfo
  {author} {\bibfnamefont {G.}~\bibnamefont {Angelatos}},\ and\ \bibinfo
  {author} {\bibfnamefont {H.~E.}\ \bibnamefont {Türeci}},\ }\href
  {https://doi.org/10.48550/ARXIV.2110.13849} {\bibinfo {title} {Physical
  reservoir computing using finitely-sampled quantum systems}} (\bibinfo {year}
  {2021}),\ \Eprint {https://arxiv.org/abs/2110.13849} {arXiv:2110.13849
  [quant-ph]} \BibitemShut {NoStop}%
\bibitem [{\citenamefont {Lukoševičius}\ and\ \citenamefont
  {Jaeger}(2009)}]{LUKOSEVICIUS2009127}%
  \BibitemOpen
  \bibfield  {author} {\bibinfo {author} {\bibfnamefont {M.}~\bibnamefont
  {Lukoševičius}}\ and\ \bibinfo {author} {\bibfnamefont {H.}~\bibnamefont
  {Jaeger}},\ }\bibfield  {title} {\bibinfo {title} {Reservoir computing
  approaches to recurrent neural network training},\ }\href
  {https://doi.org/https://doi.org/10.1016/j.cosrev.2009.03.005} {\bibfield
  {journal} {\bibinfo  {journal} {Computer Science Review}\ }\textbf {\bibinfo
  {volume} {3}},\ \bibinfo {pages} {127} (\bibinfo {year} {2009})}\BibitemShut
  {NoStop}%
\bibitem [{\citenamefont {Konkoli}(2017)}]{Konkoli2017}%
  \BibitemOpen
  \bibfield  {author} {\bibinfo {author} {\bibfnamefont {Z.}~\bibnamefont
  {Konkoli}},\ }\bibinfo {title} {On reservoir computing: From mathematical
  foundations to unconventional applications},\ in\ \href
  {https://doi.org/10.1007/978-3-319-33924-5_23} {\emph {\bibinfo {booktitle}
  {Advances in Unconventional Computing: Volume 1: Theory}}},\ \bibinfo
  {editor} {edited by\ \bibinfo {editor} {\bibfnamefont {A.}~\bibnamefont
  {Adamatzky}}}\ (\bibinfo  {publisher} {Springer International Publishing},\
  \bibinfo {address} {Cham},\ \bibinfo {year} {2017})\ pp.\ \bibinfo {pages}
  {573--607}\BibitemShut {NoStop}%
\bibitem [{\citenamefont {Adesso}\ \emph {et~al.}(2014)\citenamefont {Adesso},
  \citenamefont {Ragy},\ and\ \citenamefont {Lee}}]{adesso2014}%
  \BibitemOpen
  \bibfield  {author} {\bibinfo {author} {\bibfnamefont {G.}~\bibnamefont
  {Adesso}}, \bibinfo {author} {\bibfnamefont {S.}~\bibnamefont {Ragy}},\ and\
  \bibinfo {author} {\bibfnamefont {A.~R.}\ \bibnamefont {Lee}},\ }\bibfield
  {title} {\bibinfo {title} {Continuous variable quantum information: Gaussian
  states and beyond},\ }\href {https://doi.org/10.1142/s1230161214400010}
  {\bibfield  {journal} {\bibinfo  {journal} {Open Systems \& Information
  Dynamics}\ }\textbf {\bibinfo {volume} {21}},\ \bibinfo {pages} {1440001}
  (\bibinfo {year} {2014})}\BibitemShut {NoStop}%
\bibitem [{\citenamefont {Serafini}(2017)}]{serafini2017}%
  \BibitemOpen
  \bibfield  {author} {\bibinfo {author} {\bibfnamefont {A.}~\bibnamefont
  {Serafini}},\ }\href@noop {} {\emph {\bibinfo {title} {Quantum continuous
  variables: a primer of theoretical methods}}}\ (\bibinfo  {publisher} {CRC
  Press},\ \bibinfo {year} {2017})\BibitemShut {NoStop}%
\bibitem [{\citenamefont {Genoni}\ \emph {et~al.}(2016)\citenamefont {Genoni},
  \citenamefont {Lami},\ and\ \citenamefont
  {Serafini}}]{conditional_unconditional_gaussian}%
  \BibitemOpen
  \bibfield  {author} {\bibinfo {author} {\bibfnamefont {M.~G.}\ \bibnamefont
  {Genoni}}, \bibinfo {author} {\bibfnamefont {L.}~\bibnamefont {Lami}},\ and\
  \bibinfo {author} {\bibfnamefont {A.}~\bibnamefont {Serafini}},\ }\bibfield
  {title} {\bibinfo {title} {Conditional and unconditional gaussian quantum
  dynamics},\ }\href {https://doi.org/10.1080/00107514.2015.1125624} {\bibfield
   {journal} {\bibinfo  {journal} {Contemporary Physics}\ }\textbf {\bibinfo
  {volume} {57}},\ \bibinfo {pages} {331} (\bibinfo {year} {2016})},\ \Eprint
  {https://arxiv.org/abs/https://doi.org/10.1080/00107514.2015.1125624}
  {https://doi.org/10.1080/00107514.2015.1125624} \BibitemShut {NoStop}%
\bibitem [{\citenamefont {Eisert}\ \emph {et~al.}(2002)\citenamefont {Eisert},
  \citenamefont {Scheel},\ and\ \citenamefont
  {Plenio}}]{PhysRevLett.89.137903}%
  \BibitemOpen
  \bibfield  {author} {\bibinfo {author} {\bibfnamefont {J.}~\bibnamefont
  {Eisert}}, \bibinfo {author} {\bibfnamefont {S.}~\bibnamefont {Scheel}},\
  and\ \bibinfo {author} {\bibfnamefont {M.~B.}\ \bibnamefont {Plenio}},\
  }\bibfield  {title} {\bibinfo {title} {Distilling gaussian states with
  gaussian operations is impossible},\ }\href
  {https://doi.org/10.1103/PhysRevLett.89.137903} {\bibfield  {journal}
  {\bibinfo  {journal} {Phys. Rev. Lett.}\ }\textbf {\bibinfo {volume} {89}},\
  \bibinfo {pages} {137903} (\bibinfo {year} {2002})}\BibitemShut {NoStop}%
\bibitem [{Note1()}]{Note1}%
  \BibitemOpen
  \bibinfo {note} {The notation $\protect \text {Tr}_{\protect \mathbf
  {p}}\left [ \bullet \right ]$ does not actually stand for the usual partial
  trace of a matrix, it is just a way of writing we are tracing out the
  components of the covariance matrix and first-moment vector which contain
  information of the $p$-quadratures of every mode. For a generic
  $2N$-dimensional covariance matrix, $\sigma $, tracing out these components
  would yield a $N$-dimensional matrix with components: \begin {equation} \left
  [ \protect \text {Tr}_{\protect \mathbf {p}}\left ( \sigma \right ) \right
  ]_{ij} = \left \langle \protect \hat {x}_{i} \protect \hat {x}_{j} \right
  \rangle - \left \langle \protect \hat {x}_{i} \right \rangle \left \langle
  \protect \hat {x}_{j} \right \rangle \ , \end {equation} where the mean
  values stand for the quantum expected values of the observables for a given
  quantum state.}\BibitemShut {Stop}%
\bibitem [{\citenamefont {Braunstein}(2005)}]{bloch-messiah}%
  \BibitemOpen
  \bibfield  {author} {\bibinfo {author} {\bibfnamefont {S.~L.}\ \bibnamefont
  {Braunstein}},\ }\bibfield  {title} {\bibinfo {title} {Squeezing as an
  irreducible resource},\ }\href {https://doi.org/10.1103/PhysRevA.71.055801}
  {\bibfield  {journal} {\bibinfo  {journal} {Phys. Rev. A}\ }\textbf {\bibinfo
  {volume} {71}},\ \bibinfo {pages} {055801} (\bibinfo {year}
  {2005})}\BibitemShut {NoStop}%
\bibitem [{\citenamefont {Cariolaro}\ and\ \citenamefont
  {Pierobon}(2016)}]{PhysRevA.93.062115}%
  \BibitemOpen
  \bibfield  {author} {\bibinfo {author} {\bibfnamefont {G.}~\bibnamefont
  {Cariolaro}}\ and\ \bibinfo {author} {\bibfnamefont {G.}~\bibnamefont
  {Pierobon}},\ }\bibfield  {title} {\bibinfo {title} {Reexamination of
  {Bloch-Messiah} reduction},\ }\href
  {https://doi.org/10.1103/PhysRevA.93.062115} {\bibfield  {journal} {\bibinfo
  {journal} {Phys. Rev. A}\ }\textbf {\bibinfo {volume} {93}},\ \bibinfo
  {pages} {062115} (\bibinfo {year} {2016})}\BibitemShut {NoStop}%
\bibitem [{\citenamefont {Mujal}\ \emph
  {et~al.}(2021{\natexlab{b}})\citenamefont {Mujal}, \citenamefont {Nokkala},
  \citenamefont {Mart{\'{\i}}nez-Pe{\~{n}}a}, \citenamefont {Giorgi},
  \citenamefont {Soriano},\ and\ \citenamefont
  {Zambrini}}]{Mujal_nonlinearity}%
  \BibitemOpen
  \bibfield  {author} {\bibinfo {author} {\bibfnamefont {P.}~\bibnamefont
  {Mujal}}, \bibinfo {author} {\bibfnamefont {J.}~\bibnamefont {Nokkala}},
  \bibinfo {author} {\bibfnamefont {R.}~\bibnamefont
  {Mart{\'{\i}}nez-Pe{\~{n}}a}}, \bibinfo {author} {\bibfnamefont {G.~L.}\
  \bibnamefont {Giorgi}}, \bibinfo {author} {\bibfnamefont {M.~C.}\
  \bibnamefont {Soriano}},\ and\ \bibinfo {author} {\bibfnamefont
  {R.}~\bibnamefont {Zambrini}},\ }\bibfield  {title} {\bibinfo {title}
  {Analytical evidence of nonlinearity in qubits and continuous-variable
  quantum reservoir computing},\ }\href
  {https://doi.org/10.1088/2632-072x/ac340e} {\bibfield  {journal} {\bibinfo
  {journal} {Journal of Physics: Complexity}\ }\textbf {\bibinfo {volume}
  {2}},\ \bibinfo {pages} {045008} (\bibinfo {year}
  {2021}{\natexlab{b}})}\BibitemShut {NoStop}%
\end{thebibliography}
\end{document}